\documentclass[lettersize,journal]{IEEEtran}
\usepackage{amsmath,amsfonts}
\usepackage{algorithmic}
\usepackage{algorithm}
\usepackage{array}
\usepackage[caption=false,font=normalsize,labelfont=sf,textfont=sf]{subfig}
\usepackage{textcomp}
\usepackage{stfloats}
\usepackage{url}
\usepackage{verbatim}
\usepackage{graphicx}
\usepackage{setspace}
\usepackage{cite}
\usepackage{multirow}
\usepackage{makecell}
\usepackage{tabularx}
\usepackage{booktabs} 
\usepackage{adjustbox}
\usepackage{threeparttable}
\usepackage[table]{xcolor}
\usepackage{enumitem}
\begin{document}

\title{Beyond the Edge: An Advanced Exploration of Reinforcement Learning for Mobile Edge Computing, its Applications, and Future Research Trajectories}

    
\author{Ning Yang, Shuo Chen, Haijun Zhang*,~\IEEEmembership{Fellow,~IEEE}, Randall Berry, \IEEEmembership{Fellow,~IEEE}

\thanks{Ning Yang is with the Institute of Automation, Chinese Academy of Sciences, Beijing, 100190, China. (e-mail: ning.yang@ia.ac.cn).

Shuo Chen is with the Department of Electrical and Electronic Engineering, Imperial College London, London, SW72BX, UK. (e-mail: shuo.chen22@imperial.ac.uk).

Haijun Zhang is with the Department of Computing and Communication Engineering, Beijing University of Science and Technology, Beijing, 100083, China. (e-mail: zhanghaijun@ustb.edu.cn).

Randall Berry is with the Department of Electrical and Computer Engineering, Northwestern University, Chicago, 60208, USA. (e-mail: rberry@northwestern.edu).

({*}Corresponding author: Haijun Zhang)
}}

\maketitle

\begin{abstract}
\textit{Mobile Edge Computing} (MEC) broadens the scope of computation and storage beyond the central network, incorporating edge nodes close to end devices. This expansion facilitates the implementation of large-scale ``connected things" within edge networks. The advent of applications necessitating real-time, high-quality service presents several challenges, such as low latency, high data rate, reliability, efficiency, and security, all of which demand resolution. The incorporation of \textit{reinforcement learning} (RL) methodologies within MEC networks promotes a deeper understanding of mobile user behaviors and network dynamics, thereby optimizing resource use in computing and communication processes. This paper offers an exhaustive survey of RL applications in MEC networks, initially presenting an overview of RL from its fundamental principles to the latest advanced frameworks. Furthermore, it outlines various RL strategies employed in offloading, caching, and communication within MEC networks. Finally, it explores open issues linked with software and hardware platforms, representation, RL robustness, safe RL, large-scale scheduling, generalization, security, and privacy. The paper proposes specific RL techniques to mitigate these issues and provides insights into their practical applications.
\end{abstract}

\begin{IEEEkeywords}
Reinforcement learning, mobile edge computing, offloading scheduling, content caching, and communication.
\end{IEEEkeywords}

\section{Introduction}

\textit{Internet of things} (IoT) has given rise to a significant number of applications and integrates a wide range of heterogeneous wireless-enabled devices, leading to exponential growth in network traffic and data volumes. However, IoT devices suffer from limited resources and finite battery capacity, which poses requirements to new communication technologies and computing paradigms.
On the other hand, the current communication networks have the benefits of high-quality network transmission with peak rate at Tbit/s, 10-100 Gbit/s experienced rate, and sub-millisecond level latency \cite{tong_zhu_2021}, which can support seamless connectivity of a more significant amount of IoT devices \cite{yang2020artificial}. Therefore, wireless communication technologies can better fulfill the users' \textit{quality of service} (QoS) and can support the demand for more applications. Cloud computing places computing, storage, and resource management in the cloud in a centralized manner. However, additional latency arises from various stages, such as the communication of mobile backhaul and the mobile core operator \cite{hu2015mobile}. Therefore, a new computing paradigm is required.

\subsection{The Road towards Mobile Edge Computing}
 \textit{Mobile edge computing} (MEC) is driven by the development of IoT and wireless communication technologies, whose vision is to move servers to edges that are close to edge users. Edge servers conduct computing and caching at the network's edge to reduce transmission latency and congestion. The \textit{European Telecommunications Standard Institute} (ETSI) first proposed a concept of the MEC in 2014 \cite{MEC2014white}. In 2017, ETSI \textit{Industry Specification Group} (ISG) renamed it as \textit{multi-access edge computing} \cite{porambage2018survey}. MEC, combined with next-generation networks, provides real-time, low-latency, and high-throughput servers, supporting many novel services such as autonomous driving, stream gaming, \textit{virtual reality} (VR), \textit{augmented reality} (AR), remote healthcare, etc. 

The requirements of applications are low latency (sub-millisecond level), high data volumes ($>$1 Gbps), scalability, traffic loads, and security \cite{6gmec} in MEC networks. For instance, vehicles send data collected from sensors to process and ensure the driver's safety in autonomous driving scenarios. VR/AR applications with high-definition stream video occupy vast amounts of the uplink/downlink bandwidth, which easily causes network congestion and high latency. Remote surgery shows demands for real-time operation, or delayed instructions can threaten patients' lives. These emerging applications and the new requirements motivate us to consider changes for model designing, analyzing, and optimization in the MEC network. Although \textit{the 5th generation mobile communication technology} (5G) and other wireless network infrastructures are fixed, it needs to integrate them harmoniously, instilling intelligent resource management techniques across both the core and the edge of the network.
\begin{figure*}
    \centering
    \includegraphics[width=160mm]{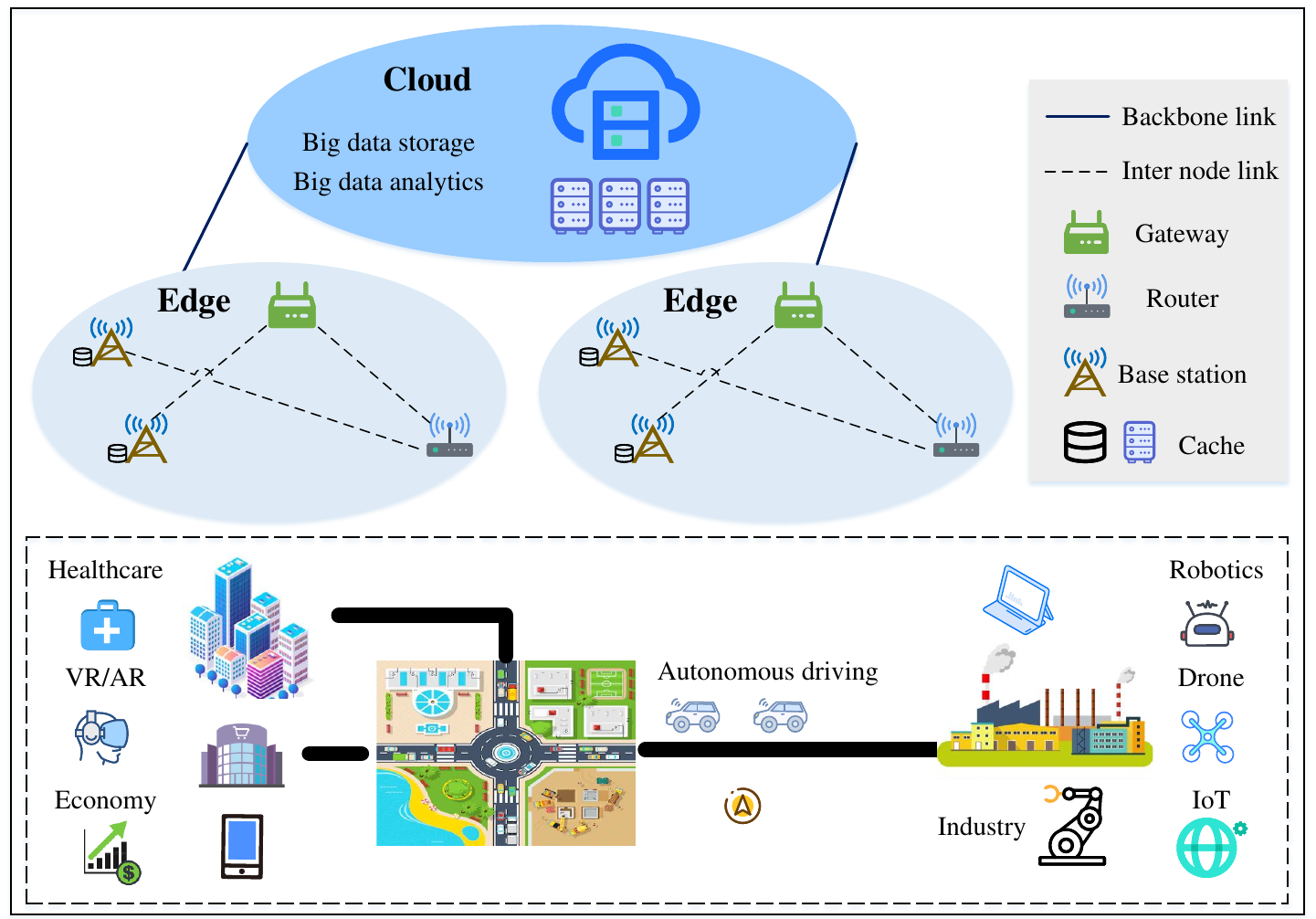}
    \caption{The network architecture and applications of MEC.}
    \label{fig:system model}
\end{figure*}
This work mainly focuses on the following three aspects where intelligent resource management techniques can be implemented in MEC networks: (1) task offloading; (2) content caching; and (3) communication.

\begin{itemize}
    \item Task offloading refers to transferring computing tasks to remote edge servers to process. An efficient offloading scheduling strategy significantly reduces edge devices' latency and energy consumption. The main issues for task offloading include what to offload, where to offload, how to offload, and when to offload. The decision to offload will result in local execution and edge computing. Tasks can either be fully offloaded to an edge server or divided into two parts, and computation runs locally and at the edge server or the cloud server. 
    \item
    Caching at the mobile edge uses edge servers' storage to cache popular contents that are requested by users at the edge. Due to the proximity of edge users to mobile servers, caching reduces backhaul network traffic, transmission latency, and energy consumption for edge users while also improving data rates. This paper examines various caching schemes and caching problems, including caching location, content, timing, and decision, which can impact hit probability, spectrum efficiency, energy efficiency, and retrieval delay.
    \item
    Communication technologies play a crucial role in the MEC network, enabling MEC applications to benefit from low latency and high reliability through effective communication resource scheduling. This paper examines communication between edge users and network servers, as well as communication among users. Primary tasks for communication resource scheduling include spectrum access, spectrum allocation, and power control and allocation. The high number of users accessing the spectrum results in spectrum scarcity and network congestion, negatively impacting real-time system performance. Scheduling policies for spectrum access and power allocation significantly influence communication efficiency in MEC networks. Spectrum allocation, as a specific form of spectrum access, requires effective schemes that can adapt to the highly dynamic environment. Power control aims to meet the demands of high-throughput applications and to guarantee QoS.

\end{itemize}

A MEC network architecture can realize MEC services' intelligence, automation, and criticality, as illustrated in Fig. \ref{fig:system model}. It consists of a cloud network and multi-edge networks. In the cloud network, it deploys large-scale caching and computing resources that perform computationally intensive tasks. Edge networks are equipped with \textit{base stations} (BSs), \textit{radio access networks} (RANs), routers, switches, \textit{WiFi access points} (WAPs), and gateways, which provide computing and caching access for devices. 

\subsection{Role of Reinforcement Learning in MEC}

Motivated by the needs of the MEC network, wireless communication technologies envision the integration of edge AI, involving \textit{machine learning} (ML) and \textit{deep neural networks} (DNN), into the MEC network. This integration enables proactive prediction of the complex environment and enhances network performance. The integration of AI technology into the MEC network facilitates network resource monitoring and management and enables high energy efficiency in intelligent applications. This convergence holds great promise and is of significant importance in emerging research.

MEC has become an important research area due to the increasing demand for low-latency and high-bandwidth applications, and \textit{reinforcement learning} (RL) is one of the crucial techniques to make intelligent decisions in the MEC networks \cite{Liu2022IRS}. RL has shown great potential in improving resource utilization efficiency in MEC networks by learning from experience and adapting to changing conditions. While fast learning dynamics cannot always be expected when solving non-convex optimization problems with RL, it has shown promising results in various applications and scenarios. RL can handle complex and dynamic environments, which may be challenging for classical black-box optimization methods, such as genetic algorithms, simulated annealing, semidefinite relaxation, etc. Additionally, RL can learn from experience and adapt to changing conditions. It is essential to compare the performance of RL with classical black-box optimization methods for non-convex optimization problems in MEC networks. RL can handle complex and dynamic environments that are common in MEC networks, making it a valuable tool for solving optimization problems in this context. While both approaches have their strengths and weaknesses depending on the specific problem being solved, RL has shown great potential in handling complex and dynamic environments that are common in MEC networks. The motivations for applying RL to MEC networks are as follows.

Firstly, optimization methods that derive policies using a pre-planning mode may not suit real environments. RL agents can conduct decisions adaptively and apply them to dynamic and uncertain MEC networks (i.e., dynamic task requests and time-varying channel gain). There are three main reasons: (1) Agents interact with the environment and enhance dynamic awareness and predictive environment uncertainty; (2) RL methods generally maximize a long-term reward, which estimates the influence of a dynamic environment on the reward; (3) Certain RL methods rely on precise mathematical models of the environment, whereas, in contrast, model-free RL methods do not, rendering them valuable for handling environments characterized by high dynamics or uncertainty. Nevertheless, implementing model-free RL in practical settings can be challenging due to suboptimal decisions made during the training process prior to convergence. Within Section VI-A, various potential solutions to address this issue are explored, such as utilizing simulators or progressively integrating the RL algorithm into the actual system.
 
Secondly, classic methods like dynamic programming and stochastic optimization make it impossible to obtain a real-time decision. They utilize value recursion that is computationally intensive and memory-consumptive. Besides, dynamic programming needs the exact information of the whole environment and gives the optimal policy by retrospecting each possible state after the current states. However, online implementation is an essential trend in intelligent MEC networks. RL algorithms obtain sequential decisions solely on the current states of agents. RL methods have relatively fast inference speeds that can support real-time applications with stringent low-latency requirements. 
 
Thirdly, most computing and communication resource allocation problems in MEC networks are non-convex problems. Current methods first reformulate the non-convex problem to a convex problem and then apply the conventional optimization algorithms to find the suboptimal solution. However, once the number of devices or edge servers exceeds a specific number, it is challenging to obtain a closed-form solution. RL executes the end-to-end optimization solving non-convex optimization problems directly. It uses a black-box solver integrating state information blocks into neural network architectures as input layers so that the output of the resource allocation policy. 

Therefore, RL plays an essential role in the applications of MEC networks:
\begin{itemize}
    \item RL methods lead mobile users and servers to actively learn and predict the complex wireless network environment through a large amount of generated data while keeping real-time decision-making;
    \item RL-empowered mobile devices can adapt to the changing environment and dynamically optimize resource allocation with excellent performance.
\end{itemize}

\subsection{Existing Surveys}

RL has become a necessary tool for edge AI in 6G networks. AI-enabled MEC research focuses on network modeling, edge training, edge inference, resource management algorithms, and implementations and applications for MEC networks, in which many opportunities exist that lead to an active area. However, few surveys have full coverage of RL-empowered MECs and their potential.

Several surveys have examined {\bf ML and DL} methodologies as tools to address challenges within MEC networks.
In the survey \cite{shakaramiSurveyComputationOffloading2020a}, the authors scrutinized ML-based strategies for computation offloading in MEC, crafting a detailed taxonomy of these strategies based on recent research and delving into their applications. Nevertheless, the authors mainly focused on challenges linked to task offloading, constraining the scope of applications and prospective research directions. Furthermore, the limited space prevented an exhaustive exploration of research enabled by recent RL methods.
The survey conducted by \cite{mldlMEC} presents a thorough investigation into the application of ML and \textit{deep learning} (DL) techniques within the realm of wireless communication, with a specific focus on MEC. Covering a diverse range of methodologies, the study explores various ML and DL approaches, introducing tailored strategies for optimizing different stages of wireless communication. The authors delve into ML/DL-based techniques for task offloading, scheduling, and joint resource allocation in MEC, providing a comprehensive overview of recent works. The paper's notable contributions include tutorials showcasing the advantages of ML and DL in MEC, insights into enabling technologies for efficient ML/DL training and inference in MEC, and an in-depth survey of challenges and future directions in ML/DL-based resource allocation. Despite the acknowledged significance of RL in addressing resource allocation issues, the paper focuses on a limited selection of RL-based schemes, leaving room for further exploration in this area.
In \cite{letaief2021edge}, it presented a comprehensive examination of edge AI, highlighting its key attributes such as the integrated design of scalable, decentralized, and trustworthy AI technologies within MEC systems for 6G. Furthermore, emphasis was placed on advancing the productization and commercialization of edge AI within the prospective domains of future MEC. Nevertheless, there is a need for additional details regarding AI technologies and a more thorough exploration of MEC tasks to enhance the depth and specificity of the discussion.

Some surveys focused on the variety of {\bf RL techniques} in wireless networks. The investigation documented in \cite{RL6GMECSurvey} revealed strategies rooted in RL within the context of MEC, with a specific focus on challenges arising from the unrestricted mobility of connected devices, time-varying channels, and distributed service requirements. In addition to presenting fundamental RL principles, the survey also offered insights into approaches for mitigating challenges posed by network constraints and dynamics. Nevertheless, there remains an opportunity for a more expansive discourse on the unresolved challenges within this domain.
Both studies \cite{liApplicationsMultiAgentReinforcement2022a,ferianiSingleMultiAgentDeep2021b} primarily concentrated on the challenges of complex multi-agent MEC systems using \textit{multi-agent reinforcement learning} (MARL) strategies, underlining the importance of cooperative MARL strategies and showcasing comprehensive research on game theory-based and online RL methods. These studies addressed several open issues, future challenges, and the practical implementation difficulties of MARL strategies. However, the central focus of these articles is RL strategies, as opposed to a comprehensive examination of MEC applications.

Others delved into the {\bf applications} of RL within specific domains of MEC.
The research conducted by \cite{RLIoTSurvey} systematically assessed the utilization of \textit{deep reinforcement learning} (DRL) methodologies in various IoT scenarios. The study initially provided a comprehensive categorization of RL techniques rooted in a single-agent paradigm, with subsequent incorporation of innovative RL algorithms. Given the expanding influence of RL in MEC, there is a necessity to enhance the existing classification and delve deeper into additional facets of forthcoming challenges.
Luong \textit{et al.} \cite{DRL8714026} furnished an overview of the applications of deep RL strategies in wireless networks, introducing diverse DNN architectures and addressing several issues related to wireless network scheduling. However, additional recent advancements in novel neural network architectures have been applied to wireless network scheduling. Moreover, focusing solely on communication does not yield a comprehensive understanding of MEC systems.
Several surveys explore specific aspects of MEC. Uprety \textit{et al.} \cite{uprety2020reinforcement} conducted a meticulous examination of integrating RL to address security challenges in IoT devices. The authors detailed primary attack types within the IoT landscape, encompassing denial-of-service attacks, jamming attacks, and spoofing attacks. The survey provided a comprehensive review of existing research utilizing RL to mitigate these attacks, with a notable emphasis on its application in real-world scenarios such as smart grids and intelligent transportation systems, addressing a diverse array of practical challenges. However, it is important to state that the discussion on future directions was concise, lacking an in-depth exploration of potential techniques and solutions to guide future research in this dynamically evolving domain.
Lei \textit{et al.} \cite{lei2020deep} focused their study on the utilization of RL in autonomous IoT systems, systematically addressing challenges across three distinct layers: the perception layer, network layer, and application layer. The authors proceeded to present a comprehensive RL formulation tailored to each specific layer, establishing a structured framework for analysis. The survey introduced various works aligned with the evolution of RL techniques, enabling a thorough comparison across diverse problems and RL schemes. Toward the paper's conclusion, the authors engaged in discussions on practical scenarios and outlined potential future research directions, providing valuable insights. However, it is worth noting that the exploration of potential solutions was only briefly touched upon, indicating an opportunity for more in-depth considerations in advancing RL applications within autonomous IoT environments.

Table \ref{papercompare} compares the concentration among the surveys above.

\begin{table*}[t]

    \centering
\caption{ \textcolor{black}{Symmary of Existing Surveys on RL Applications in MEC.}}\label{papercompare}
    \begin{tabular}{m{1.5cm}<{\centering} |c|c| c|c|c |m{1.3cm}<{\centering}|m{4.5cm}}
    \hline\hline
    \multirow{2}{*}{\textbf{Ref.}} & \multicolumn{2}{c|}{\textbf{RL techniques}} &\multicolumn{3}{c|}{\textbf{Resource allocation aspects}} &\textbf{Future directions} &\textbf{Focus of discussion}\\
    \cline{2-7}
    &SARL&MARL&Offloading&Caching&Communication&Technical analysis&\\
    \hline
    \cite{shakaramiSurveyComputationOffloading2020a} &\checkmark&&\checkmark&&\checkmark&&Give a detailed taxonomy of ML methods on offloading problems \\
    \hline
    \cite{mldlMEC} &\checkmark&&\checkmark&\checkmark&\checkmark&\checkmark& ML/DL edge resource allocation and ML/DL empowered resource allocation in MEC.\\
    \hline
    \cite{letaief2021edge} &\checkmark&\checkmark&&&\checkmark&& Scalable and trustworth edge AI methods. Wireless communication and distributed learning are widely discussed within 6G proposals. \\
    \hline
    \cite{RL6GMECSurvey}&\checkmark&&&&\checkmark&\checkmark&RL-enabled MEC. Focusing on the capability of dealing with dynamics and uncertainty.\\
    \hline
    \cite{liApplicationsMultiAgentReinforcement2022a}&\checkmark&\checkmark&\checkmark&\checkmark&\checkmark&&Utilization of MARL in resource allocation, vehicle, and UAV, and its efficiency compared with SARL.\\
    \hline
    \cite{ferianiSingleMultiAgentDeep2021b}&\checkmark&\checkmark&&&\checkmark&& RL applications in wireless network. Vision for model-based RL and cooperative MARL in future networks.\\
    \hline
    \cite{RLIoTSurvey}  &\checkmark&\checkmark&\checkmark&\checkmark&\checkmark&& DRL approaches for IoT in control, computing, caching, and communication. \\
    \hline
    \cite{DRL8714026}&\checkmark&&&&\checkmark&\checkmark&DRL-empowered IoT and UAV in communication and networking. \\
    \hline   
    \cite{uprety2020reinforcement}&\checkmark&\checkmark&&&\checkmark&&Layered classification for RL-empowered security IoT addressing network attacks.\\
    \hline
    \cite{lei2020deep}&\checkmark&\checkmark&\checkmark& &\checkmark& & RL for autonomous IoT. The general RL models are discussed from different layers.\\
    \hline
    
    This paper&\checkmark&\checkmark&\checkmark&\checkmark&\checkmark&\checkmark&\\
    \hline
    \end{tabular}
\end{table*}

\subsection{Key Contributions and Organization of This Paper}
The exploration of how novel RL strategies can be employed to address future challenges is seldom detailed, and the discussion on future directions requires further expansion. This study is distinguished by three main contributions:

\begin{itemize}

\item[$\bullet$]
Firstly, we provide a comprehensive introduction to RL. We initially introduce the fundamental concepts like \textit{Markov decision processes} (MDP), \textit{partially observable Markov decision processes} (POMDP), and Decentralized POMDP, offering a mathematical model and problem formulation for general decision-making problems. We then present traditional strategies such as \textit{dynamic programming} (DP), \textit{Monte Carlo} (MC), and \textit{temporal difference} (TD) to address these problems. Additionally, we offer various taxonomies of RL methods and discuss the advantages and limitations of each class. We also introduce the concepts of \textit{single-agent reinforcement learning} (SARL), MARL, and recent advancements in RL, elaborating on their motivation, characteristics, advantages, and disadvantages. This comprehensive overview aims to advance the application of RL in MEC networks in both academic and industrial settings.

\item[$\bullet$]
The principal sections of Sections IV and V delve into RL-based algorithms from the viewpoint of three typical resource allocations: task offloading, content caching, and communication (discussed in Section IV). We explore their applications in industrial IoT, autonomous driving, robotics, VR/AR, healthcare, and the tactile internet. Unlike existing surveys, we summarize the advancements and limitations of RL applications in each of these aspects individually while also providing a comparative analysis across these aspects. We answer critical questions such as why a state space is designed, how a reward function is set, why a neural network is included, and why certain modifications are made. By concentrating on essential problems, this study offers a comprehensive and detailed reference for practitioners and researchers.

\item[$\bullet$]
Lastly, most studies primarily explore the underlying communication challenges and opportunities concerning future directions without proposing specific RL solutions. In contrast, this study introduces additional open issues and future challenges from a broader range of perspectives, covering real-world software and hardware platforms, demands for improvements in RL methods, high adaptability of RL applications in MEC, and issues concerning security and privacy. We also propose RL strategies to tackle these problems and provide insights into the practicality of these methods. Thus, this study aims to bridge the gap between simulation and reality (sim-to-real) in the deployment of RL in MEC.
\end{itemize}

This article surveys the literature over the period 2012-2023 on RL for MEC. The paper is organized as follows:
\begin{itemize} 
   \item[$\bullet$]
    The vision of MEC and the role of RL in MEC (i.e., advantages and motivations) are introduced in Section I.

    \item[$\bullet$]
    The main challenges in the MEC network are presented in Section II, such as massive connections, latency, limited resources, dynamic uncertainty, and privacy. 

    \item[$\bullet$]
    A comprehensive overview of RL techniques is presented in Section III. The basic concepts, such as the mathematical formulation of single-agent and multi-agent decision problems, the architecture of SARL and MARL approaches, and the advantages and motivations for introducing these methods are discussed.

    \item[$\bullet$]
    In Section IV, we examine RL solutions for three problems in the MEC network: offloading scheduling, content caching, and communication. We propose a concluding subsection that provides a summary of the application of RL in these MEC tasks and distills information across these aspects.

    \item[$\bullet$]
    Several popular application scenarios for edge RL systems are proposed in Section V, including industrial IoT, autonomous driving, robotics, VR and AR, healthcare, and tactile internet. 
    
    \item[$\bullet$]
    Finally, in Section VI, the paper discusses the challenges and various future research prospects. It provides a detailed analysis of each subfield and guidance on leveraging RL methods to address these issues. This facilitates practitioners and the market for deploying RL in the MEC network.

\end{itemize}

\begin{figure*}
    \centering
    \includegraphics[width=180mm]{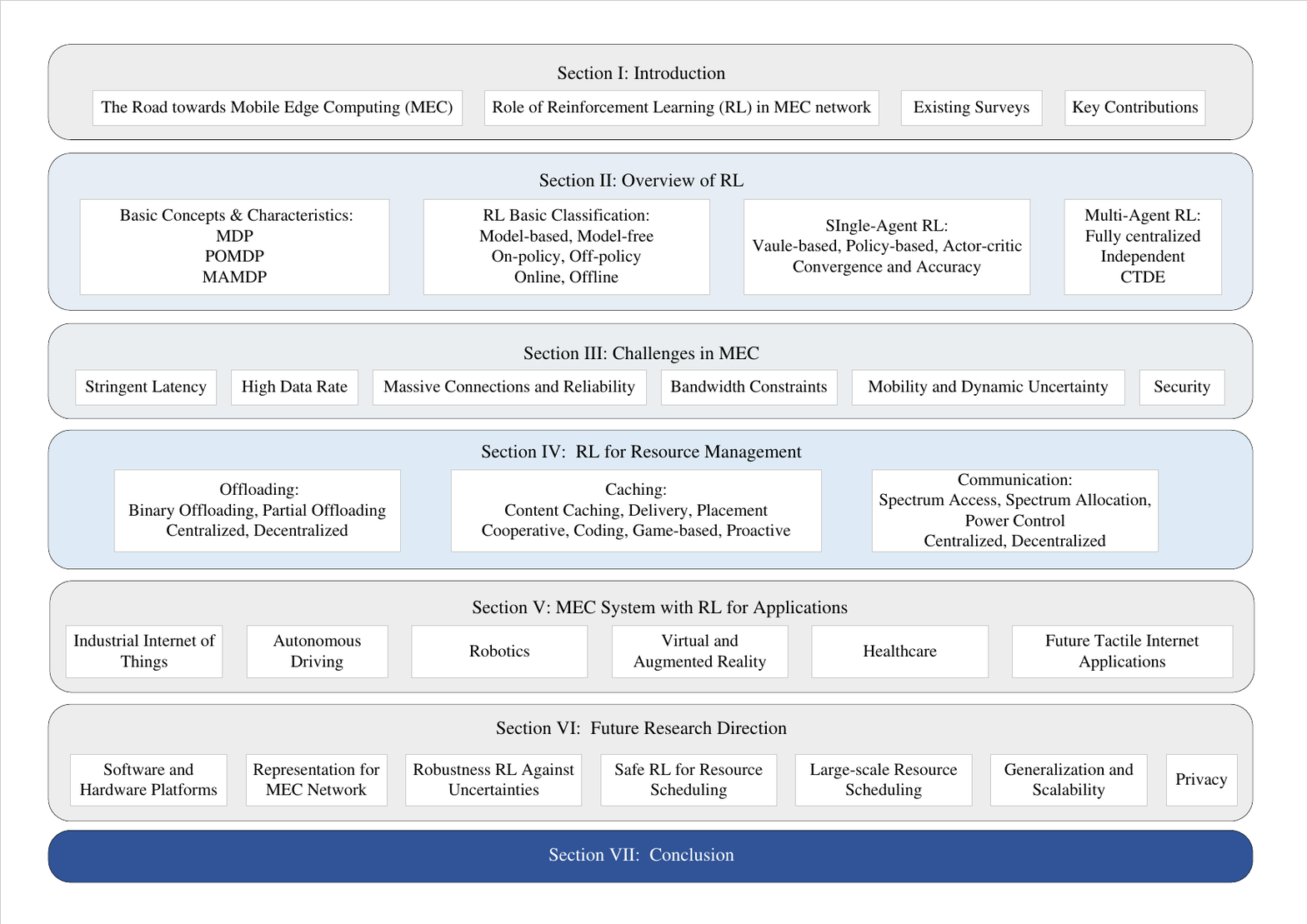}
    \caption{\textcolor{black}{The outline of the paper.}}
    \label{fig: outline of paper}
\end{figure*}

This outline of this paper is organized as Fig. \ref{fig: outline of paper}. The main acronyms in this paper are listed in Table \ref{table:acronyms}.

\renewcommand\arraystretch{1.2}
\begin{table*}[t]
\normalsize
\caption{List of Acronyms.}
\begin{center}
\begin{tabular}{m{2cm}m{13em}|m{2cm}m{13em}}
\hline\hline
{\bf Acronyms} & {\bf Definitions } & {\bf Acronyms}  & {\bf Definitions } \\
\hline
BicNet & bidirectionally-coordinated nets & MDP & Markov decision process\\
BSs & base stations & MEC & mobile edge computing \\
CommNet & communication neural net & meta-RL & meta reinforcement learning \\ 
CoMP & coordinated multipoint & ML & machine learning \\
CTDE & centralized training distributed execution & NOMA & non-orthogonal multiple access\\
D2D & device-to-device & OFDMA & orthogonal frequency division multiple access \\
DDPG & deep deterministic policy gradient & PEC & pervasive edge computing\\
DDQN & double DQN & POMDP & partially observable Markov decision process \\
DNN & deep neural network & PPO & proximal policy optimization\\
DP & dynamic programming & RANs & radio access networks \\  
DPS & direct policy search & RL & reinforcement learning \\
DQN & deep Q-network & RSU & roadside units\\
DRL & deep reinforcement learning & SAGIN & space-air-ground integrated network \\
DSA & dynamic spectrum access & SDN & software defined networking\\
GCN & graph convolutional network & SGD & stochastic gradient descent\\
GNN & graph neural network & TD & temporal difference\\
HetNets & heterogeneous network & TRPO & trust region policy optimization\\
IoT & internet of things & TS & Thompson sampling \\
IQL & independent Q-learning & UAV & unmanned aerial vehicle\\
LSTM & long short-term memory & UCB & upper confidence bound \\
MAB & multi-arm bandit & UDN & ultra-dense network\\
MADDPG & multi-agent deep deterministic policy gradient & VCG & Vickrey-Clarke-Groves\\
MC & Monte Carlo & WAPs & WiFi access points\\

\hline \hline

\end{tabular}
\end{center}\label{table:acronyms}
\end{table*}

\section{An Emerging Era of RL in MEC Systems}
RL has emerged as a powerful paradigm for training intelligent agents to make sequential decisions in dynamic environments. Before delving into the specific application of RL in MEC networks, it is essential to provide a brief overview of RL concepts and their diverse applications. RL encompasses critical mechanisms such as unsupervised learning, transfer learning, and hierarchical RL \cite{li2017deep}. Specifically, SARL and MARL represent RL paradigms where a solitary intelligent agent and multiple intelligent agents interact with their respective environments, respectively. The latter is commonly employed to model intricate scenarios involving multiple decision-makers, including multi-agent systems, game theory, and collective decision-making. In the context of MEC, RL, through dynamic learning and adaptive decision-making, achieves intelligent optimization across various aspects, including resource allocation and optimization, edge caching optimization, communication resource management, and task offloading.

In light of the aforementioned facets, the subsequent subsections within this survey expound upon the conceptual framework and characteristics of RL, as well as delineate the distinctive features, limitations, and advantages of SARL and MARL.
 
\subsection{\textcolor{black}{A Classification of RL Algorithms}}
\textcolor{black}{In this section, we will furnish fundamental knowledge and offer a general classification of RL algorithms, with a specific emphasis on MEC networks.}

\subsubsection{Basic Model}
\paragraph{MDP}
\textcolor{black}{To solve a decision-making problem, we need a generic mathematical model to describe the problem. The MDP is the basic model to characterize an RL problem. The next state depends only on the current state and action. We also introduce several some variant of MDP which covers more complicated conditions.}

\paragraph{MAB}
\textcolor{black}{\textit{Multi-arm bandit} (MAB) problems are a class of reinforcement learning problems that model the dilemma of choosing between exploiting known rewards and exploring potentially larger but unknown rewards.}

\paragraph{POMDP}
\textcolor{black}{POMDP extends MDPs to situations where the state of the environment is not fully observable.}

\paragraph{MAMDP}
\textcolor{black}{\textit{Multi-Agent Markov Decision Process} (MAMDP) extends MDPs to situations with multiple agents. They are used to model multi-agent systems, such as traffic systems and robotic teams.}

\subsubsection{Basic classification for RL}

\paragraph{Model-Based and Model-Free}
\textcolor{black}{Model-based reinforcement learning methods build a model of the environment and use it to plan future actions. Model-free methods, on the other hand, learn directly from experience without assuming knowledge of the environment. As most existing studies consider a model-free policy to learn the optimal policy, the classification of RL is based on a model-free baseline.}

\paragraph{On-Policy and Off-policy}
\textcolor{black}{On-policy methods learn the value of the policy being followed, while off-policy methods learn the value of a different policy, using transitions that are not under the current policy.}

\paragraph{Online and Offline}
\textcolor{black}{Online reinforcement learning methods learn while interacting with the environment, while offline methods learn from a fixed dataset of experience.}

\subsubsection{Single-Agent RL}

\paragraph{Value-Based}
\textcolor{black}{Value-based methods, such as Q-learning, estimate the value of each action in each state and use these estimates to make decisions.}

\paragraph{Policy-Based}
\textcolor{black}{Policy-based methods, such as policy gradients, directly optimize the policy without maintaining a value function.}

\paragraph{Actor-Critic}
\textcolor{black}{Actor-critic methods combine value-based and policy-based methods. The actor updates the policy in the direction suggested by the critic, which estimates the value function.}

\paragraph{Other Techniques}
\textcolor{black}{Here we also introduce some other techniques utilized in SARL.}
\begin{itemize}
    \item \textcolor{black}{\textit{Meta-RL:} Meta-RL is a framework that is designed to learn how to learn across a distribution of tasks. It can be applied to both single-agent and multi-agent settings and can use value-based, policy-based, or AC methods depending on the specific algorithm used. It’s not strictly a type of RL, but rather a way of applying RL to a certain kind of problem.}

    \item \textcolor{black}{\textit{Hierarchical RL:} Hierarchical RL is an approach that structures the policy space in a hierarchical manner to handle complex tasks more efficiently. Like Meta-RL, it can be applied to both single-agent and multi-agent settings and can use value-based, policy-based, or AC methods depending on the specific algorithm used. It’s not strictly a type of RL, but rather a way of applying RL to a certain kind of problem.}
\end{itemize}

\subsubsection{Multi-Agent RL}

\paragraph{Fully Centralized}
 \textcolor{black}{In fully centralized methods, all agents share the same policy and value function. They cooperate to maximize a common reward.}
 
\paragraph{Independent}
\textcolor{black}{In independent methods, each agent learns its own policy and value function. The agents may compete or cooperate, depending on the environment.}

\paragraph{Centralized Learning Distributed Exectution}
\textcolor{black}{In CTDE methods, the agents are trained together but act independently. This allows for coordination during training but scalability during execution.}

\paragraph{Other Techniques}
\textcolor{black}{Here we also introduce some other techniques utilized in MARL.}

\begin{itemize}
    \item \textcolor{black}{\textit{Federated RL:} Federated RL is an emerging field in RL that leverages the basic idea of Federated Learning to improve the performance of RL while preserving data privacy. It’s a new method with great potential. According to the distribution characteristics of the agents in the framework, FRL algorithms can be divided into two categories: Horizontal Federated Reinforcement Learning (HFRL) and Vertical Federated Reinforcement Learning (VFRL). The existing works on FRL are summarized by application fields, including edge computing, communication, control optimization, and attack detection1.}

    \item \textcolor{black}{\textit{Game-Theory-Based RL:} Game-Theory-Based RL is a perspective of MARL that uses game theory to understand and design algorithms. Game theory provides a mathematical framework for modeling and analyzing situations where multiple individuals interact. In the context of RL, game theory can help understand how multiple learning agents interact and influence each other’s learning processes. For example, one framework casts Model-Based RL as a game between a policy player, which attempts to maximize rewards under the learned model, and a model player, which attempts to fit the real-world data collected by the policy player.}
\end{itemize}

\textcolor{black}{To conclude the classification of RL methods, we reference Table \ref{RLClass}.}

\renewcommand\arraystretch{1.2}
\begin{table*}[t]
\small
\caption{\textcolor{black}{RL Classifications.}}
\begin{center}
\resizebox{\linewidth}{!}{\begin{tabular}{m{0.6cm}|m{1.1cm}||m{1.6cm}|c|m{1.6cm}||c|m{1.6cm}<{\centering}|m{1.5cm}||c|m{1.5cm}<{\centering}|m{1.5cm}}
\hline\hline
\multicolumn{2}{c||}{\bf Basic Model} & \multicolumn{3}{c||}{\bf Basic Classification}  & \multicolumn{3}{c||}{\bf SARL} & \multicolumn{3}{c}{\bf MARL} \\
\hline
\multirow{6}{*}[-15ex]{MDP}&\multirow{2}{*}[-4ex]{MAB}&\multicolumn{2}{c|}{Types} &Approaches&Types&Approaches& Other Techniques&Types&Approaches&Other Techniques\\
\cline{3-5}\cline{6-8}\cline{9-11}
 & &\multirow{2}{*}[1ex]{Model-Based}& & Planning& Value-Based& Q-Learning, SARSA, DQN, DDQN&\multirow{5}{1.5cm}[-12ex]{Meta-RL, Hierarchical RL}& Independent& IQL, Dec-POMDP, MADDPG, MAPPO& \multirow{5}{1.5cm}[-12ex]{Federated RL, Game-Theory-Based RL}\\
 \cline{2-7}\cline{9-10}
  &\multirow{2}{*}[-6ex]{POMDP}&\multirow{4}{*}[-12ex]{Model-Free}&On-Policy& SARSA, A2C, PPO&\multirow{2}{*}[-5ex]{Policy-Based}&\hspace{-0.4cm}\multirow{2}{1.5cm}[-5ex]{\begin{tabular}{p{1.5cm}<{\centering}}
  {DPG, TRPO, PPO}
  \end{tabular}}& &\multirow{2}{1.5cm}[-5ex]{\centering{Fully Centralized}}& \hspace{-0.4cm}\multirow{2}{1.5cm}[-2ex]{\begin{tabular}{p{1.5cm}<{\centering}}
  Global critic network and action selection
  \end{tabular}}&\\
  \cline{4-5}
   & & &Off-Policy&Q-Learning, DQN, DDQN, Dueling DQN& & & & & &\\
\cline{2-2}\cline{4-7}\cline{9-10}
 & \multirow{2}{*}[-2ex]{MAMDP} & &Online &SARSA, Q-Learning, TD-Based Approaches &\multirow{2}{*}[-2ex]{Actor-Critic}&\hspace{-0.4cm}\multirow{2}{1.5cm}[-2ex]{\begin{tabular}{p{1.5cm}<{\centering}}
  {A2C, A3C, DDPG, TD3, SAC}
  \end{tabular}}& &\multirow{2}{*}[-2ex]{CTDE}& \multirow{2}{1.5cm}[-2ex]{\begin{tabular}{c}
   COMA, \\
   QMIX
 \end{tabular}}& \\
\cline{4-5}\
 
  & & &Offline &Importance Sampling, Inverse RL& & & & & \\

\hline\hline
\end{tabular}}
\end{center}\label{RLClass}
\end{table*}

\subsection{An Overview of RL: The Basic Concepts and Characteristics}\label{subsec:An Overview of RL: The Basic Concepts and Characteristic}

\subsubsection{Markov Decision Process}

\paragraph{Concept}
RL has shown great potential to address sequential decision-making problems. RL algorithms are based on the MDP framework, which mathematically formulates sequential decision-making problems. An MDP is represented as a tuple $\langle \mathcal S, \mathcal A, \mathcal T, \gamma, \mu, \mathcal R \rangle$ \cite{sutton2018reinforcement} in which:
\begin{itemize}
\item $\mathcal  S$ is a state space.
\item $\mathcal  A$ is an action space.
\item $ \mathcal {T :  S \times  A \to  S} $ is a probabilistic transition process.
\item $\gamma \in [0, 1]$ is a discount factor.
\item $\mu :\mathcal  S \to [0, 1]$ is a probability distribution over initial states.
\item $\mathcal {R: S \times A \times S} \to\mathbb{R}$ is a reward function.
\end{itemize}

During the process of decision-making, an entity engages with its surrounding context. This interaction can be represented in two different ways: as an episodic MDP that features a definite endpoint or as an infinite-horizon MDP that possesses an undefined temporal boundary, potentially evolving to infinity. At step $t$, the agent observes the current environment state $s_t \in \mathcal{S}$ and takes action $a_t \in \mathcal{A}$. The environment provides an immediate reward $r_t$ to the agent according to the reward function $\mathcal{R}$. After performing action $a_t$, the environment moves to the next step, and the state changes to $s_{t+1}$ according to the transition probability $\mathcal{T}(s_{t+1} \mid s_t, a_t)$. In the framework of episodic MDPs, an agent iteratively undertakes the decision-making process until a terminal point is achieved. Every episode commences from an initial state and terminates upon reaching a terminal state. Conversely, in the context of infinite-horizon MDPs, the agent perpetuates the decision-making process indefinitely without a predetermined termination point. Two components of an MDP model are the state space design and the reward function. A carefully designed state space incorporates comprehensive information to fully characterize the environment, which is crucial in assisting the agent in discerning the system dynamics accurately. The reward should directly reflect the utility of action $a_t$ in state $s_t$, denoted as reward $r_t$. The agent refines its policy through interaction trajectories and experiences, aiming to maximize a discounted cumulative reward\cite{mahadevan1996average}.

The goal of RL is to learn an optimal policy $\pi^*(a|s)$ that maximizes the cumulative reward. The cumulative reward can be expressed as:
\begin{equation} G_t=\sum^{\infty}_{k=0}\gamma^kr_{t+k+1}, \end{equation}
where $k$ represents the number of steps, and $\gamma\in[0,1]$ is the discount factor that balances the importance of immediate and future rewards. A discount factor closer to 1 indicates that the agent places greater emphasis on long-term rewards, while a discount factor closer to 0 means the agent emphasizes more on short-term rewards.

Maximizing the cumulative reward can be expressed as maximizing the state value function $V^*(s)$ or the action value function $Q^*(s,a)$:
\begin{equation}
    V^*(s)= \max\limits_\pi \ \mathbb E[G_t|s_t=s,\pi],
\end{equation} 
\begin{equation}
  Q^*(s,a)= \max\limits_\pi \ \mathbb E[G_t|s_t=s,a_t=a,\pi],  
\end{equation}
where $V^*(s)$ represents the maximum cumulative reward at state $s$, and the state value function $V^*(s)$ describes the expected cumulative reward when following the optimal policy from state $s$. The action value function $Q^*(s,a)$ describes the expected cumulative reward when taking action $a$ at state $s$ and following the optimal policy afterward, and $Q^*(s,a)$ represents the maximum cumulative reward when taking action $a$ at state $s$.

We now review a specific category of MDPs, known as the MAB. Notably, MAB is characterized by a state space comprised of a single state, effectively representing a simplification of the MDP framework with respect to the state space $\mathcal{S}$. In the context of the MAB problem, the agent grapples with decision-making amidst multiple options, each associated with an uncertain probability distribution, endeavoring to identify the optimal choice that maximizes cumulative rewards. Within MEC networks, agents may confront decisions regarding diverse edge servers or communication channels, each possessing uncertain performance attributes. Bandit algorithms, such as the \textit{upper confidence bound} (UCB), offer a means to strike a balance between exploration (i.e., venturing into unexplored options) and exploitation (i.e., utilizing the best-known option) to maximize long-term rewards. Another variant of the MAB paradigm, contextual MAB, owns a state space the same as a normal MDP. Nevertheless, it is worth noting that the state transition in a contextual MAB is independent of both the prior state and action, leading to its classification as a simplification of MDP with regard to transition probabilities.

\paragraph{Traditional Approaches}
In the field of RL, the credit assignment problem stands as a central challenge. This issue relates to how each action taken by an agent contributes to the final outcome or cumulative reward. Three core aspects make up the credit assignment problem: The first aspect is reward delay. RL agents often perform a series of actions before receiving rewards. It becomes a challenge to identify which specific action is primarily responsible for triggering the reward. This delay in reward receipt adds complexity to the situation. In many instances, the long-term impact of an action is not immediately discernible, further complicating the credit assignment process. The second aspect revolves around the short-term and long-term effects of actions. To effectively evaluate the contribution of each action, it is crucial to consider both their immediate and future implications. The credit assignment problem, therefore, seeks to establish a balance between optimizing immediate rewards and those that will be received in the long term. This balance is essential for the optimization of the overall policy. The third aspect involves the balance of exploration and exploitation. A significant challenge in RL is to achieve an equilibrium between exploration, which involves trying out new actions to gather more information, and exploitation, which involves making optimal decisions based on existing information. The credit assignment problem needs to consider the best method to weigh these trade-offs.
To address the credit assignment problem, researchers have developed several techniques, such as TD learning, MC methods, and \textit{Q-learning}. These approaches attempt to tackle the credit assignment problem to some extent, allowing agents to learn more effective policies. The credit assignment problem is an essential challenge in the field of RL, dealing with how to determine the contribution of each action to the final outcome or cumulative reward. Solving the credit assignment problem helps improve the agent's policy, enabling better performance in various environments.

DP \cite{bertsekas1996neuro} is one classic method to derive the optimal policy for an MDP. They require perfect information about an environment, and the optimum is guaranteed. However, the system dynamics can not be fully characterized in most environments. MC algorithms, on the other hand, do not require perfect system information. They sample experiences from the environment and update value estimators and policies based on new experiences after one episode. TD algorithms \cite{tesauro1995temporal} also need to sample experiences from an environment but update value estimators and policies after each step within an episode. In practice, the key difference between MC and TD algorithms lies in their update frequency. MC algorithms update value estimators and policies after completing an entire episode, while TD algorithms update them after each step within an episode. This difference in update frequency leads to several pros and cons. MC algorithms tend to have a lower estimation variance as they use the complete return from an episode. However, they can be slow to converge and may not be suitable for continuous or very long tasks. On the other hand, TD algorithms can learn online and in real-time, which makes them more suitable for continuous tasks. They can also converge faster than MC algorithms but may have higher variance in their estimates due to their reliance on single-step updates.

\subsubsection{Partial Observable MDP}

\paragraph{Concept}
In MDP problems, the agent observes the full environment state information. However, this is non-trivial to achieve in real-world applications. POMDP can generalize MDP frameworks \cite{shani2013survey}. POMDP considers the incompleteness of state information. A POMDP is denoted as a tuple $\langle \mathcal S, \mathcal A, \mathcal T, \gamma, \mu, \mathcal R, \mathcal O, \mathcal Z \rangle$, where $\mathcal S, \mathcal A, \mathcal T, \gamma, \mu$, and $ \mathcal R$ are the same as MDPs. The observations space is $\mathcal O$ and $\mathcal Z: \mathcal S \times \mathcal A \times \mathcal O \to [0,1]$ is the probability distribution of observations in a POMDP. Specifically, it represents the probability of receiving an observation $o \in \mathcal O$ given a state $s \in \mathcal S$ and an action $a \in \mathcal A$. In a POMDP, an agent obtains observation $o_t$, which provides partial information about the underlying state $s_t$ rather than the full state information. In other words, an agent only receives an observation that reflects some aspects of the system state in each step instead of having complete knowledge of the state.

\paragraph{Traditional Approaches}
There are two main approaches to solving a POMDP problem\cite{SAMLMARL9372298}. The first approach is the history-based method. In this approach, an agent maintains an observation history or an action-observation history to learn a policy. This enables the agent to capture the temporal dependencies and improve decision-making capability. However, this method can lead to high computational complexity and memory requirements due to the increasing history length. Another approach is the \textit{predictive state representation} (PSR) method. This approach aims to predict the future according to past actions and observations. PSR models represent the environment in terms of predictions about future events rather than directly modeling the hidden states. This allows for more efficient and compact representations of the environment but may require more sophisticated learning algorithms.

\subsubsection{Multi-Agent MDP}

When there are multiple agents, the decision-making problems are generally modeled using the framework of the \textit{Markov games} \cite{littman1994markov} or \textit{decentralized partially observable Markov decision processes} (Dec-POMDPs) \cite{bernstein2002complexity}.

A Markov game, also known as a stochastic game, is a widely adopted framework for modeling scenarios involving multiple adaptive agents with interacting or competing goals. In a Markov game, the multi-agent scenario is considered as a single entity. It can be defined as a tuple $\langle \mathcal{I}, \mathcal{S}, \{\mathcal{A}^i\}, T, \{\mathcal{R}^i\} \rangle$, where $\mathcal{I}$ represents the set of agents. For each agent $i \in \mathcal{I}$, $\mathcal{A}^i$ denotes the action space specific to that agent. The joint action space for all agents is denoted as $\mathcal{A} = \times_{i \in \mathcal{I}} \mathcal{A}^i$. The reward function $\mathcal{R}^i: \mathcal{S} \times \mathcal{A} \times \mathcal{S}  \to \mathbb{R}$ determines the reward for each agent, taking into account the global state information and the actions of all agents. The system transition and rewards in a Markov game depend on the collective behavior of all agents. Markov games provide a versatile framework for addressing both competitive and cooperative relationships among agents. In certain scenarios, such as edge servers, cooperation is crucial to maximize the overall transmission rate or improve the hitting rate. On the other hand, in resource-restricted environments, agents must compete to maximize their individual utility while considering the global utility. The Markov game framework is capable of handling various settings, including fully cooperative, fully competitive, or mixed scenarios.

A Dec-POMDP is a general multi-agent modeling framework that allows each agent to make decisions independently based on local, imperfect information about the environment.
A Dec-POMDP can be defined as a tuple $\langle \mathcal{I}, \mathcal{S}, \{\mathcal{A}^i\}, \{\mathcal{O}^i\}, T, O, \mathcal{R} \rangle$, where:
\begin{itemize}
    \item $\mathcal{I}$ is the set of agents. For each agent $i \in \mathcal{I}$, we have:
    \item $\mathcal{A}^i$ is the action space of agent $i$. $\mathcal{A} = \times_{i \in \mathcal{I}} \mathcal{A}^i$ is the joint action space of all agents.
    \item $\mathcal{O}^i$ is the observation space of agent $i$. $\mathcal{O} = \times_{i \in \mathcal{I}} \mathcal{O}^i$ is the joint observation space of all agents.
    \item $\mathcal{S}$ is the global state space of the environment.
    \item $T: \mathcal{S} \times \mathcal{A} \to \Delta(\mathcal{S})$ is the state transition function, which describes the probability of the environment transitioning to a new state given the current state and actions of all agents.
    \item  $O: \mathcal{A} \times \mathcal{S} \to \Delta(\mathcal{O})$ is the observation function, which describes the probability of each agent receiving its observation given the current actions and next state.
    \item  $\mathcal{R}: \mathcal{S} \times \mathcal{A} \times \mathcal{S} \to \mathbb{R}$ is the reward function, which describes the reward received by the agents given the current state and actions of all agents.
\end{itemize}

In a Dec-POMDP, the goal of the agents is to maximize the expected cumulative discounted reward. This is typically achieved by finding an optimal policy, which is a mapping from historical observations to actions. In the finite-horizon setting, this problem is generally considered NP-hard.

\subsection{Reinforcement Learning: Basic Classification}

\paragraph{Model-Based and Model-Free RL}
Generally, RL has many types of classifications. RL techniques try to learn from the environment and make the optimal decision to maximize the accumulated reward. One basic taxonomy of RL is how they learn from the environment. Therefore, the RL algorithms are divided into model-based RL and model-free RL. Model-free RL, which is also the most popular scheme, adapts its policy by the received reward from the environment. It only relies on samples from the environment while never making predictions about the next state. However, model-based RL first learns the environment based on a pre-defined model and then optimizes the policy by manners such as optimal control or planning. The agents can learn by making predictions about the consequences of their actions \cite{kaidrlsurvey}. Model-based RL always needs some knowledge of the real environment. For instance, traditional approaches to address MDP like value iteration and policy iteration, both need exact information on the dynamics of the environment to calculate the value function, then we take them as model-based RL approaches. MAB only has one state and the dynamics are learned solely from the reward of pulling arms, so methods like UCB and \textit{Thompson sampling} (TS) can be deemed as model-free methods.

\paragraph{On-Policy and Off-Policy}

One basic classification of RL is established on how agents learn from the data. The RL algorithm has two processes: (1) collecting sample data by interacting with the environment, and (2) learning samples and improving policies. RL algorithms need to explore and find the optimal policy. There are two ways to achieve this exploratory capability: on-policy and off-policy. Here we introduce two concepts: the target policy and the behavior policy. The target policy is a learned policy that improves its performance by using the samples collected by the behavior strategy and eventually becomes the optimal strategy \cite{sutton2018reinforcement}. The behavior policy generates behavior responsible for acquiring learning data. For on-policy, the target and the behavior policies remain the same. When the agent learns a new policy, it can immediately put the policy directly into the environment, and behavior strategy will determine the execution of the target strategy. Hence, the RL method uses on-policy and should be cautious in its exploration of the environment. As mentioned above, MC sampling methods operate by evaluating and enhancing a single policy at each epoch, maintaining its consistency throughout a single step. This characteristic categorizes it as an on-policy method. In contrast, TD methods offer both on-policy and off-policy versions, providing a broader range of applicability. One of the most notable deep reinforcement learning approaches, \textit{deep Q-learning} (DQN), leverages the off-policy TD error. This approach allows DQN to be more sample-efficient as it learns from data that may not necessarily be generated by the current target policy. This feature of DQN underscores its potential for efficient learning in complex environments where data efficiency is crucial. Still, they may experience slower convergence compared to on-policy methods like \textit{proximal policy optimization} (PPO), which are more exploratory and update the policy more aggressively.

\begin{figure}
    \centering
    \includegraphics[width=90mm]{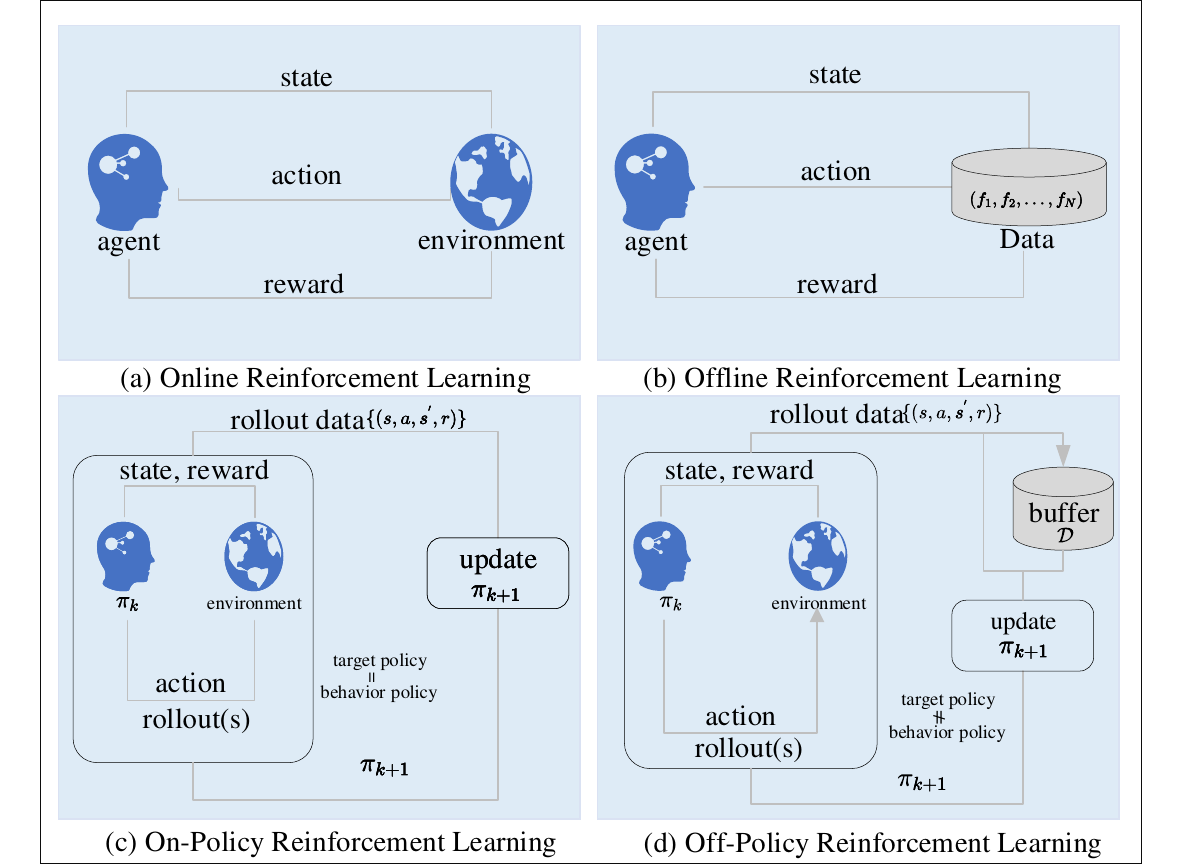}
    \caption{The illustrations of RL approaches: a) online RL; b) offline RL; c) on-policy RL; d) off-policy RL.}
    \label{fig:online and offline}
\end{figure}

\paragraph{Online Learning and Offline Learning}
It can be divided into offline RL and online RL according to whether the agent needs to interact with the environment during training. Offline RL learns directly from the dataset without interacting with the environment during the learning process, which is the data collected by other strategies. Online RL needs to interact with the environment during the learning process, including on-policy RL and off-policy RL, as shown in Fig. \ref{fig:online and offline}. The benefits of offline RL in the MEC context include sample efficiency, safety, and reusability. Offline RL reduces the need for additional data collection, providing a safer learning process by avoiding potential negative consequences during direct environment interaction. Furthermore, the pre-collected dataset can be reused to train and evaluate various RL algorithms, minimizing repeated data collection. However, offline RL has its limitations. It may struggle to adapt to changes in the environment or system dynamics due to its dependence on the pre-collected dataset. The quality of the policy it learns is tied to the dataset's diversity and relevance, and suboptimal solutions may arise if the dataset is insufficient or outdated. Additionally, offline RL does not directly address the exploration-exploitation trade-off, assuming that the dataset already contains adequate exploratory data. On the other hand, online RL continuously learns and updates its policy or value function through ongoing interactions with the environment. In terms of MEC, this offers adaptability, the ability to balance exploration and exploitation, and real-time learning. Online RL can adapt to changes in the environment or system dynamics and actively balance the exploration-exploitation trade-off, potentially leading to robust policies. It also supports real-time learning and decision-making, fitting well for dynamic MEC scenarios with changing conditions. However, online RL also has its drawbacks. It may require a large number of interactions with the environment, making it resource-intensive and time-consuming. Safety issues may also arise as online RL can make suboptimal or even harmful actions during the learning process. Moreover, online RL can take a longer time to converge to an optimal policy, particularly in large and complex state-action spaces common in MEC scenarios.

\subsection{Classic SARL: Characteristics, Weakness, and Advantages}\label{subsec:Classic Single  Agent RL: Characteristics, Weakness, and Advantages}

RL provides a mechanism for an agent to learn how to implement policies to maximize cumulative rewards through interactions with its environment. However, traditional RL methods face limitations when dealing with environments with large-scale state and action spaces or are continuous and ambiguous. To overcome these challenges, DRL emerged. DRL combines the representational learning capabilities of deep learning with the decision-making mechanisms of RL, enabling RL to be applied to more complex and realistic environments. DQN is a classic DRL algorithm that uses a DNN to approximate the Q-function and continuously updates this function through interactions with the environment, eventually learning a policy that maximizes cumulative rewards. Not only has DRL overcome some of the limitations of traditional RL methods, but it has also demonstrated superhuman performance in certain tasks, such as AlphaGo's performance in the game of Go. Therefore, research into DRL is of great value for understanding and solving complex decision-making problems.

\paragraph{Value-Based Approaches}
Depending on whether to learn the environment model, DRL algorithms are generally classified as model-free and model-based methods. Model-free methods are the mainstream DRL algorithms and are widely employed in the single-agent setting. Model-free methods consist of value-based, policy-based, and actor-critic methods.
Value-based approaches attempt to approximate a value function. The standard Q-learning method learns policy using a Q-table \cite{watkins1989learning}. Q-learning is an algorithm used to determine the best action in a given state by estimating the expected future rewards. In Q-learning, the Q-value $Q(s, a)$ represents the expected total reward when taking action $a$ in state $s$. The Q-values are updated iteratively using the following rule:
\begin{equation}
    Q(s, a) \leftarrow Q(s, a) + \alpha [R(s, a) + \gamma \max_{a'} Q(s', a') - Q(s, a)],
\end{equation}
where $\alpha$ is the learning rate ($0 \leq \alpha \leq 1$), $R(s, a)$ is the immediate reward after taking action $a$ in state $s$, $s'$ is the next state after taking action $a$ in state $s$, and $\max_{a'}$ is the maximum Q-value for the next state $s'$. However, in reality, the state and behavior are often highly complex and unpredictable, making it challenging to represent them in a finite Q-table due to the vast number of possible state-action combinations. The DQN combines Q-learning with a DNN and learns a control policy directly from a high-dimensional state of input \cite{mnih2013playing}. Experience replay, a technique used in DQN, alleviates non-stationary distributions and correlated data, smoothing the training distribution for past behaviors. It applies DNN and other technologies to efficiently extract the features of the state and improve learning efficiency. The weakness of DQN is that it may overestimate action values due to selecting and evaluating actions using the same value. To solve this common overestimation, the \textit{double DQN} (DDQN) decouples the selection stage from the evaluation stage using two Q-networks, yielding a more accurate value estimation \cite{van2016deep}. Value-function approaches have several limitations. First, while value-based algorithms are designed to find a deterministic policy that is greedy with respect to the optimal Q-value, there may be cases where an optimal stochastic policy is more appropriate, and these algorithms may not perform as well in such situations. Second, a slight change of action value can lead to this action being or not being selected, making it sensitive to small fluctuations in action values.

\paragraph{Policy-Based Approaches} Policy-based methods such as policy gradient approximate a stochastic policy with an independent function approximator \cite{sutton1999policy}. The input represents states, and the output is the probability of action selection. The policy gradient method aims to optimize the policy, represented by a parameter vector $\theta$, by ascending the gradient of the expected return $\nabla_\theta J(\theta)$. The updating rule for policy gradient is:
\begin{equation}
\theta \leftarrow \theta + \alpha \nabla_\theta J(\theta),
\end{equation}
where $\theta$ is the policy parameter vector, and $\nabla_\theta J(\theta)$ is the gradient of the expected return with respect to the policy parameters. Theoretically, this method is proven to converge to a locally optimal policy with arbitrary differentiable function approximation. However, policy-based methods learn action probabilities based on their previous estimates rather than using true action labels, which are also known as ground-truth tokens, and this limitation can hinder the training model's performance. Moreover, the policy gradient method suffers from poor robustness and data efficiency due to high variance in gradient estimates, and the quality of the learned policy may degrade as the step size (learning rate) increases.

\begin{figure}
    \centering
    \includegraphics[width=90mm]{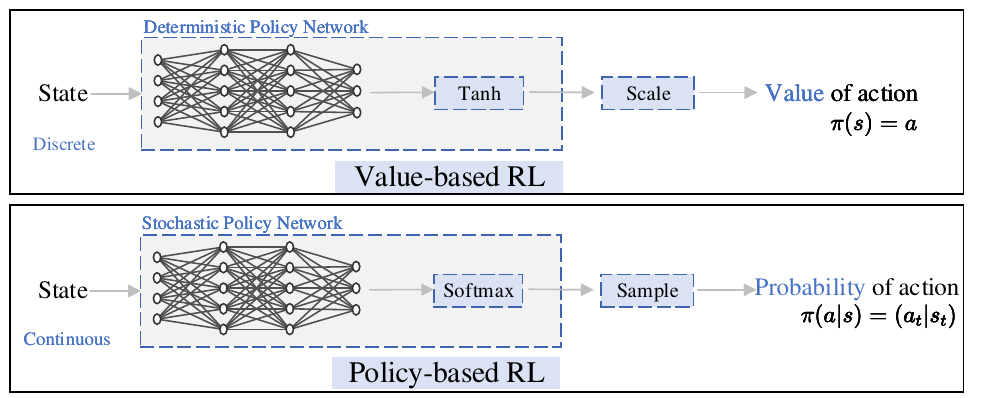}
    \caption{Value-based RL and policy-based RL.}
    \label{fig:Value-based RL and Policy-based RL}
\end{figure}

The discrepancies between the value-based and the policy-based methods are as bellow: (1) The value-based method only solves a low-dimensional or discrete problem. In contrast, the policy-based methods solve high-dimensional or continuity issues; (2) The output of the value-based method is a particular value, whereas the output of the policy-based is the probability of actions, as shown in Fig. \ref{fig:Value-based RL and Policy-based RL}.

\paragraph{Actor-Critic Framework}
The \textit{actor-critic} (AC) method integrates the value-based and policy-based methods. In AC \cite{bahdanau2016actor}, an actor-network makes an action, and a critic network predicts the action value. The actor-network updates are as follows:
\begin{equation}
  \theta \leftarrow \theta + \alpha \nabla_\theta J(\theta) \approx \theta + \alpha \nabla_{\theta} Q(s, a), 
\end{equation}
and the critic network update follows:
\begin{equation}
  Q(s, a) \leftarrow Q(s, a) + \beta [R(s, a) + \gamma \max_{a'} Q(s', a') - Q(s, a)],  
\end{equation}
where $\theta$ represents the actor policy parameter vector, $\alpha\in [0, 1]$ is the actor learning rate, $\beta\in [0,1]$ is the critic learning rate, $\gamma\in [0, 1]$ is the discount factor, $R(s, a)$ is the immediate reward after taking action $a$ in state $s$, and $s'$ is the next state after taking action $a$ in state $s$. The AC model uses a baseline method to reduce high variance and stabilize the training process.

\textit{Trust region policy optimization} (TRPO) uses trust region constraints on the old policy and new policy with \textit{Kullback-Leibler} (KL) divergence \cite{schulman2015trust}  to guarantee monotonic improvement of policy. It uses a surrogate function instead of the Hessian matrix to reduce the computation complexity. TRPO can provide stable and reliable policy updates, making it suitable for problems where monotonic improvement is crucial. However, its complexity and incompatibility with techniques like noise or parameter sharing can limit its applicability. Another family of policy gradient methods is PPO, which optimizes a ``surrogate" function with stochastic gradient ascent \cite{schulman2017proximal}. PPO-based methods give a parameter of a certain probability distribution as the output. Therefore, they suit both discrete and continuous action space by pre-selecting a probability distribution. It performs a minibatch gradient update for multiple epochs, which is simpler to implement with better sample complexity than TRPO. PPO has two classic versions, i.e., PPO-clip and PPO-penalty. PPO-clip is a popular version and more widely used due to its low computation complexity. PPO-clip is widely adopted as the baseline for SARL scenarios.
\textit{Asynchronous advantage actor-critic} (A3C) is an asynchronous variant of AC with parallel actor-learners in multiple threads \cite{mnih2016asynchronous}. A3C is suitable for situations where synchronous training is not possible or when the problem can be parallelized effectively for faster training. However, it may not be the best option for tasks that require high levels of synchronization.

Meta-learning involves utilizing meta-knowledge to dynamically adjust learning strategies, enabling adaptive decision-making on new tasks\cite{vilalta2002perspective}. RL always demands an extensive amount of data to learn the model of a particular task, resulting in low data efficiency and difficulty in adapting the model to additional new tasks\cite{beck2023survey}. \textit{Meta reinforcement learning} (meta-RL) is an extension of RL that allows agents to learn how to learn from experience \cite{wang2016learning}. In other words, meta-RL algorithms can adapt their learning strategies based on the characteristics of the task at hand. This can be particularly useful in MEC networks where task requests and channel gain vary rapidly over time.
In Table \ref{table:sarl}, we compare the advantages and disadvantages of different algorithms and applicable scenarios.

\paragraph{Convergence and Accurracy}

Convergence rate and accuracy are two crucial metrics for assessing RL models in dynamic environments. The convergence rate refers to the speed at which the algorithm approaches the optimal solution during the learning process. A faster convergence rate implies the need for fewer iterations or sample quantities to achieve the optimal solution within an acceptable margin of error. Factors such as learning rate, initial parameters, loss function, and model complexity all affect the convergence rate. It can be enhanced through adaptive learning rates, pre-training, and algorithmic improvements, among other methods\cite{azar2011reinforcement}. Accuracy is employed to evaluate the predictive capability of RL. Enhancing the accuracy of the model in decision-making for tasks involving a multitude of states and actions presents certain challenges. Methods involving probabilistic inference such as Gaussian processes and ensemble techniques can improve model accuracy. Moreover, employing model-based control for planning actions, learning the environmental dynamics through latent models, and utilizing end-to-end learning and planning directly from sensor data can further enhance the model's accuracy in decision-making processes\cite{plaat2023high}.

\renewcommand\arraystretch{1.2}
\begin{table*}[t]
\small
\caption{Common SARL Algorithms.}
\begin{center}
\begin{tabular}{m{2em}|m{3.2cm}|m{2.2cm}|m{10cm}}
\hline\hline
{\bf Ref.} & {\bf Action Space Type} & {\bf Algorithms}& {\bf Characteristics}\\
\hline
\cite{mnih2013playing}&{\begin{tabular}{@{}l@{}}
    \multirow{4}{3.2cm}[-1.2ex]{Discrete Action Space}
\end{tabular}} & DQN & DQN algorithm has a simple framework, while it suffers from the overestimated Q-value.\\
\cline{1-1}\cline{3-4}
\cite{van2016deep}&&DDQN & DDQN alleviates the overestimation of DQN, resulting in more stable training. \\
\cline{1-1}\cline{3-4}
\cite{huang2018vd}&&Dueling DQN& Dueling structure accelerates training in environments with a large number of states unaffected by actions.\\
\cline{1-1}\cline{3-4}
\cite{wang2016dueling}&&Dueling DDQN& Draw benefits from dueling and double structure.\\
\hline
\cite{silver2014deterministic}&{\begin{tabular}{@{}l@{}}\multirow{4}{3.2cm}[-4ex]{Continuous Action Space}\end{tabular}}& DPG& DPG uses a deterministic policy network to output a specific action.\\
\cline{1-1}\cline{3-4}
\cite{lillicrap2015continuous}&&DDPG& Introducing deep NN to enable continuous action spaces. The training process is slow and unstable. It is suitable for solving some simple problems.\\
\cline{1-1}\cline{3-4}
\cite{haarnoja2018soft}&&SAC& Maximizing reward and entropy at the same time. Exploration and learning speed are improved.\\
\cline{1-1}\cline{3-4}
\cite{fujimoto2018addressing}&&TD3& Utilizing target networks and target policy smoothing regularization to reduce the overestimation bias.\\
\hline

\cite{schulman2017proximal}&{\begin{tabular}{@{}l@{}}\multirow{4}{3.2cm}[+1ex]{Discrete \& Continuous Action Space}\end{tabular}}& PPO& PPO simplifies the trust region computation process of TRPO, maintaining the characteristics of stable training, simple parameter tuning, and robustness.\\
\cline{1-1}\cline{3-4}
\cite{schulman2017proximal}&&PPO2& Using the clipping technique to simplify the updating process.\\
\cline{1-1}\cline{3-4}
\cite{mnih2016asynchronous}&&A3C& Introducing asynchronous training which accelerate learning efficiency.\\

\hline\hline

\end{tabular}
\end{center}\label{table:sarl}
\end{table*}

\subsection{Classic MARL: Characteristics, Weakness, and Advantages}\label{subsec: Classic MARL: Characteristics, Weakness, and Advantages}

\paragraph{Fully Centralized Learning}
Fully centralized learning represents approaches whose decision is made by a central entity. In contrast to decentralized or distributed paradigms, fully centralized learning emphasizes the aggregation of information and control in a centralized fashion. The central actuator, armed with a comprehensive view of the multi-agent environment, undertakes the responsibility of making decisions that have a system-wide impact. The key elements of centralized learning MARL involve maintaining a unified, global state representation that provides the central entity with a holistic understanding of the environment. Furthermore, centralized policy learning is integral, where a unified policy is trained to guide the actions of all agents based on the centralized state representation. Complementary to this, centralized value estimation entails estimating value functions centrally, offering a basis for evaluating the desirability of different states or actions in the overall system.

Fully centralized learning schemes offer notable advantages. They facilitate global optimization by leveraging a comprehensive perspective of the environment, contributing to cooperative decision-making, and potential communication among agents. However, these benefits come with some drawbacks. The computational complexity associated with maintaining a centralized view may lead to increased training and decision times. Scalability challenges may emerge as the number of agents or the complexity of the environment grows. Additionally, there is a vulnerability to centralized failures, where the entire system is at risk if the central entity malfunctions.

\paragraph{Independent Learning}
In numerous MEC networks, a group of edge servers collaboratively supply caching and computational resources for interconnected devices, acting in a decentralized manner. Consequently, MARL methods prove crucial for MEC networks incorporating heterogeneous edge servers and devices. \textit{Independent Q-learning} (IQL) \cite{tan1993mul}, a notable method in this space, employs Q-learning to train a distinct action-value function for each individual agent. However, IQL has its limitations. Traditional RL methods, including Q-learning, often exhibit instability in multi-agent environments due to their non-stationarity. Additionally, the policy gradient's variance tends to increase with the growing number of agents, making these traditional methods less effective for larger systems. To mitigate these issues, recent advancements have introduced communication among agents. For instance, \textit{communication neural network} (CommNet) \cite{sukhbaatar2016learning} leverages a centralized network, and \textit{bidirectionally-coordinated nets} (BicNet) \cite{peng2017multiagent} utilize bidirectional \textit{recurrent neural networks} (RNNs). These methods facilitate inter-agent communication, enhancing coordination and decision-making efficiency in multi-agent systems.

\paragraph{Centraluzed Training Distributed Execution}

The \textit{centralized training and distributed execution} (CTDE) has also gained attention, with methodologies like \textit{value decomposition network} (VDN) and QMIX being prime examples. VDN \cite{sunehag2017value} introduces the concept of the action-value function factorization method, providing a novel way of managing multi-agent action values. Meanwhile, QMIX \cite{rashid2018qmix} offers a strategy to learn decentralized value functions or policies using a value-based approach in an off-policy manner. It estimates the action value as a non-linear combination of each agent's value, and the joint action values are monotonic, which guarantees the existence of its maximization. These methods are called value function factorization schemes. However, it's worth noting that both VDN and QMIX are primarily suitable for cooperative scenarios.

In response to the need for algorithms suitable for both cooperative and competitive systems, the \textit{multi-agent deep deterministic policy gradient} (MADDPG) \cite{lowe2017multi} was proposed. Unlike other methods, MADDPG learns policies using local information during execution without assuming the environment dynamics to be differentiable or the presence of any specific communication structure between agents. This adaptability makes MADDPG suitable for a wider range of multi-agent environments. Federated learning is a distributed learning approach where multiple devices collaborate to train a shared model. In MEC networks, edge devices can collect data and train models locally, which can then be aggregated at a central server to improve the system's overall performance. Federated learning provides a way to train models in a privacy-preserving manner while leveraging the computational resources of edge devices. \textit{Heterogeneous agent proximal policy optimization} (HAPPO) \cite{gu2021multi} also plays crucial roles in MARL. HAPPO operates within a CTDE framework, while, different from MADDPG, agents are deemed as heterogeneous and conduct different policies. This approach can achieve better performance in complicated conditions. These approaches are called methods with fully centralized critic.

In centralized training, the agents collaborate by sharing a global, centralized critic. This critic evaluates the joint actions taken by all agents, providing a comprehensive assessment of the overall system performance. The shared critic facilitates effective learning and coordination among agents during the training phase. Conversely, during distributed execution, each agent utilizes its learned policy independently, without relying on a centralized critic. This decentralized approach allows agents to make decisions autonomously based on local information, promoting adaptability to dynamic environments.

We list the characteristics of common MARL algorithms in Table \ref{table:marl}.   

\renewcommand\arraystretch{1.2}
\begin{table*}[t]
\small
\caption{Common MARL Algorithms.}
\begin{center}
\begin{tabular}{m{2em}|p{1.5cm}|p{1.8cm}|m{12cm}}
\hline\hline
{\bf Ref.} & {\bf Type} & {\bf Algorithms}& {\bf Characteristics}\\
\hline
\cite{yu2022surprising}&{\begin{tabular}{@{}l@{}}
    \multirow{6}{1.5cm}[+1.5ex]{On-policy}
\end{tabular}} & MAPPO & MAPPO addresses continuous action space problems using a centralized training and distributed execution framework for cooperative MARL. It offers stable training and easy parameter tuning, but its performance is limited in complex environments.\\
\cline{1-1}\cline{3-4}
\cite{gu2021multi}&&HAPPO & While the algorithm is slightly complex and requires greater implementation difficulty, it achieves superior performance with fast convergence. It outperforms algorithms such as MAPPO, MADDPG, and IPPO in SMAC and Multi-Agent MuJoCo environments. \\
\cline{1-1}\cline{3-4}
\cite{gu2021multi}&&HATRPO& HATRPO performs better than HAPPO on most tasks.\\

\hline
\cite{lowe2017multi}&\multirow{9}{1.5cm}[-1.2ex]{Off-policy}& MADDPG& MADDPG is a pioneering algorithm with a ``centralized training and distributed execution" framework, suitable for cooperative, adversarial, and mixed environments. \\
\cline{1-1}\cline{3-4}
\cite{foerster2018counterfactual}&&COMA& COMA is a foundational algorithm for discrete action space problems, utilizing the ``centralized training and distributed execution" framework. \\
\cline{1-1}\cline{3-4}
\cite{ackermann2019reducing}&&MATD3& MATD3 combines TD3 with MADDPG to address cooperative, adversarial, and mixed environments. Despite challenges posed by hyperparameters, high training difficulty, and variance, MATD3 performs well with fine-tuning and simple implementation.\\
\cline{1-1}\cline{3-4}
\cite{sunehag2017value}&&VDN& VDN utilizes value function decomposition to address multi-agent credit assignment. Although VDN has a simple implementation and concise logic, its effectiveness is limited.\\
\cline{1-1}\cline{3-4}
\cite{rashid2018qmix}&& QMIX& QMIX is an advanced algorithm for cooperative MARL in discrete action spaces. It has a concise and easy-to-implement structure, achieving excellent performance across diverse tasks. However, QMIX is exclusively tailored for cooperative environments. \\
\cline{1-1}\cline{3-4}
\hline\hline
\end{tabular}
\end{center}\label{table:marl}
\end{table*}

\section{New Challenges in MEC Systems}
The emerging massive connected devices introduce many challenges in MEC networks. The 6G system needs to provide real-time and high-quality service. However, several key performance indicators  \cite{ series2017minimum}, such as latency, data rate, reliability, efficiency, and security, present a roadblock to existing technologies. 
These subsections discuss the fundamental requirements and technical challenges for deploying MEC.

\subsection{Stringent Latency Requirements}\label{subsec:Stringent Latency Requirements}
Many MEC applications may require latency within tens of milliseconds, such as the \textit{internet of vehicles} (IoV), industrial IoT, drone flight control, VR/AR, and financial trading. To meet these requirements, the workload of incoming traffic needs to be offloaded from remote cloud servers to a network of edge servers. As a result, the users' latency has been remarkably reduced to below 100 $\mu s$, or even 10 $\mu s$ with advanced communication technologies \cite{Jiang2021delay}. However, the MEC network incurs delays consisting of task transmission and processing. Optimizing total latency demands offloading, caching, and communication technologies.

\textcolor{black}{Employing RL algorithms in MEC networks is a new paradigm to reduce network delay. RL agents can learn optimal policies that specifically minimize latency. By continuously interacting with the environment, they can make real-time decisions that reduce computation and transmission times. Unlike non-RL algorithms, RL can adapt to network changes, and unlike supervised learning, RL does not require a labeled dataset that represents all possible latency scenarios, which is often infeasible to obtain. Additionally, distributed computing models ensure that time-critical MEC systems maintain QoS-compliant latency.}

\subsection{High Data Rate Requirements}\label{subsec:High Data Rate Requirements}
The proliferation of connected devices has resulted in a significant increase in data rate requirements for various real-world applications. One notable example is the tactile internet, which necessitates ultra-high reliability and real-time decision-making capabilities for delivering multimedia content and facilitating interactive communication involving video streams, voice, data files, emails, and control signals \cite{simsek20165g}. Likewise, autonomous driving requires a high transmission rate and timely information feedback to enable communication among vehicles as well as between vehicles and \textit{Roadside Units} (RSU). The IoV monitors the entire transportation system, provides location information for individuals and vehicles, and offers safety, emergency, and entertainment services.

The future communication technologies deployment of MEC networks for dense urban areas aims to offer a 1 Gbps data rate or even higher for downlink and uplink. Users should have at least a 95$\%$ probability of satisfying the experience data rate at any time or place. The emerging advanced technologies, such as THz communications, improve the peak data rate to 1 Tbps, which is ten times the data rate of 5G. Meanwhile, combining intelligent technologies with wireless resource scheduling is a promising way to improve data rates. \textcolor{black}{RL-based methods enjoy the benefit of handling high data rate requirements. RL can dynamically allocate resources such as bandwidth and computing power to maximize data throughput. Unlike rule-based systems that may not scale well, RL can prioritize data rate requirements in its reward function and adjust its policies in complex environments where the availability of resources fluctuates.}

\subsection{Massive Connections and Reliability}\label{subsec:Massive Connections and Reliability}
For \textit{massive machine-type communications} (mMTC) scenarios, it is expected that the connection density of the minimum number of devices will far exceed $10^6$ per square kilometer of 5G\cite{pokhrel2020towards}. Therefore, the minimal requirement for traffic capacity is 1${\rm{Gbps/}}{{\rm{m}}^{\rm{2}}}$ per unit area, regarding the spectral efficiency, available bandwidth, and network densification.
Reliability refers to the success probability of transmitting a signal with a given amount of traffic. In the massive \textit{ultra-reliable low-latency communication} (URLLC) scenario of MEC networks, the minimum reliability requirement is a success probability $1-10^{-7}$ when a 32-byte packet is transmitted in 1 $ms$. \textcolor{black}{In these scenarios, RL can manage resources efficiently by learning from the patterns of device demands. Traditional methods might become overwhelmed by the scale, but RL can scale by clustering similar types of devices and applying policies that manage connections more effectively.}

\subsection{Network Bandwidth Constraints}\label{subsec:Network Bandwidth Constraints}
Edge users transmit less information to the cloud and send most data to nearby edge servers for processing with the limited wireless communication bandwidth. Single or multiple radio frequency carriers provide the signal transmission bandwidth through the maximum aggregated system bandwidth. 5G requires a bandwidth of at least 100 MHz, while future communication technologies satisfy frequency bandwidth up to 1 GHz for operation. Future communication technologies will explore a broad bandwidth, adapting the mmWave technique, THz communication, and optical wireless communication channels. Spectral efficiency is a vital performance representing the bandwidth utilization and measuring the advance of wireless communication technologies. Besides, peak spectral efficiencies can reach 90 bps/Hz and 45 bps/Hz for downlink and uplink in future communication technologies. \textcolor{black}{However, due to the full coverage of massive devices in IoV and industrial IoT, the shortage of bandwidth resources and low spectral efficiencies still need to be addressed. Reliability in network systems requires robust and fault-tolerant policies. RL can incorporate reliability into its reward structure, penalizing downtime and errors, thus learning to take preemptive actions to avoid failures. Heuristic-based solutions may not have the ability to learn and adapt from past failures as RL does.}

\subsection{Mobility and Dynamic Uncertainty}\label{subsec:Mobility and Dynamic Uncertainty}
Mobile terminals exist in many MEC applications, such as vehicles in the IoV, drones, and wearable devices in healthcare. To support the high-speed trains scenario, the requirement of maximal mobility speed is 500 km/h in 5G. However, commercial airline systems must support the highest mobility speed of even 1000 km/h. In addition, MEC networks are complex, dynamic, and uncertain. For example, from the perspective of the user terminal, users' preferences and locations are changing on a real-time scale. For edge nodes, the network dynamics include covered users, cached content, available cache capacity, computing resources, communication resources, etc. Furthermore, large-scale fading is caused by blocking like buildings or mountains, and small-scale fading is due to multi-path transmission in wireless communication links \cite{pan2015physical}. 

\textcolor{black}{RL is well-suited to environments with high mobility and dynamic uncertainty because it can adapt policies based on the observed outcomes of previous actions. Non-RL methods might rely on predictions from static models that become quickly outdated, whereas RL continuously updates its understanding of the environment.}

\subsection{Cyber-Physical Security}\label{subsec:Cyber-Physical Systems}
The cyber-physical system is essential to ensure secure communication \cite{zhang2018secure}. Massive cyber-physical systems are seamlessly connected to MEC networks, such as industrial control networks, smart homes, healthcare, and vehicles. Security and privacy of MEC networks are essential to protect infrastructure, mobile devices, transmission data, and assets \cite{wang2019sdn}. The edge network provides a platform to securely collect sensor data in a distributed system. It prevents sensitive information from being exposed to unauthorized entities and guarantees integrity and authentication. In particular, the security of healthcare IoT is vital as misusing patient-generated data will lead to legal issues \cite{sun2018security}.

Existing cyber security methods focus on perimeter-based protections. A secure system or individual device is placed behind a firewall, which works with intrusion detection to prevent security threats from breaching the protected perimeter. These existing solutions will no longer be competent for addressing emerging challenges in MEC networks with future communication technologies. Thus, it presents several challenges: (1) How to keep security credentials on large-scale devices; (2) How to protect resource-constrained devices; (3) How to assess distributed systems in a trustworthy manner; (4) How to design a novel disruptive incident response paradigm. 

\textcolor{black}{RL possesses the potential to enhance security in cyber-physical systems of learning to detect and respond to anomalies and potential threats. While traditional security solutions might rely on predefined rules that could be circumvented by novel attacks, RL can evolve its defensive strategies over time, learning from each interaction to improve security measures.}
 
\subsection{Summary}

\textcolor{black}{In all these areas, the key advantages of RL over non-RL solutions are its scalability, ability to handle uncertainty, and the continuous improvement of its decision-making policies through trial and error. RL's capability to learn from the environment and to optimize complex, multidimensional reward functions makes it suitable for handling the dynamic challenges of modern network systems. Non-RL methods often require complicated computational loads, have limited scalability, are incapable of unseen scenarios, and may not perform well in environments that change over time or are not well understood.}

\section{RL Methods for Computing, Caching, and Communication}
This section overviews the implementations and innovations of RL technologies in MEC networks. It focuses on the solutions to the three problems: offloading scheduling, content caching, and communication. RL is a promising way to enable online learning in a dynamic communication environment, making resource utilization more efficient. Thus, the following subsection will present resource management problems solved by specific RL learning models.

To provide clarity on the organization of the surveyed work, the problem is divided into two main categories: single-agent MEC problems and multi-agent MEC problems.

For single-agent problems, methods can be classified into traditional RL approaches (such as MAB, tabular Q-learning, multi-objective Q-learning, and Q-learning with approximation) and deep RL approaches (including standard DRL methods and their modifications). Traditional RL methods prioritize low complexity and fast inference speed, while DRL methods emphasize the representative capacity of neural networks. However, both approaches aim to improve the approximation of the value function.

For multi-agent problems, research has focused on three main areas: centralized policy, centralized training with decentralized execution, and decentralized policy. In contrast to the centralized policy, the policies of CTDE and decentralized training structures are both distributed, with agents independently executing policies based on local information. CTDE policies enhance access to global information by introducing a central controller. Decentralized training employs various techniques, such as specific schemes and game models, to gather global information. Distributed policies help mitigate the challenges posed by high dimensionality.

\begin{figure*}
    \centering
    \includegraphics[width=180mm]{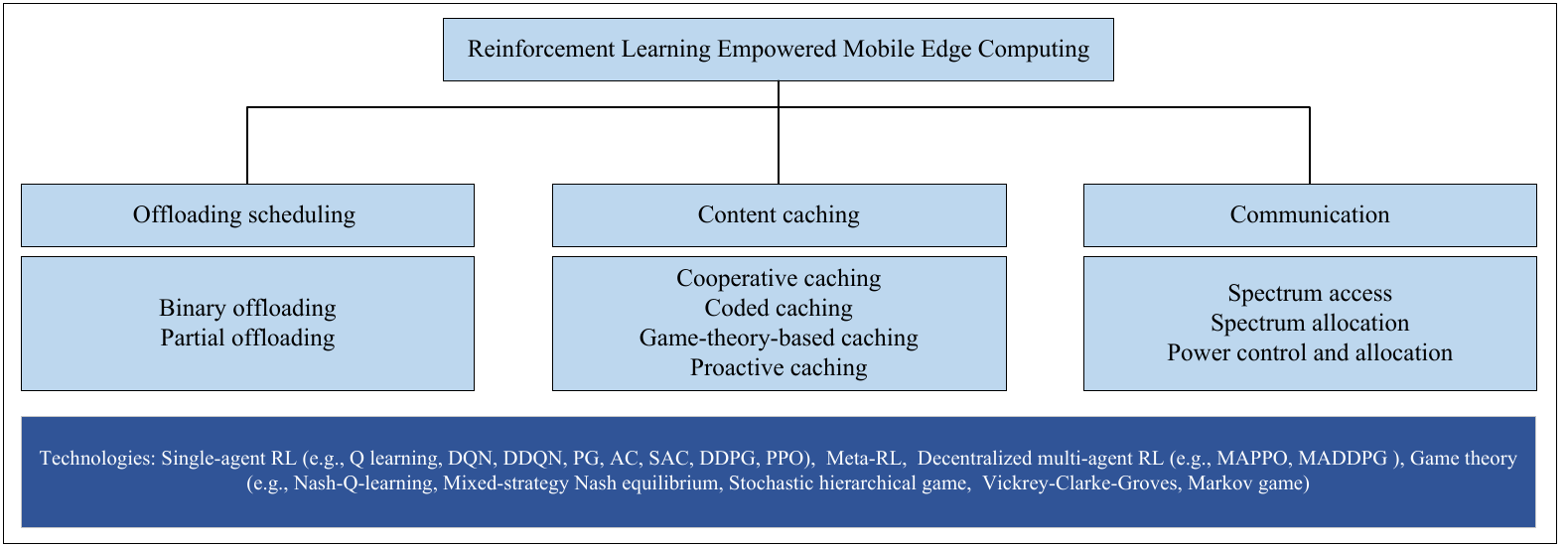}
    \caption{Reinforcement learning for mobile edge computing.}
    \label{fig:Reinforcement learning for mobile edge computing}
\end{figure*}

 A summary of this section is provided in Fig. \ref{fig:Reinforcement learning for mobile edge computing}. The following subsections review these related works.

\subsection{Offloading Scheduling}\label{subsec:Offloading Scheduling}

Driven by the tremendous traffic, the MEC system manages computing tasks by offloading them to edge servers. Generally, offloading scheduling has two types: partial offloading and binary offloading. The partial offloading model can offload and execute a part of a task at a cloud server. The rest of this task is processed locally on the device or a particular edge server. The binary offloading model makes a binary offloading decision, and a task is offloaded to a cloud server or an edge server.

\subsubsection{Performance Optimization Utilizing RL}

Intelligent offloading is a promising technology to satisfy enormous computing requirements in MEC networks. Therefore, offloading scheduling is significant and is evaluated by the following performances:
\begin{itemize}
    \item \textit{Task latency}: Task latency contains the delay of transmission and execution. The latency performance determines the efficiency and security in some latency-sensitive systems, such as industrial production and autonomous driving.
    \item \textit{Energy consumption}: Some MEC systems have strict limits on energy consumption. For example, the \textit{unmanned aerial vehicle} (UAV) systems require considering each UAV device's electricity. Therefore, high energy consumption affects the regular operation of systems.
\end{itemize}

The latency and energy consumption are essential to measuring the performance of an MEC system. RL approaches show colossal potential for the improvement of these indices. Furthermore, RL schemes learn information through interacting with the varying MEC environment and maximizing long-term rewards. Some work \cite{8552454,8377343,9522334} designs a reward function by weighting a normalized latency and energy consumption cost. In \cite{8552454}, the authors studied an offloading problem and struck a balance between latency and energy consumption in the vehicular network using a DQN-based method and dynamic RL.

\begin{figure}[t]
    \centering
    \includegraphics[width=0.5\textwidth]{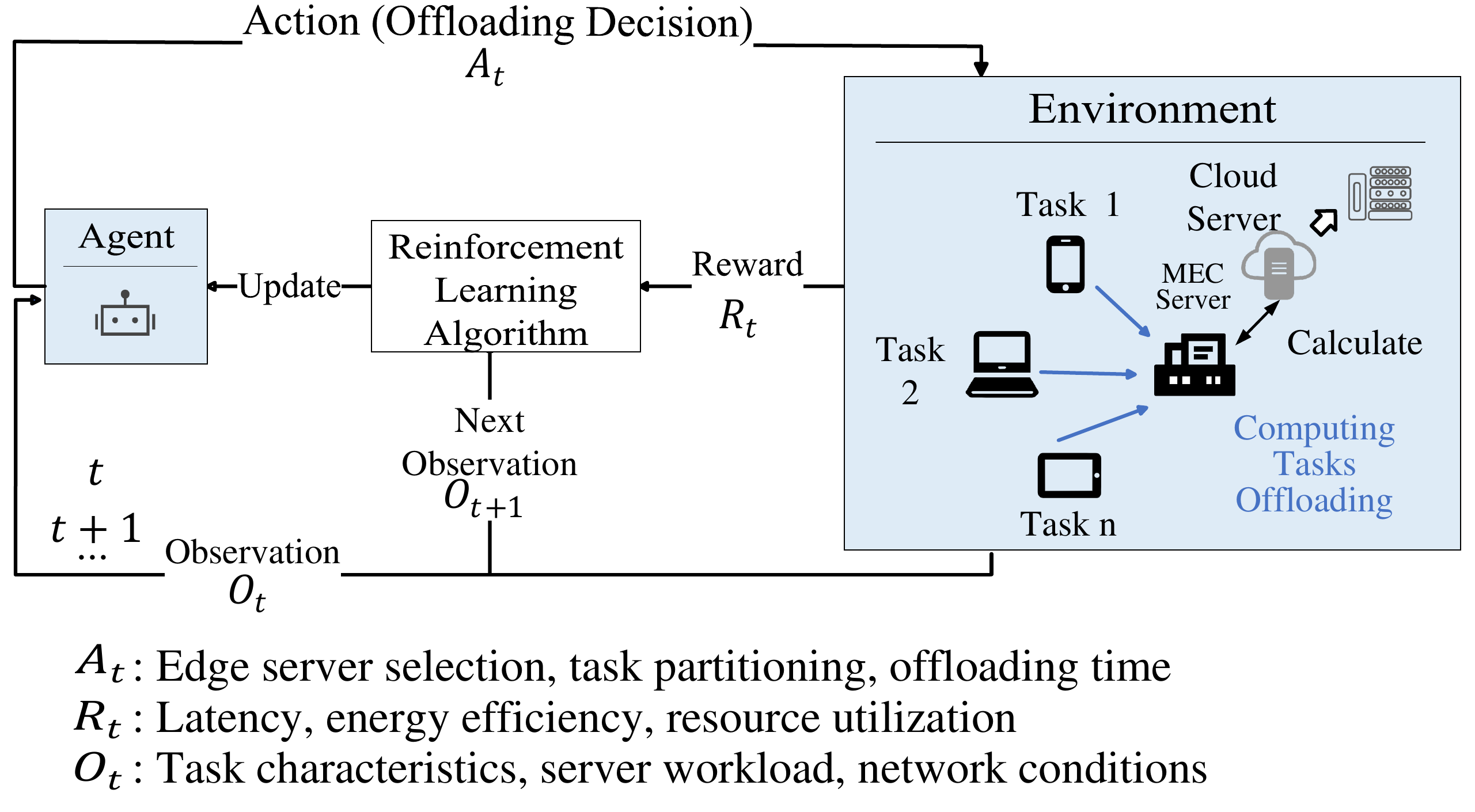}
    \caption{MDP for task offloading.}
    \label{bs_offload}
\end{figure}

The delayed time was usually divided into two parts: transmission time $T_{tran}$ and computation time $T_{comp}$, where the transmission time is decided by the size of the task and transmission rate, and the computation time is determined by the volume of the task and the CPU level at the edge server. The total execution time is $T_{total}=\mathbb{E}[T_{tran}+T_{comp}]$. Similar to the execution time, the power consumption during offloading is also divided into two parts. The power consumption for computation is denoted as $E_{comp}$ which is proportional to the square of task size, and the power consumption of transmission is proportional to transmission time. The total energy consumption is defined as $E_{total} = \mathbb{E}\left[E_{comp}+E_{tran}\right]$. Task offloading aims to optimize the delay of energy efficiency. Therefore, the basic formulation of task offloading problems considers the minimization of long-term average delay, the improvement of the average energy efficiency, or joint optimization of these two objectives. Therefore, the reward of the corresponding problem at each time slot is defined as the current execution time, the energy consumption, or a combination of these two values, i.e., $R=\delta T_{total}(t)+ (1-\delta)E_{total}(t)$, where $\delta\in (0,1)$ is a scalar to balance multiple objectives. The scheduler, which is always modeled as the agent, takes the size of tasks, transmission time, computing time, and task size from the acknowledgment as the state, and seeks to maximize the accumulated reward. A description of a generic model for RL-based offloading structure is shown in Fig. \ref{bs_offload}.

To establish the state transition, the attributes of edge users and edge servers have to be included in the state representation. In some complicated scenarios, the topology and mobility of edge nodes should also be considered. Dynamic RL iteratively updates the policy at each step until the estimated value function converges. In \cite{8377343}, the authors applied the Q-learning and the DQN approach to make offloading decisions in a multi-user MEC system. In \cite{9522334}, the authors used an \textit{advantage actor critic} (A2C) algorithm to jointly optimize the offloading and power allocation strategies. The energy cost is reduced by 80\% compared with a random policy. In some work, energy consumption is not directly considered in the reward function but is implicitly indicated by regular terms. For example, in \cite{9631180}, the authors denoted an energy consumption function as a reward function and decomposed it into communication and computation payments. The cost was converted to the payment of the MEC service. In the simulation, DQN outperformed game-theory-based policy which considered individual interest. In \cite{8493155}, they studied a wireless charging model with constrained energy consumption. The authors calculated the reward considering execution delay, queuing delay, task dropping, task failure, and the payment of accessing MEC services. According to these works, weighting latency and energy consumption were widely used to make a tradeoff between these two objectives. The additive structure of the reward function associated with a DDQN framework makes it a faster convergence and higher accumulated return compared with classic DDQN. Some works also considered the economic benefits, energy limits, and other factors \cite{9631180,8493155}.

\subsubsection{Centralized Offloading Scheduling with RL Approaches}

Most studies choose mature single-agent algorithms to solve the offloading problem. These algorithms describe a centralization framework, which receives global observation of the MEC system and makes a joint decision by one agent. For example, in \cite{8377343}, the authors provided a DQN-based method to minimize offloading costs. The RL agent scheduled tasks for all users at each time slot. As an early attempt at the utilization of RL schemes in offloading, the offloading cost is predominantly good compared with full offload and local computing. The authors in \cite{8406881} implemented a DQN-based method to conduct offloading scheduling in a cellular network. DQN-based approaches are suitable for offloading decisions since DQN outputs the best possible discretized action based on the current state. In most offloading problems, the offloading decision can be represented in one-hot variables or integers, and the optimal action for a given state is deterministic. \textcolor{black}{The DQN-based offloading schemes function in various conditions with different arrival rates. However, DQN prefers to overestimate action values because it has a greedy choice of actions.}

\textcolor{black}{For a better value estimation, the work \cite{9253665} used the DDQN method \cite{van2016deep} to solve the offloading problem.} The authors considered a decentralized way of offloading scheduling. Since partial observation and limited communication among edge servers, they incorporated \textit{long short-term memory} (LSTM) into dueling DDQN, which improved the long-term average reward estimation. The LSTM could supplement global information through past observations, which contributed to the policy implementation for independent users. The integration of LSTM, dueling, and double structure contributes to a precise estimation of the Q-value and a long-term accumulated reward, which leads to a decrease of 35.7\% in total delay. The DDQN method yields more accurate value estimation by introducing a target neural network. Both the DQN and DDQN algorithms belong to the value-based methods, and they can solve the problem with discrete action space, such as binary offloading scheduling problems. However, the partial offloading scheduling problems call for a continuous solution, which leads unavailability of DQN-based algorithms. 

\textcolor{black}{Some works apply policy gradient methods to deal with power allocation problems in continuous action space.} For example, the work \cite{8513863} focused on the general fog-enabled IoT offloading problem using an AC algorithm \cite{barto1983neuronlike} to get a joint offloading decision and power allocation scheme. The studies \cite{8513863} assumed that each user raised only one request simultaneously, and the simulation showed a prominent improvement compared with local computation and fully offloading schemes. Differently, the work \cite{8672604} considered a multi-user multi-task system in which each user conducted all computation applications simultaneously. The authors designed an AC-based RL method to cope with an offloading problem in \textit{space-air-ground integrated network} (SAGIN). The AC-based computing offloading approach effectively handled the multi-dimensional SAGIN resources by analyzing the information, such as remaining tasks and path loss of the dynamic network. The simulation showed the AC-based scheme obtained a better average delay and energy consumption compared with greedy and random offloading manners.
The AC-based manners own a faster update speed than policy gradient approaches; however, they suffer from a slow convergence of both critic and actor networks.

\textcolor{black}{The development of meta-RL has attracted significant attention. Some works try to introduce the meta-RL method for MEC offloading.} Existing RL methods have poor adaptability, which means that a well-trained scheme cannot adapt to a new or unseen environment. In \cite{9448034}, the authors proposed a meta-RL-based offloading scheme with inner and outer models. The inner model focused on offloading decision-making. The outer model focused on environmental changes. Compared with DQN, the meta structure increased the offloading effect by 17.6\%. Based on meta-RL, the authors \cite{chen2022cache} proposed a cache-assisted ofﬂoading scheme for the MEC network. The meta-policy initialized the model of a speciﬁc MDP to improve the adaption and accuracy. Results showed that this approach achieved rapid convergence and had better \textit{quality of experience} (QoE) than a PPO-based \cite{schulman2017proximal} scheme. 
It modified the initial hyperparameters of the underlying PPO algorithm to make it quickly adapt to a new environment. The simulation results showed that the meta-learning-based PPO ran a lower latency and energy consumption compared with classic PPO as the task size and number varied, which was approximate to the optimal solution derived by exhaustive search.
Meta-RL methods improve the generalization performance of the scheduling scheme. However, it requires extensive exploration of data and intensive computation to converge to a stable policy.

\subsubsection{Decentralized Offloading Scheduling with RL Approaches}
\textcolor{black}{A centralized framework can obtain global observation, making coordinating all users easier. However, it leads to a large action space and high training computation when dealing with many users. A decentralized framework does not suffer from the problem of action space but guarantees that all agents operate coordinately. Thus, another way to solve the MEC offloading problem is using decentralized frameworks to model the problem.} Decentralized frameworks lead to a multi-agent architecture and scale to different edge server models. In \cite{9522334}, the authors proposed a multi-agent method to split the decision action into three sub-actions which are the target device, channel, and node power allocation. They also extended the double Q-learning with a delay-sensitive replay memory algorithm, in which the weight of the sample was calculated. The agent would choose samples with the highest weight to train. The work \cite{9606690} proposed a distributed approach in the SAGIN environment, which allowed every single device to process its own ofﬂoading decision.

\textcolor{black}{Most existing works model a decentralized framework, yet they apply traditional SARL methods to learning each offloading agent.} For instance, in the IQL method \cite{tampuu2017multiagent}, each agent used a SARL algorithm independently and treated others as a part of the environment. However, these works ignored the inﬂuences caused by other agents' decisions, which was not conducive to converging to a globally optimum solution. The work \cite{9825954} applied a standard MARL algorithm with the CTDE framework to optimize the offloading scheme, which regarded mobile devices as agents. Based on the framework, the authors combined the DQN method and the QMIX \cite{rashid2018qmix} method to propose a MARL-based ofﬂoading scheme. The method introduced a collaboration mechanism, which led an agent to take individual actions according to local observations and used a mixed neural network to estimate a Q-value of the joint action. The simulation showed a 5.4\% reduction in execution latency compared with traditional manners.

The MADDPG algorithm derives from the \textit{deep deterministic policy gradient} (DDPG) method \cite{lillicrap2015continuous} and belongs to an AC method. It adaptively learns each agent's best policy and cooperates with others for optimal joint actions. The authors \cite{9485089} proposed a MADDPG-based scheme and a federated-DRL-based scheme, which tackled the collaboration offloading issue among \textit{small base station} (SBS) agents. In the MADDPG method, each SBS was modeled as a DDPG agent, which could take action by considering itself and other agents. However, the MARL method needed to exchange local information among SBS agents, which may lead to privacy overhead. In the federated-DRL-based method, each SBS had a local DDPG model. The agent only considered its utility and was independently trained with local data, which reduced the communication overhead. After training, agents uploaded their local model to coordinators (e.g., a \textit{macro base station} (MBS)) to perform model aggregation. Then, each SBS agent received the averaged global model from the coordinator and updated the local model. The proposed scheme had a faster convergence speed and outperformed either federated RL or independent learning.
In \cite{9409695}, it proposed an AC-based MARL method jointly optimizing latency and energy efﬁciency in the UAV swarms network. Each UAV agent made an offloading decision individually and communicated with neighbors to share the value estimate to achieve a consensual assessment.

The authors \cite{9583941} proposed a MARL method with an AC framework in UAV networks to optimize the offloading and the motion jointly. In\cite{zhang2022user}, for the multi-tier computation offloading problem in the vehicular networks, the authors proposed a \textit{primal-dual deep deterministic policy gradient} (PD-DDPG) algorithm for minimizing transmission delay and computation delay under multiple discrete variables and energy consumption constraints. Furthermore, it improved this method by embedding the multi-head attention technology \cite{vaswani2017attention}. Multi-head attention enabled the scheduling model to extract more information, and it enhanced the convergence performance of the critic neural network. The proposed AT-MARL approach demonstrated a remarkable 27.2\% increase in accumulated reward compared to MADDPG methods. Notably, MADDPG exhibited a performance similar to that of the genetic algorithm, a commonly employed technique in random optimization. However, it is crucial to highlight that the genetic algorithm faces challenges stemming from computational complexity and limited adaptability to dynamic environments.

\textcolor{black}{Game theory has been applied in various ﬁelds of MEC as a valuable tool to optimize ofﬂoading scheduling and power allocation.}
Game-based model formulation considered that mobile users were selfish and they pursued their own goals. The authors in \cite{9295872} combined game theory and RL to propose a Nash-Q-learning method. In the multi-user game, Nash equilibrium could effectively reduce interference and achieve stability for offloading decisions. In \cite{9838691}, the authors proposed a MARL-based Nash-Q-learning method in the multi-user MEC system, which sets each user as an agent. Considering the general Markov game, the authors analyzed the current stage payoff matrix of joint action. They used the proposed method toward the Nash equilibrium point in the general Markov game.
The results showed that the MARL-based Nash-Q-learning method outperformed the DQN-based and MARL-based \cite{8385143} methods with a 30\% improvement in the total utility.

The work \cite{9590352} focused on the vehicular network offloading problem. In this work, the task ofﬂoading process consisted of the local-or-ofﬂoading stage and the RSU-edge-cloud stage. The authors applied a potential game-based method to solve a local-or-ofﬂoading stage problem. The potential-game-based manner could map the utility generated by changing any user's decision to a potential function. Optimizing the potential value achieved a globally optimal solution. 
In the RSU-edge-cloud stage, the potential game-based method selected the offloading target node and then used a Q-learning method to allocate power resources. Most studies assume that all servers share their information. Thus, servers generally needed more coordination and communication \cite{asheralieva2019hierarchical}.
Then, this work proposed a two-layer hierarchical game model in the UAV-enabled MEC network, in which a coalition formation method for the upper layer and an RL-based offloading method for the lower layer were established. The upper layer was a cooperative game in that BSs formed coalitions with shared computing resources. The lower layer contains several stochastic subgames. In addition, the authors proved that the proposed methods converged to a mixed-strategy Nash equilibrium. The authors modeled the vehicular network offloading as the second-price combinatorial auction with \textit{Vickrey-Clarke-Groves} (VCG) mechanism \cite{9796717}. This approach implemented a MARL method and regarded a vehicle as a bidder agent. The authors described a vehicle's service request as a bid in an auction, and the bid included service details, bidding price, and the vehicle's estimated resource needs. The work proved that the VCG mechanism led to a Nash equilibrium and a social welfare maximization in the static case. For the dynamic case, the proposed MARL method learned a best-response strategy updated in a ﬁctitious self-play approach to pursue lower cost or better payoff. Nash Q in this research contributed to a 40\% reduction in offloading failure rate and a 32\% decrease in communication overhead. Jin \textit{et.al.} \cite{jin2024fractional} studied the joint optimization of task updating and offloading with fractional form objective measured by average AoI. This problem posed challenges of hybrid continuous-discrete action space and fractional objective of average AoI which conventional RL approaches maximizing the accumulated reward can not address. They proposed a novel fractional Q-learning method to overcome these issues. They turned to the Dinkelbach’s reformulation to obtain an equivalent problem with a summation-formed reward by introducing a quotient coefficient. They utilized two different DRL networks to deal with hybrid action space: DDPG for continuous waiting time decision and D3QN for discrete edge node choice. They proved that the fractional Q-learning method showed a linear convergence rate and their method reduced the average AoI by up to 57.6\% compared with other baselines.

Some classical RL-based offloading scheduling schemes are listed in Table \ref{table:offloading}. For convenience, this table uses some abbreviations. SA and MA are single-agent and multi-agent, respectively, and Cen., Dec., CTDE are centralized, decentralized, and centralized training decentralized execution, respectively. The FL is the abbreviation of federated learning, and the NE is the abbreviation of the Nash equilibrium. The AC is the abbreviation of the AC, and the AM is the abbreviation of the attention mechanism. 

\renewcommand\arraystretch{1.2}
\begin{table*}[t]
\normalsize
\caption{The RL-based Task Offloading Approaches in MEC Networks.}
\begin{center}
\begin{tabular}{m{1.5em}|p{3.3cm}|p{1.4cm}|p{1.4cm}|m{9cm}}
\hline\hline
{\bf Ref.} & {\bf Objectives}& {\bf SA or MA}&{\bf Cen. or Dec.}  &{\bf RL Approaches and Improvements}\\
\hline
\cite{8552454}&\begin{tabular}{@{}l@{}}
    {\footnotesize $\bullet$} Response delay\\
    {\footnotesize $\bullet$} Energy consumption
\end{tabular}& \multirow{5}{1.4cm}[-13ex]{SA} & Cen.& Both dynamic RL and DDQN are utilized. An accurate approximation of the value function guarantees excellent performance.  \\
\cline{1-2}\cline{4-5}
\cite{9631180} & \begin{tabular}{@{}l@{}}
    {\footnotesize $\bullet$} Energy consumption
\end{tabular} & & Cen.&DQN is applied and the state is discretized into several levels. DQN can output multi-dimensional actions. It obtains a higher utility function than rule methods.\\

\cline{1-2}\cline{4-5}
\cite{8493155}& \begin{tabular}{@{}l@{}}
    {\footnotesize $\bullet$} Long-term utility
\end{tabular}  & & Cen. & A DDQN-based algorithm breaks the curse of high dimensionality in state space. The structure of the objective function motivates a linear combination of Q functions.\\
\cline{1-2}\cline{4-5}

\cite{8513863}  &\begin{tabular}{@{}l@{}}
    {\footnotesize $\bullet$} Average delay
\end{tabular} & & Cen.  & An AC approach with a fixed target network and experience replay buffer is implemented. The natural policy gradient to avoid local minima.\\

\cline{1-2}\cline{4-5}
\cite{9448034} &\begin{tabular}{@{}l@{}}
    {\footnotesize $\bullet$} Average delay\\
    {\footnotesize $\bullet$} Energy consumption
\end{tabular} & & Cen. & A DMRO policy is composed of a DQN-based with multiple parallel DNNs to output single-dimensional actions. A group of meta-learning parameters is set to each DNN, which enables adaptability to varying environments.\\
\hline
\cite{9825954}&\begin{tabular}{@{}l@{}}
    {\footnotesize $\bullet$} Average delay\\
    {\footnotesize $\bullet$} Energy consumption
\end{tabular}& \multirow{4}{1.4cm}[-13ex]{MA} & CTDE & Combining DQN and QMIX in training. QMIX estimates the global value function given the local observation and actions of every single agent.\\

\cline{1-2}\cline{4-5}
\cite{9485089} & \begin{tabular}{@{}l@{}}
    {\footnotesize $\bullet$} Energy consumption
\end{tabular} &  & Dec.
 & A standard MADDPG is utilized. Each user has an individual actor and critic network, and the training is performed in a decentralized manner, compared with the centralized training of QMIX in \cite{9825954}. Federated learning helps to reduce computational complexity and signal overhead.\\

\cline{1-2}\cline{4-5}
\cite{9583941} &\begin{tabular}{@{}l@{}}
    {\footnotesize $\bullet$} System utility
\end{tabular} &  & CTDE  & A multi-agent AC-based structure is utilized. Due to the high variance of the critic network (especially in a multi-agent environment), a multi-head attention mechanism is involved, which improves convergence and scalability.\\

\cline{1-2}\cline{4-5}
\cite{asheralieva2019hierarchical} & \begin{tabular}{@{}l@{}}
    {\footnotesize $\bullet$} Average delay\\
    {\footnotesize $\bullet$} Energy consumption
\end{tabular} & & Dec.& A 2-level coalition game is modeled: upper level for cooperation and lower level to maximize self-interest. Deriving Nash Q-value is another way to estimate the global value function compared with QMIX, MADDPG, etc.\\

\cline{1-2}\cline{4-5}

\cite{jin2024fractional}&\begin{tabular}{@{}l@{}}
     {\footnotesize $\bullet$} Age of information  \\
\end{tabular}& &Dec.& Solving problems of fractional objectives by introducing quotient coefficient. Utilizing two DRL networks to address hybrid action space.\\

\hline\hline

\end{tabular}
\end{center}\label{table:offloading}
\end{table*}

\subsection{Content Caching}

\subsubsection{Performance Optimization Utilizing RL}
In MEC networks, emerging applications, such as IoT, IoV, healthcare, and VR/AR, motivate computational capacity, storage, bandwidth, and energy demands. Mobile edge caching provides a way to satisfy these demands by storing an incredible amount of popular content. 
The advantages of mobile edge caching are as follows \cite{MECsurvey}:
\begin{itemize}
    \item The caching nodes (i.e., edge servers) are equipped near the edge devices rather than cloud servers. Thus, the latency of retrieving contents is reduced;
    \item The traffic in backhaul links can be significantly alleviated by avoiding frequent connections between cloud servers and edge devices;
    \item Requesting caching contents consumes less energy;
    \item Caching enhances the efficiency of the spectrum.
\end{itemize}

All these benefits motivate the study of edge caching in the MEC network. Regarding the caching process, there are four phases to be considered: content request, content exploration, caching delivery, and caching update. First, edge users generate requests and then search the contents throughout the mobile network. Then, based on the result of the exploration, contents are delivered to the user from an edge node or a remote node. If the popularity of content changes, the caches are updated. This process of mobile edge caching has four problems to solve: (1) content delivery; (2) content placement; (3) content updating; and (4) joint delivery and placement.

\subsubsection{RL in Various Caching Phases}
\begin{figure}[t]
    \centering
    \includegraphics[width=0.5\textwidth]{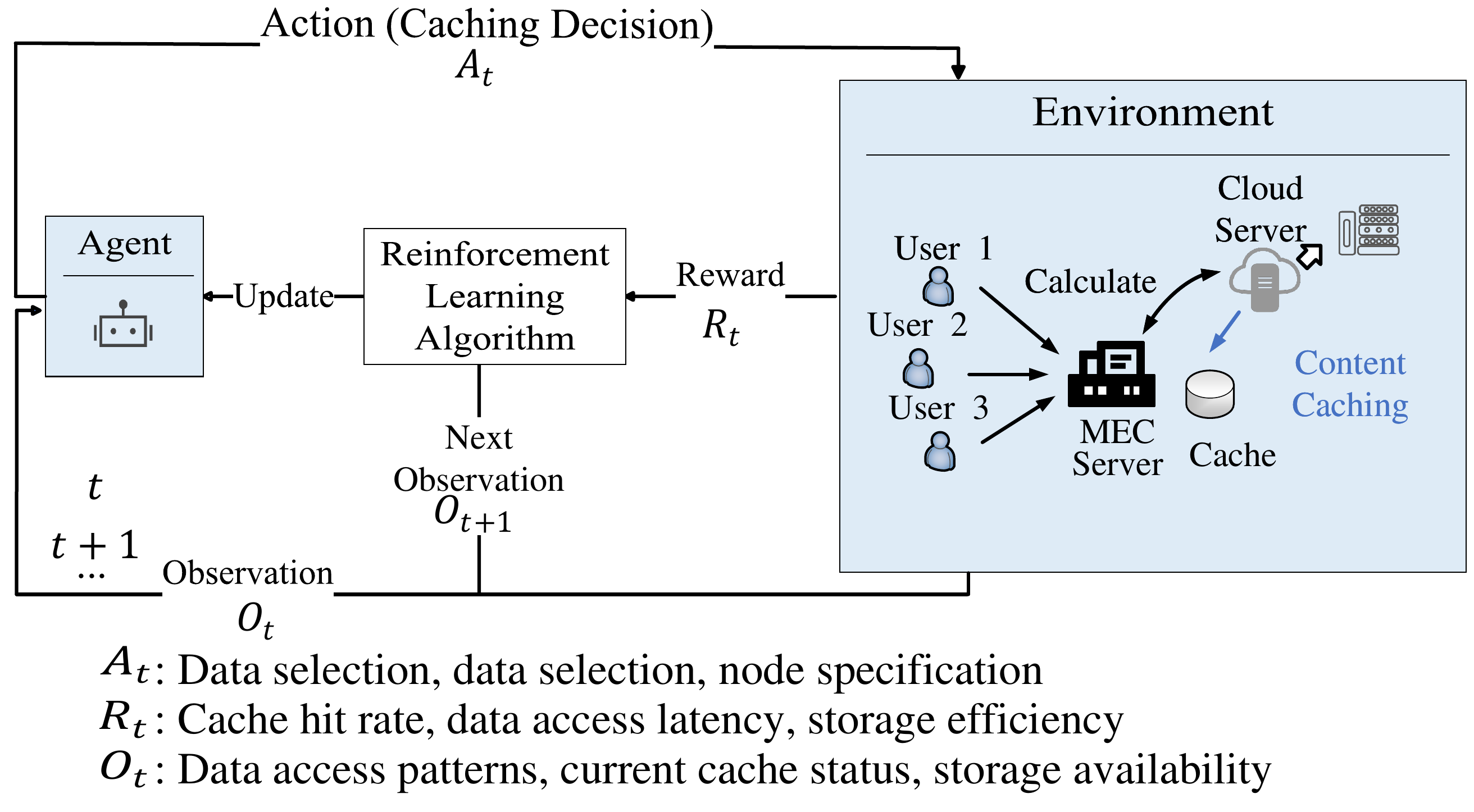}
    \caption{MDP for content caching.}
    \label{bs_cache}
\end{figure}

 We begin by presenting the baseline problem formulation within the context of a classic caching task in the content delivery scenario. This scenario involves multiple edge servers and users, each with diverse requirements for caching content. Both the edge servers and content files are indexed, and the limited storage capacity of each edge server necessitates a strategic selection of content files for caching. The simplest caching condition stipulates that each edge server can cache only one content file at a time, resulting in a cache matrix that signifies the specific type of content file stored by each edge server. The caching indicator is represented as
 \begin{equation}
     x_{n,c}=\left\{\begin{array}{cc}
        1,& \rm{cache},  \\
        0,& \rm{otherwise},
     \end{array}\right.
 \end{equation}
 where $x_{n,c}=1$ means whether the file $c$ is cached at edge server $n$. Similarly, the user association indicator and the request indicator are denoted as $y_{u,n}$ and $z_{u,c}$. The user $u$ successfully request content $c$ from server $n$ when $x_{n,c}y_{u,n}z_{u,c}=1$.
 Mobile users exhibit varying content preferences, characterized by content popularity, often modeled using the Zipf distribution \cite{tatar2014survey}. One of the primary objectives of the caching problem is to minimize content delivery delays. In contrast to offloading problems, caching issues prioritize data efficiency over energy considerations. A crucial metric in caching problems is the hit probability, representing the ratio of cached files requested by edge users to the total number of files cached. Within a given time window of length $L$, the reward of hit probability increased by $\frac{1}{L}$ if the content is requested at each time slot, or remains unchanged otherwise. A higher hit rate indicates a more efficient data flow, contributing to optimized network throughput. The data rate can be represented by
\begin{equation}
    r_{n,c}=\log_2\left( 1+\frac{x_{n,c}y_{u,n}z_{u,c}p_{n,c}|h_{n,c}|^2}{I_{n}+\sigma^2}\right),
\end{equation}
where $p_{n,c}$ is the average transmmit power of edge server $n$ when delivering content indexed by $c$, $|h_{n,c}|$ is the corresponding channel gain, and $I_{n}$ is the interference at server $n$. In summary, a standard caching problem takes into account the cached content, edge user requirements, and user mobility as the state. The goal is to maximize the accumulated average hit rate or network throughput by allocating the transmit power, caching location, and the selection of cached content. Fig. \ref{bs_cache} illustrates a generic model for an RL-based caching structure, providing a visual representation of the interplay between various components in the caching system.

\textcolor{black}{Content delivery refers to the pathway bringing the cached content to edge users and content placement scheduling the locations of the content.} In \cite{nanReinforcementLearningBasedOptimizationContent2019}, they considered a content delivery problem for which the contents were cached at MBSs and SBSs. The content placement was fixed, and edge users requested different content from MBSs or small BSs to reduce the transition expense. They modeled the mobility of users as an MDP utilizing Q-learning to minimize the delivery cost. Uncertainty existed in the mobility trajectory of mobile devices, and classic Q-learning was able to capture the varying information. The ability to learn the changing environment made RL reduce the cost by 14.2\% compared with the greedy caching scheme, which performed poorly when users were not located in the same SBS. 

In \cite{hanplacementmab}, they developed an extended MAB to estimate the density of users and the popularity of the contents, i.e., two correlated features that affected the decision of the content placement. Therefore, they combined a standard MAB with a generalized global bandit miscalculation when edge servers owned overlapped service regions. The reward of each arm to a bandit is characterized by the combination of content popularity and global reward distribution. The extended MAB was then designed to maximize the newly defined accumulated reward and lower the complexity of the joint action-state space. This approach also worked in decentralized cooperative ways by exchanging information with adjacent edge servers. The proposed scheme reached the lowest regret bound among approaches such as UCB, epsilon greedy, and LFU. This work is the first to consider both user density and content popularity within an MAB framework, which extends the use of MAB manners into a more complicated environment.

\textcolor{black}{Content delivery also involves the consideration of hit probability, delay, throughput, energy consumption, etc., and is always coupled with each other.}
Kirilin \textit{et al.} \cite{kirilinRLCacheLearningBasedCache2020} designed an RL-cache policy that determined the content placement manners to maximize the hit rate. The instantaneous reward was obtained only when the next request was the same as the previous one. Hence the Q-learning or TD methods could not give a precise estimation of the Q-value of the state-action pairs. To address the problem of noisy reward, they turned to the DPS approach, which recorded the sequence-long return. At the same time, they utilized individual returns instead of average returns to build a subset with high returns in which a better policy was searched. Such RL-cache approach outperformed Q-learning, TD methods, and MC methods and is also robust when implemented in other CDN servers of the same geographic region. This paper highlighted the utilization of DPS in a noisy reward environment, which built a possible connection with policy-based RL approaches like PPO. In this particular scenario, the noise is not white and has something to do with the next state and the total reward. This inspires using long-term return or training in epoch to mitigate the impact of noisy reward. They also highlighted the impact caused by the hyperparameters. The trade-off between the hit rate and the number of samples should be considered since a longer sequence of subsequent requests may deteriorate the performance.

Ji \textit{et al.} \cite{jitrajectorycache} considered the content transmission in a dense multi-access cellular network assisted by UAVs. There were multiple objectives that involved content acquisition delay, transmission power, and UAV trajectories. Since the environment was highly complicated, they modeled the problem into a partially observable stochastic game where the MBs and UAVs were deemed as agents, and decisions were made based on partial environmental states and agent actions. They formulated the optimization problem as a stochastic game and a Dual-Clip PPO manner was deployed to each agent in a cooperative way. The problem was not tractable with classic RL methods like DQN and DDPG since the state variables were continuous and the variance of value estimation could be extremely high. Therefore, a Clip-PPO was implemented, which fed on continuous state space and gave the distribution of action as the output. To encourage exploration, UAVs were given an intrinsic reward when the agent visited regions beyond the explored areas. PPO takes the advantage function as the objective to be estimated and uses the clip technique to limit the update of parameters, which derives a lower variance. On the other hand, the exploration efficiency and speed of convergence will improve by designing a suitable intrinsic reward during training. The proposed scheme improved the content acquisition delay against classic PPO and dual-clip PPO without intrinsic reward.


\subsubsection{RL-Enabled Caching Scheme}

There are also many caching criteria and schemes to deal with caching problems. Caching schemes aim to increase the hit probability, enhance spectrum efficiency and energy efficiency performance, and minimize latency. 
The typical caching problem includes cooperative caching \cite{linCooperativeCachingTransmission2020}, coded caching \cite{gaoReinforcementLearningBased2020}, game-theory-based caching \cite{sunGameTheoreticApproachCache2019a}, and proactive caching \cite{liuDeepReinforcementLearning2019}. 

\renewcommand\arraystretch{1.2}
\begin{table*}[t]
\normalsize
\caption{The RL-based Content Caching Approaches in MEC networks.}
\begin{center}
\begin{tabular}{m{0.8cm}<{\centering}|m{4cm}|m{4cm}|m{7.3cm}}
\hline\hline
{\bf Ref.} & {\bf Caching Phases} & {\bf Objectives}& {\bf RL Approaches and Improvements}\\
\hline
\cite{nanReinforcementLearningBasedOptimizationContent2019}&{{\footnotesize$\bullet$} Content delivery}&{{\footnotesize$\bullet$} Delivery cost}  &For fixed content placement, standard Q-learning is able to capture the dynamic environment.  \\
\hline
\cite{kirilinRLCacheLearningBasedCache2020}&{{\footnotesize$\bullet$} Content placement}&{\begin{tabular}{@{}l@{}}
     {\footnotesize$\bullet$} Hit rates\end{tabular}}&Monte Carlo sampling and Direct Policy Search. The hit-rate-related reward depends on the state of the next time slot. Therefore RL approaches like DQN and DDQN may not perform well. DPS searches for the best policy among the trajectories that have high long-term accumulated rewards, which addresses the problem of the noisy reward. \\

\hline
\cite{sunGameTheoreticApproachCache2019a} & \multirow{3}{3.5cm}[-7ex]{\begin{tabular}{@{}l@{}}
    {\footnotesize$\bullet$} Content delivery\\
    {\footnotesize$\bullet$} Content placement
\end{tabular}}& {\begin{tabular}{@{}l@{}}
     {\footnotesize$\bullet$} Long-term utility \end{tabular}}& 
     \hspace{-0.25cm}\multirow{2}{7.3cm}{\begin{tabular}{p{7.3cm}}
    A Stackelberg game is modeled and a leader-follower structure is proposed. Stateless MARL which does not use the value function of action-state pairs is utilized. This reduces the complexity.\end{tabular}}\\
\cline{1-1}\cline{3-3}
\cite{al-abiadJointReinforcementLearningEnabled2022a}&&\multirow{1}{3.5cm}{\begin{tabular}{@{}l@{}}
     {\footnotesize$\bullet$} System sum rate
\end{tabular}}& \begin{tabular}{@{}l@{}}
   \\ \\ \\ 
\end{tabular}\\ 
\cline{1-1}\cline{3-4}
\cite{cpdA2C}&&{\begin{tabular}{@{}l@{}}
     {\footnotesize$\bullet$} QoS\\
     {\footnotesize$\bullet$} Backhaul load
\end{tabular}}&A2C, instead of AC, reduce the variance of the actor-network. Using Dirichlet distributions instead of Gaussian distributions for action output fulfills the constraints of caching decisions.\\
\cline{1-1}\cline{3-4}

\hline

\cite{vehicle_cache}&{\begin{tabular}{@{}l@{}}
     {\footnotesize$\bullet$} Content placement  \\
     {\footnotesize$\bullet$} Request update  
\end{tabular}}&{{\footnotesize$\bullet$} Long-term overhead}& Markov game formulation, and distributed Q-learning with Nash Q-function. What's different from the Stakelberg game is the agents take actions at the same time, and the global optimum is not guaranteed. \\ 
\hline

\cite{liuDeepReinforcementLearning2019} &{\begin{tabular}{@{}l@{}}
    {\footnotesize$\bullet$} Content delivery \\
    {\footnotesize$\bullet$} Content recommendation
\end{tabular}}& {\begin{tabular}{@{}l@{}}
    {\footnotesize$\bullet$} Net  profit
\end{tabular}}& Dueling DDQN owns faster convergence speed and more appealing policies compared with A2C and PPO. When the desired policy is deterministic, utilizing AC or policy-gradient-based approaches may result in a stochastic policy.\\
\hline\hline

\end{tabular}
\end{center}\label{table:caching}
\end{table*}

\paragraph{Cooperative Caching}

Cooperative caching balances the popularity of contents and the limited storage space of BSs. BSs share content with each other, reducing the latency of searching and retrieving content. Some current work focuses on joint cooperative caching and transmission schemes. In \cite{hierarchical_cache}, the authors studied mobile network caching with \textit{coordinated multipoint} (CoMP) techniques to improve data rates in networks, considering both the caching and the transmission process. The dynamic request and location distribution of edge devices were constantly changing, which made the popular content of the edge cache also change. They assumed that the content popularity obeyed a Zipf distribution and that the dynamics of edge devices connecting to an edge server followed the Poisson process. The authors designed an MDP for this problem with the request distribution and the content request preference as the input. Each edge server could determine the type and size of the contents it caches. The caching strategy combined with a standard Q-learning method. The Q-table is an efficient method to record the value of the Q-function with low-dimensional spaces and guarantees real-time performance compared with the DQN.

Cooperative caching models were constructed with a hierarchical caching structure \cite{hierarchical_cache}, i.e., parent nodes connected to many leaf nodes. The behavior of leaf nodes exerted a significant effect on neighbor nodes. The network dynamics, including the interaction between parent and leaf nodes and content requests, made formulating the caching model challenging. They designed a scalable RL approach in \cite{hierarchical_cache}, assuming those leaf nodes stored anticipated popular contents locally and parent nodes provided contents not served by leaf nodes. The authors divided the time horizon into two scales: fast time slots and slow slots. The leaf nodes determined what to store with a fast time slot, while parent nodes decided on slow slots. The policy of leaf nodes used existing methods, and the policy of the parent node utilized a DQN method to approximate optimal Q-functions. They also used a hyper-DQN which executed multiple DQNs in parallel to address the hierarchical caching problem with multiple parent nodes. In the IoV, unlike common edge nodes, vehicles suffered from limited storage capacity and tight content delivery deadlines, obtaining a caching policy for IoV non-trivial. The authors formulated hierarchical architecture and introduced MDP to minimize the overhead \cite{vehicle_cache}. They also extended the MDP into a multi-agent system and proposed a MARL caching approach to alleviate mobile traffic and reduce the latency of accessing content. This structure witnessed a 19\% improvement in hit rate compared with independent RL, and at least 36\% better than simple rules like LRU or LFU.

\textcolor{black}{Some works focus on cooperative caching and content delivery.}

\textcolor{black}{\it Joint content placement and delivery:} In \cite{cpdA2C}, they utilized A2C to find an optimal caching placement and delivery policy. The authors applied a probabilistic model to characterize the content placement process and formulated a cache placement and delivery problem as an MDP. Compared with an AC-based method, the A2C-based policy reduced the variance of the actor network by using the advantage function. They also utilized the Dirichlet distribution to generate the action vector instead of the Gaussian distribution. If caching strategies use outdated information to schedule the caching process, the performance of MNC will be significantly damaged \cite{cachingsurvey}. In \cite{cacheMDP}, the authors proposed the caching problem in a slotted structure, and each slot contains three stages: first, content delivery happened at the beginning; next, the BSs (or edge server, edge nodes) stored the content would exchange information to obtain the popularity of contents; finally, the BS determined contents to hold for the next time slot. They defined the ratio of local requests to local requests as the state. The BS would determine a subset of files to be stored. The cost function considered the update of caching contents, the satisfaction, and the mismatch of actions. They derived optimal caching policy through Q-learning.

\textcolor{black}{\it Joint content caching and transmission:} The authors in \cite{mi2021joint} proposed a joint transmission and cooperative caching approach in a CoMP network, where joint and single transmission was allowed for edge users. It aimed to minimize the transmission delay for all devices by a bi-level optimization for caching and transmission. The authors designed a MADDPG-based MARL caching mechanism to cache popular content. Based on these cached contents, they combined a Bayesian learning automaton method to present a transmission scheme for multiple agents. Bayesian learning automaton generates the Bayesian estimation of the reward probability of actions based on the beta distribution, which has provable convergence and low computing complexity and is suitable to implement in a decentralized edge server. However, the stability and convergence of this joint caching strategy may be damaged in a large-scale non-stationary environment.

\paragraph{Coded Caching}

Coded caching is a distributed caching approach that reduces network load by exploiting and creating coded multicasting opportunities among users for different demands. Coded caching integrates two or more packets into one single message, and then edge users separate the encoded message to the original form. However, coded caching involves significant processing overheads when decoding packets at the user. Therefore, they need to consider the equilibrium between the encoded messages and the processing time. 

Cooperative coded caching technology was considered in the \textit{ultra-dense network} (UDN), which was the key feature of 5G networks. The UDN could significantly improve the data throughput by deploying SBSs, which coexisted with MBSs. However, transmission data burdens the caching network pressure. Therefore, the authors studied a cooperative coded caching strategy \cite{ultradense_cache} in UDN. Each edge device could be served by multiple SBSs simultaneously. The agent observed the number of requested contents during peak hours and recorded decisions made during these off-peak hours at the previous time slot. The action was the type and amount of contents cached at SBSs. They reformulated the problem as an MDP and utilized the Q-learning approach to maximize traffic offloading. In \cite{gaoReinforcementLearningBased2020}, they investigated a cooperative coded caching between MBSs and SBSs in a UDN. SBSs cooperatively delivered content to users, which greatly increased the efficiency \cite{cooperativeefficiency}. The MBS decided how much content to be stored at SBSs. The authors utilized Q-learning to solve the small-scale caching problem. For a large-scale problem, they designed a parameterized function to approximate a real Q-value and updated the parameter of the approximation function through \textit{stochastic Gradient descent} (SGD). The simulation showed optimal and near-optimal performance for small and large-scale conditions.

\paragraph{Game-Theory-based Caching}
Many parts co-work in MEC systems, including the service providers, mobile network operators, and mobile users \cite{cachingsurvey}. However, different parts may conflict in maximizing the system's profit. Game theory is utilized to balance various aspects of the caching process. The authors in \cite{sunGameTheoreticApproachCache2019a} studied a joint cache resource and radio resource optimization scheme. They established a hierarchical resource management model. The cache resources were optimized in the upper layer according to the channel gain and the content request. In the lower layer, the \textit{access points} (APs) were decomposed into clusters to mitigate the interference among each other. They modeled the competition between cache resources and radio resources as a Stackelberg game, where the resource manager acted as the leader, and APs acted as followers. The strategy also had two stages in different time scales: In the cluster formation stage, APs cooperated to form clusters that reached the Nash stable. In the caching optimization stage, each leader chose the feasible caching action by the SARL or the MARL algorithms. While SARL algorithms suffered from the curse of dimensionality, the MARL algorithm showed a faster convergence and near-optimal performance.

Some works study game-based content placement and delivery schemes. The requested content affected the content placement process, while the content distribution determined the content delivery schemes. The authors in \cite{al-abiadJointReinforcementLearningEnabled2022a} the resource allocation and content delivery were optimized with \textit{cross-layer network coding} (CLNC), game theory, and MARL approaches for multiple leaders and multiple followers. For the first stage, they used a Stackelberg game to divide the user clustering with several leaders, and each group utilized the CLNC resource allocation. Then, each leader conducted content delivery by the MARL algorithm with partial observation.
Furthermore, the proposed MARL-based caching and delivery algorithm had low computation complexity due to being stateless. In \cite{vehicle_cache}, the authors studied the game-based cooperative caching in the IoV network with RSUs and MBS. When a vehicle requested content, the connected RSU searched if contents were cached in their storages. The RSU could also explore the relevant content at the adjacent RSU or request the content from the MBS. The authors modeled the content delivery process as a hierarchical network and established an MDP to minimize the system's overhead. They utilized a Q-learning method to solve the origin problem. The authors also extended the MDP to a Markov game, which considered the interaction with other agents. The authors utilized the Nash Q-function to conduct distributed Q-learning, obtaining the maximal value function.

\paragraph{Proactive Caching}

Proactive caching policies determine what and where to cache specific content before it has been requested. By predicting the demands of users, the BSs determine what contents are the most popular and cache them to the storage in advance. Proactive caching usually happens during off-peak times and serves edge users with pre-downloading contents during peak hours. In \cite{liuDeepReinforcementLearning2019}, the authors considered a joint content recommendation and delivery problem. The caching process was divided into two stages: the BS would recommend content to the associated user according to the pre-determined store policy, and the user could either accept or decline the recommended content. If the content was declined, the user would request new content from the BS. Therefore, they divided this problem into two RL subproblems, which dealt with the recommendation and pushing process. They assumed that the impacts of the recommendation on the pushing process outweighed that of the pushing process on recommendation, and the recommendation policy fed on the pushing history as input. They utilized a dueling DQN structure for training two agents, and the parameters of this training network were updated at each step. It had a better performance than the centralized A2C and the PPO architectures which suffered from the huge state space. The near-optimal performance also proves the importance of the decomposition of MDP and content recommendation. It is worth noticing that a desirable caching recommendation should be deterministic, while the AC and PPO may derive a stochastic policy that is less efficient than dueling DDQN. Therefore, the selection of RL approaches should match the specific structure of the optimal solution.

Based on previous research, some classical RL-based content caching schemes are listed in Table \ref{table:caching}. 

\subsection{Communication}\label{subsec:Communication}

\subsubsection{Performance Optimization Utilizing RL}
The growing number of edge devices adds numerous mobile data to the MEC system and affects communication efficiency. It brings higher-quality communication demands, such as lower service latency and energy consumption. Efficient wireless communication resource management in a limited-resource MEC network is essential to guarantee stable data transmission and satisfy the requirements of massive and heterogeneous users \cite{zhang2020power}. Communication resource management generally includes spectrum access, spectrum allocation, and power control and allocation. Some advanced communication techniques are required due to the following reasons\cite{goldsmith2005wireless}:
\begin{itemize}
    \item \textit{Multi-path fading}\cite{puccinelli2006multipath}: The status of channels is highly time-varying, which results in great \textit{inter-symbol-interference} (ISI) \cite{lathi1995modern}. This occurs when a pulse spreads out to adjacent at the sampling instant;
    \item \textit{Spectrum access collision}: The spectrum is a kind of transmission resource composed of a narrow band of frequencies. If two users transmit signals through a channel occupying the same spectrum, the signals will interfere with each other and reduce the \textit{signal to interference plus noise ratios} (SINRs). The spectrum access collision will significantly reduce the transmission efficiency and impair the real-time performance of the system;
    \item \textit{Spectrum shortage}: High bandwidth requires wider bands of frequencies, and mobile devices cannot occupy the same spectrum simultaneously. The demand for bandwidth grows, and the available spectrum will become scarce;
    \item \textit{Energy limitation}: Edge devices are designed to connect to edge servers for real-time tasks. However, due to the limited energy, it is vital to manage the power allocation of wireless channels.
\end{itemize}

This has driven the development of wireless communication resource management in MEC networks. However, some dynamic communication environments, such as interfered channels, channel fading, and spectrum access collision, lead to the traditional operation optimization methods with given system parameters being unable to find an efficient solution. RL approaches are designed to make intelligent sequential decisions in unknown or non-deterministic environments, helping to develop spectrum sharing, prediction, aggregation techniques, and power allocation. As 6G envisions edge intelligence, there is a vast potential to introduce the RL method to MEC to allocate wireless communication resources dynamically. 

\begin{figure}[t]
    \centering
    \includegraphics[width=0.5\textwidth]{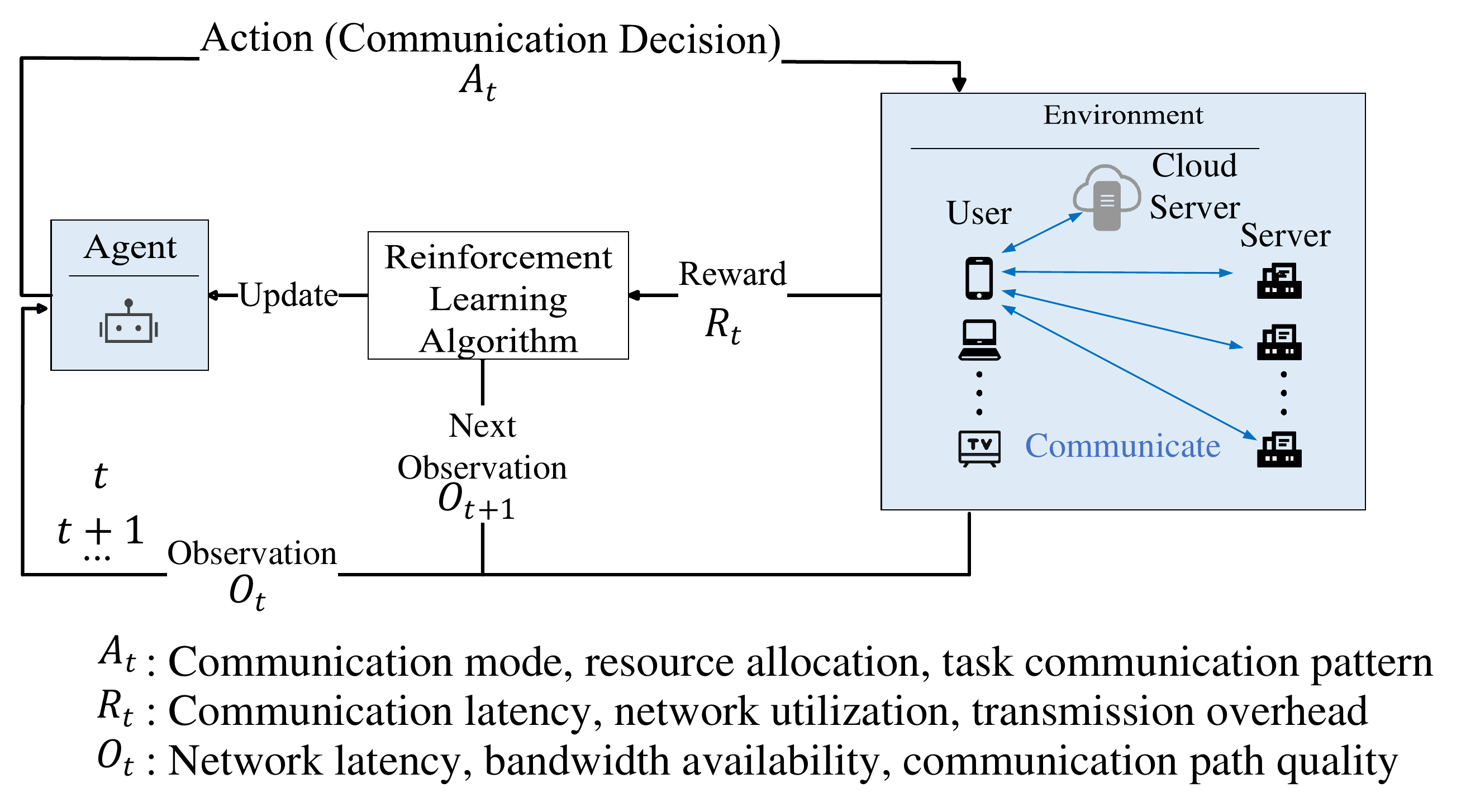}
    \caption{MDP for communication.}
    \label{bs_communicate}
\end{figure}

Before delving into specific papers, we provide a comprehensive overview of the standard elements involved in communication optimization through a standard problem formulation. In general, the optimization criteria of communication problems, including energy efficiency, data rate, and delay, are established as the reward for the actuator, which endeavors to maximize the accumulated reward. The distinction in communication problems lies in the configuration of the state. While offloading or caching tasks primarily concentrate on application-layer scheduling, communication problems emphasize physical-layer scheduling, encompassing spectrum access, transmission power, and channel selection. Concerning this, the objectives are usually defined as the data rate and energy efficiency. The data rate for user $n$ through channel $i$ is characterized by
\begin{equation}
    r_{n,i}=\log_2\left( 1+\frac{p_{n,i}|h_{n,i}|^2}{I_{n,i}+\sigma^2}\right),  
\end{equation}
where $p_{n,i}$ is the average transmit power of edge user $n$ through channel $i$, $|h_{n,i}|$ is the corresponding channel gain, and $I_{n,i}$ is the inter-cell interference. The \textit{energy efficiency} (EE) is defined as
\begin{equation}
    EE=\frac{\sum_n\sum_i B_i\log_2(1+\gamma_{n,i})}{\sum_nP_n},
\end{equation}
where $B_i$ is the bandwidth allocated for channel $i$, $\gamma_{n,i}$ is the SINR for user $n$ via channel $i$, and $P_n$ is the total power consumption of user $n$. Typically, the state is represented as a matrix detailing the SINR within each user-to-user pair. In the context of delay optimization, two conditions are considered. The first involves transmitting a packet with a specified size, and the calculation of delay aligns with approaches used in offloading problems. The second scenario arises when contending with unreliable channels. Multiple users must coordinate channel scheduling to transmit a packet and receive an acknowledgment, aiming to minimize the average delay. In such instances, the state shall encompass the selection of other edge users and the current delay experienced by those users. Fig. \ref{bs_communicate} provides a visual representation of a generic model for an RL-based communication structure, illustrating the intricate interplay among different components within the communication system. This visual aid enhances our understanding of the complexities involved in optimizing communication processes through RL methodologies.

\subsubsection{Centralized Communication Resource Allocation Using RL}
RL-based approaches obtained near-optimal performance in most single-user scenarios. \textit{Dynamic spectrum access} (DSA), or dynamic spectrum management, is a critical technology to enhance spectrum efficiency. Spectrum allocation is a typical dynamic spectrum access, which optimizes spectrum assignment via exploiting the temporal and spatial traffic statistics of various services to improve spectrum efficiency \cite{zhao2007survey}. Specifically, given a fixed region and a special time, the divided spectrums (i.e., subchannel, subcarriers) are allocated to provide user services.

\textcolor{black}{The spectrum allocation problem is also widely studied in \textit{device-to-device} (D2D) communication mode for the MEC system. The communications are mainly between edge servers and edge users, and the direct connection is from D2D.} Edge servers are always equipped at the wireless APs (i.e., wireless routes and BSs) to reduce the site rental \cite{MECsurvey}. In D2D communication, edge users cannot connect to edge servers and can only communicate with the neighbor users \cite{D2Dsurvey}. Thus, some research focuses on decentralized spectrum allocation in D2D communication.

\textcolor{black}{Some research focuses on RL-based DSA problems in MEC networks.} There are two typical schemes to deal with DSA problems: listen before talk \cite{LBT} and spectrum sharing \cite{spectrumsharing}. RL approaches handle these two problems by integrating the current observation and the past learned information. In \cite{1SUDQN}, they studied a multi-channel selection problem for a single-edge user. The user was modeled as the agent, and it transmitted packets through multiple correlated channels, which was modeled as a Markov process. To optimize the number of long-term transmission success rates, they formulated a POMDP for the spectrum access problem with unknown parameters and utilized a classic DQN algorithm. The DQN showed brilliant performance in an unseen environment with large spaces, which approximates the optimal solution given a system with known statics. DQN is a value-based model-free RL method, which learns the value of state-action pairs without knowing the statics of the environment. The long-term performance of DQN outperforms myopic greedy policy or heuristic policy like the Whittle index. The authors derived the optimal policy given system dynamics, which had a round-robin structure. This structure also implied that the optimal policy is deterministic, and it was not even necessary to know the transition probability. The DQN chose actions that had a maximum Q-value (by $\epsilon$-greedy), which has the potential to approximate the optimal policy through interaction with the environment. The proposed method reached an increase of 37.5\% in accumulated reward compared with traditional approaches like myopic policy and whittle index policy and approximated the optimal solution. However, under conditions that include more stochastic behavior such as multi-agent scenarios with competition, methods like DQN and DDPG may not behave well, and we will discuss it later in this section.

\textcolor{black}{The energy capacity of mobile edge terminals is limited, and energy constraints power allocation.} Therefore, energy limitation is an essential factor that needs to be considered for MEC devices. Guo \textit{et al.} \cite{guoResearchReinforcementLearningBased2018} considered the tradeoff between service performance and energy consumption. The tasks to offload were modeled as a FIFO queue which satisfied the Markovian property. The power manager utilized DQN to learn the optimal timeout threshold to reduce power consumption. Since there were two objectives (QoS and energy consumption), a weighted-sum reward function was designed. DQN is also capable of addressing multi-objective problems by modifying the reward function. In this study, the dynamic strategy gave an edge to the improvement of the total reward compared with a fixed optimal timeout threshold and could save 6.53\% energy consumption compared with the expert-based method. However, there should be other deterministic policies that reach optimal results, and there is only slight randomness within the system. The improvement of DQN is very limited.

\textcolor{black}{RL methods outperform some non-RL methods in dynamic environments with uncertainties.} Huang \textit{et al.} \cite{huangAutonomousPowerManagement2020} also considered the balance of performance and energy consumption due to limited battery capacity. Since static policy can not capture the varied environment, they utilized an RL approach to improve the adaptivity. In the study, apart from static policies, it derived better performance from predefining multi-level processing energy for the application. However, it is worth noticing that these static or scaling policies could not adapt to system changes. DQN can, to some extent, capture the variation, while it overestimates the Q-value and damages the performance. They utilized Double Q-learning that contained two Q-tables, and the bias on Q-value would be largely reduced. Compared with a standard Q-learning baseline, the energy efficiency was reduced by up to 18\%. It is highlighted that they also implemented the algorithm on a Linux kernel, and the performance and speed were both guaranteed. Kim \textit{et al.} \cite{kimHarvesting} considered an energy harvesting edge device. The arrival energy, which would further be fed into the battery, was random, and the channels to transmit packets were also time-varying. Therefore, the environment was highly dynamic. They established an online Q-learning to learn the dynamics of energy harvesting and channel changes. In the update of the Q-function, they proved the partially monotonic property of the value function through the Bellman equation and set constraints on the update of parameters. Simulations showed a better performance compared with offline policies and were close to the theoretical upper bound with rapid-varying channels. The monotonic property of the value function suggests a deterministic policy structure, and it served the algorithm as prior knowledge, which improved the performance of Q-learning.
 
\textcolor{black}{Power control and allocation are efficient ways to satisfy the high throughput demand of edge AI applications.} In \cite{wang2021joint}, they controlled the power level and channel selection with Q-learning to maximize the total network capacity. Although they considered the scenario of multiple BSs and D2D pairs, they were divided into separate groups, and each contained only one D2D pair which was deemed as one agent. The action of the agent was multi-dimensional, which included channel selection and power level. By utilizing the neural network, i.e., DQN, it could solve problems with a larger state and action space due to the representative capability. In \cite{zhangEnergyEfficientModeSelection2021}, the energy efficiency was considered because of the limited battery life of some edge devices \cite{batterylifesurvey}. The authors studied the DDPG-based communication model selection and power control approach maximizing the energy efficiency in the system. Edge users could choose to work in D2D or traditional communication modes, and a certain fraction of bandwidth resources would be allocated to this D2D communication. To simplify the state space, they designed a new continuous variable to characterize the degree of QoS to substitute the complicated whole network information. The action was designed to represent different behavior with specific values without increasing the dimension of the action space, which improved the efficiency. The DDPG-based scheduling scheme earned a higher accumulated reward compared with dueling DQN, which approximated the result of the exhaustive search. Li \textit{et al.} \cite{Lijointpowersurface} considered the problem of joint optimization of resource allocation and task offloading. The action space was extended to a 4-dimensional variable, which increased the complexity of the approximation of the value function. The authors compared the performance of DQN and DDQN, where DDQN provided a more accurate estimation and higher sum rate, and the gap broadened as the number of devices increased. However, it was a little bit slower than DQN in training.

\subsubsection{Decentralized Communication Resource Allocation Using RL}

When there are multiple users or edge nodes to schedule, the AP, BS, etc., can act as a single controller, and the single-agent RL approaches can be extended to multi-agent scenarios in a centralized way. Zhao \textit{et al.} \cite{zhaocontextmab} modeled a TDD-OFMDA wireless network with multiple users. The eNodeB co-located with a MEC server would carry out the uplink and downlink transmission among mobile devices, which was modeled as the agent. The agent minimized the total cost which was a weighted sum of computing time and energy consumption by allocating the spectrum and CPU resources and MEC server connections. They designed an online contextual MAB to solve the optimization problem, which could adapt to the changing information lag. The decisions were made in two time scales: each period contained a specific number of time slots, where the total cost was minimized by contextual UCB within this period. At the start of the next period, a new TDD configuration was set, and a different minimization was carried on sequentially. The two-time-scale structure and contextual UCB adapted the algorithm to changing network traffic and enabled the tradeoff between exploration and exploitation. This algorithm guaranteed an asymptotic convergence to the categorization solution with polynomial time complexity and close performance with the optimal solution in simulation. MAB-like approaches are suitable for learning problems whose optimal solution can be analytically derived from given parameters, and they will obtain an ideal tradeoff between exploitation and exploration with provable regret bound, hence are classic and rigorous. Multiple time scales are also common in online learning scenarios where the parameters are time-varying.

\textcolor{black}{Currently, some spectrum allocation with RL methods were studied in D2D-enabled networks, enhancing the scalability while reducing the system overhead.} The authors in \cite{moussaidDeepReinforcementLearningbased2018} considered a dedicated spectrum allocation while building the D2D pairs between each user and D2D transmitter. Therefore, they defined the action as a matrix that contained the binary decision for each possible link. They utilized a centralized DQN-based algorithm to derive a spectrum allocation policy, and the sum rate was improved by 2.5 times compared with DQN. The authors of \cite{DSAD2DDDQN} studied spectrum allocation and the selection of D2D pairs for both  \textit{orthogonal frequency division multiple access} (OFDMA) access and \textit{non-orthogonal multiple access} (NOMA) manners. The central actuator did not have information about users' location and spectrum utilization. The incumbent user determined which IoT devices it connected to. They derived the decision of central by estimating access probability, which was the parameter of the Bernoulli random variable. In the learning process, the central unit gathered spectrum information based on the action of the previous stage and updated the parameter using the gradient descent method. They modified the reward as a nonlinear function, which guaranteed the fairness of each D2D node. They utilized DDQN in the algorithm, which gave a more accurate estimation of the Q-value and showed a better convergence performance and higher reward in the simulation which was close to the optimal result.

\textcolor{black}{Except for value-based methods, the AC method was also studied to solve the spectrum allocation problem.} In \cite{liMultiAgentDeepReinforcement2020}, the authors formulated the spectrum allocation problem as a Markov game for the first time and executed a centralized training with an AC structure. Each actor and critic network shared the information of all other agents during the training. This framework made the network easier to converge and learn the spectrum allocation policy faster. The proposed manner greatly improved the sum rate and reduced the outage probability by 37\% compared with standard centralized AC or Q-learning.
In \cite{vehiclepower}, they investigated the RL-based user association and the power management approach in a high-speed vehicle network with energy constraints. A vehicle user communicated to multiple APs simultaneously, which needed accurate power allocation. This paper first incorporated a user-centric design into the software-defined environment. Then, they discussed the performance of SARL and MARL. The SARL suffered from the increase in dimension, while the MARL needed more episodes of training to reach acceptable performance. Therefore, the authors incorporated distributed learning by dividing the action set into several sets, and each agent individually derived an action via a Q-learning-based approach.
The authors in \cite{wangmachinelearning} also considered joint scheduling of task offloading, transmit power, and sub-carriers. They recorded the historical scheduling schemes and the state of the current users in stacks to avoid learning the same information, thus contributing to the convergence. There were multiple BSs in the model which shared all the information, thus was a centralized policy. This method reduced the number of iterations for convergence by 18\% and decreased the average delay by 11.1\% compared with the standard Q-learning. Since the environment was highly complicated, they modeled the problem into a partially observable stochastic game where the MBs and UAVs were deemed as agents, and decisions were made based on partial environmental states and agent actions. Typically, POSG involves more intricate reward functions, and intricate dynamics of collaboration and competition, and necessitates the use of more sophisticated algorithms to handle partial observability and stochasticity.

\textcolor{black}{In a real-world environment, edge devices suffer from limited spectrum, battery capacity, communication, or partial observation. Implementing a centralized policy for all edge users is not applicable.} Therefore, edge users should carry on distributed policy through centralized/decentralized training. Existing work has taken multi-agent RL approaches to the problem of DSA. In \cite{yeDeepReinforcementLearning2019}, DQN was used to schedule spectrum allocation and power control. The DQN needed global information to train, which was unsuitable for partial observation. Therefore, agents took actions asynchronously, and the actions of all other agents could be observed. The simulation results showed that the probability of satisfaction increased by 5\% and the sum capacity was improved by up to 50\% compared with the baseline. However, the gap diminished with the increase in the number of vehicles.
In \cite{multiuserDQN}, the authors proposed a spectrum access scheme with the DQN-based method for multiple agents in the offline learning model. Since the users were isolated and no communication was allowed, the policy was distributed. The state of the environment was partially observable to each user. Hence the problem was not Markovian. Therefore, they designed a CTDE algorithm, where an LSTM layer was introduced in the training stage to aggregate information over time that gave an estimation of the global information for users. They utilized the dueling DDQN structure, which contained a value function to estimate the average value of the current state and a second Q-network for accurate Q-value estimation. The LSTM technique and dueling DDQN structure enhanced the communication system performance while slowing down the convergence speed compared with the DQN network. The proposed policy reached 80\% of successful packet delivery probability and about twice the channel throughput as compared with slotted-Aloha.
In \cite{ziaDistributedMultiAgentRLBased2019}, they studied Q-learning-based spectrum allocation in multi-tire \textit{heterogeneous networks} (HetNets) in a decentralized manner. Each user updated its Q-table through the latest action of other users. The action and state spaces were relatively small because the system model was simple, and users' locations were static. Therefore, using a Q-table to record the Q-value is practical. However, as transmission links and devices increase, DQN is introduced to solve larger spectrum allocation problems.
In \cite{changDistributiveDynamicSpectrum2019}, they studied a DSA problem with multiple \textit{primal users} (PUs) and \textit{secondary users} (SUs). They proposed a DSA scheme with a decentralized DQN based on the RNN structure. Each SU was assigned a DQN plus a \textit{reservoir computing} (RC) and was trained individually.

The decentralized DQN showed a much faster convergence than the Q-learning and maintained similar communication performance. The DQN architecture is used in the DSA problem because DQN is a mature technique to solve decision problems with finite actions without considering the specific model of the system (i.e., model-free).

\textcolor{black}{Furthermore, the end-to-end architecture reduces the complexity of the training network.}
In \cite{destributedDSA}, the authors designed a DQN-based framework to allow each primary user to perform spectrum management individually. The occupation probability of channels was unknown, and the primary user learned the decision from the occupation status of the current user. The simulation showed that DQN outperformed some model-based RL approaches and traditional approaches, including slotted-Aloha and Whittle index policy, which achieved 87\% of the optimal channel access.
In \cite{tanDeepReinforcementLearning2021}, they considered the power allocation for multiple D2D pairs. The D2D communication pairs and available channels were predefined. Each D2D pair determined whether to transmit and power allocation through independent DQN.
The authors in \cite{DSA_RNN} investigated spectrum access and power allocation in a more complicated condition. The users assigned the transmission power to each accessing channel. The power was divided into several values, thus satisfying the DQN structure. They also used a particular RNN to enhance the temporal performance of the DQN. In \cite{hassanjointthroughput}, Hassan \textit{et al.} considered the problem of joint optimization of throughput and power allocation in the F-RAN network. The objective was to maximize the sum rate and minimize transmission power at the same time. Because of the interference among user devices and the partial observation, they designed a hierarchical structure to address this issue. In the upper layer, the user devices were divided into clusters by multi-agent RL approaches, through which the average data rate of each cluster was maximized. The received data rate of each user device was unknown. The user devices selected clusters repeatedly with a mixed strategy profile, which was modeled as a game. The user devices took cluster selection sequentially and updated its utility estimation and cluster selection policy by the instantaneous reward. The transmission power was optimized by classic optimization approaches.

The most common difficulty for distributed users to learn a decentralized policy is the partial observation and limited communication among each other, which ruins the Markovian property. By modeling the process as a game with a certain strategy profile, the agents can still learn global information and thus improve the existing RL policies.
In \cite{saadatrlassistedfl}, the authors utilized a DDQN structure to optimize the total throughput over a D2D-enabled IoT network. The D2D pairs shared the spectrum resources with cellular users. Hence they considered utilizing both orthogonal and non-orthogonal manners, which depended on how much the spectrum interfered. Besides the total throughput, they also focused on the fairness of each user. They modified the single Q-function into a nonlinear combination of Q-functions, through which the throughput of a single user would be taken into consideration, and 95\% accuracy was obtained.
In \cite{relaypower}, they proposed a decentralized RL-based transmit and mobility mechanism to minimize the outage in the IoT architecture. Each source could send data to the destination via neighborhood relays. They denoted the number of neighbor relays as the state. Each edge relay could observe the outage cost, defined as the probability of SINRs falling below a given threshold. The reward was set to 1 if the agent achieved the predefined goal. Otherwise, it was set to 0. This approach achieved considerable improvement compared with the centralized baseline with less number of MFRAs. The proposed approach reduced the overall energy cost by up to 88.03\% while ensuring reliable delivery of data.

Federated learning in MEC also takes advantage of RL. In \cite{saadatflrl}, Saadat \textit{et al.} considered an energy minimization framework in IoT systems, where each edge user trains an ML model by the local data. The minimization problem involved the joint optimization of the allocation of IoT devices and bandwidth resources, which was relaxed and decomposed into two sub-problems. As the environment varied, both of the sub-problems should be resolved. Therefore, they designed a virtual centralized controller that conducted the allocation of IoT devices for clients and solved the bandwidth scheduling problem using the original optimization approach. The RL-assisted policy reached the best tradeoff between energy consumption and KL Distance among other State-of-the-Art approaches.

With the rapid development of the IoT, an increasing number of diverse services have emerged, leading to heightened flexibility demands on the network infrastructure. \textit{Network function virtualization} (NFV) involves utilizing virtualization technology to decouple network functions from hardware and deliver services through software modules on standardized hardware, effectively addressing the aforementioned challenges. In \cite{fu2019dynamic}, the authors introduced a DRL-based embedding technique for heterogeneous \textit{virtual network functions} (VNFs) and device to IoT network \textit{service function chains} (SFCs). Their approach enhances decision-making efficacy by decomposing VNFs into smaller \textit{virtual network function components} (VNFCs) and dynamically embedding SFCs using the DRL method. In \cite{chen2021drl}, the authors proposed an aware adaptive online orchestration with a focus on QoS/QoE, leveraging a DRL-based approach to design an orchestration mode that addresses the challenge of NFV networks failing to adapt to fluctuations in network traffic. Notably, the scheme optimizes QoE while ensuring adherence to QoS constraints. In \cite{jalodia2019deep}, they proposed a model for real-time VNF resource prediction in a dynamic NFV environment, which integrated DRL and \textit{graph neural network} (GNN) techniques. The model incorporates topology awareness by capturing the topological relationships among nodes in the SFC. In \cite{he2021ddpg}, the authors proposed an algorithm that combined DRL and attention mechanism to determine the optimal placement of VNFs in dynamic networks and address the routing problem. The algorithm employs the MDP model to depict the network state evolution and introduces the attention mechanism to enhance the smoothness of network behavior transitions.

\textit{Software-defined network} (SDN) is an emerging network architecture that decouples the control plane and data plane of network devices, thereby enhancing the flexibility and convenience of network management and configuration. In \cite{zhang2020cfr}, the authors proposed a learning scheme that combines \textit{critical flow rerouting} (CFR) and RL to identify critical flows within the network. Subsequently, they formulated the problem as a \textit{linear programming} (LP) problem and employed it to make routing decisions. This approach involves rerouting selected critical flows to achieve a more balanced utilization of network links. In \cite{chen2020rl}, they proposed an RL-based routing algorithm to address the challenge of limited throughput and delay in \textit{traffic engineering} (TE) within SDN networks. The algorithm employs a one-to-many network configuration during routing. The reward function incorporates throughput and delay, which are optimized accordingly. In \cite{akbari2020atmos}, they proposed a comprehensive framework for efficiently designing RL applications and leveraging SDN for network management. This framework not only enables network autonomy and security management but also effectively mitigates the risk of attacks. In \cite{al2021innovative}, they proposed an RL framework tailored to the SDN environment for multimedia-related applications, aiming to intelligently manage traffic while ensuring QoS. This framework enhances QoS by selecting the optimal routing algorithm.

Network slicing refers to dividing a network into multiple end-to-end virtual networks through logical isolation, enabling the integration and efficient utilization of resources to cater to diverse user requirements. Each network slice can be independently managed to fulfill distinct functions. In \cite{sciancalepore2019rl}, the authors proposed the adoption of RL in the network slice broker to construct network slices, encompassing traffic management, learning, prediction, decision control, and other aspects. This approach effectively addresses the challenges associated with resource allocation and isolation among slices. This method effectively meets the \textit{service level agreements} (SLAs) of users by adapting the prediction scheme. In \cite{raza2019reinforcement}, they proposed an RL-based slice admission strategy for 5G RAN aimed at maximizing profits. This strategy enables the allocation of different slices on the same infrastructure. In \cite{koo2019deep}, the authors presented an RL-based solution to address real-time variations in traffic and resource demands within dynamic networks. This scheme enhances resource utilization and reduces latency. In \cite{wang2019data}, the authors presented a dynamic resource scheduling scheme in network slicing employing DRL. By leveraging previous network interaction data to infer users' requirements, this scheme safeguards user privacy while ensuring QoS.

Some classical RL-based communication schemes are listed in Table \ref{table:communication} and Table \ref{table:communication2}.

\renewcommand\arraystretch{1.2}
\begin{table*}[t]
\normalsize

\caption{The RL-based single-agent Communication Resource Management manners}
\begin{center}
\begin{tabular}{m{0.8cm}|m{3cm}|m{4.5cm}|m{7cm}}
\hline\hline
{\bf Ref.} & {\bf \begin{tabular}{@{}l@{}}
    Optimization\\
    Variables
\end{tabular}} & {\bf Objectives}& {\bf RL Approaches and Improvements}\\
\hline
\cite{1SUDQN}&{\begin{tabular}{@{}l@{}}
   {\footnotesize $\bullet$} Spectrum allocation
\end{tabular}}&  {\begin{tabular}{@{}l@{}}
    {\footnotesize $\bullet$} Transmission success rate
\end{tabular}} & POMDP formulation and standard DQN, which can catch unseen environments. Better performance than a heuristic policy like Whittle's Index.\\

\hline
\cite{guoResearchReinforcementLearningBased2018}&{\begin{tabular}{@{}l@{}}
    \multirow{6}{7cm}[-3.2ex]{{\footnotesize $\bullet$} Transmission power\\
     {\footnotesize $\bullet$} Spectrum allocation}
\end{tabular}}&  {\begin{tabular}{@{}l@{}}
   \multirow{3}{7cm}[-8.2ex]{{\footnotesize $\bullet$} Service performance\\
    {\footnotesize $\bullet$} Power consumption}
\end{tabular}} & Establish FIFO queue model which fulfills Markovian property. A standard DQN showed a better performance than the baseline. The improvement is slight when the optimal policy is deterministic. \\
\cline{1-1}\cline{4-4}
\cite{huangAutonomousPowerManagement2020}&&&Double Q-learning reduces the overestimation of Q-value compared with Q-learning and maintains a rapid inference process compared with Double DQN.\\
\cline{1-1}\cline{4-4}
\cite{kimHarvesting}&&&The partial monotonic property of the value function can be proven. An online Q-learning is enough to learn the dynamics of a time-varying environment.\\
\hline
\cite{wang2021joint}& {\begin{tabular}{@{}l@{}}
    \multirow{2}{7cm}[-6.2ex]{{\footnotesize $\bullet$} Transmission power\\
     {\footnotesize $\bullet$} Channel selection}
\end{tabular}} &{\footnotesize $\bullet$} Total network capacity&Thanks to the representative of NNs, the output of DQN can be designed to represent decisions on both power level and selected channel. \\
\cline{1-1}\cline{3-4}
\cite{zhangEnergyEfficientModeSelection2021}&&{\footnotesize $\bullet$} Energy efficiency& A DDPG-based approach is proposed, and it can represent continuous transmission power and channel decisions. \cite{wang2021joint,zhangEnergyEfficientModeSelection2021} have limitations on the number of agents and is non-trivial to adapt to multi-agent scenarios.\\
\hline
\cite{Lijointpowersurface}&{\begin{tabular}{@{}l@{}}
    {\footnotesize $\bullet$} Resource allocation\\
    {\footnotesize $\bullet$} offloading
\end{tabular}}&  {\begin{tabular}{@{}l@{}}
    {\footnotesize $\bullet$} Service performance
\end{tabular}} & The 4-dimensional action variables highly increase the complexity of the problem. DDQN is utilized to derive a precise estimation of Q-value and also improve the convergence performance.\\

\hline\hline
\end{tabular}
\end{center}\label{table:communication}
\end{table*}

\begin{table*}[t]
\normalsize

\caption{The RL-based Multi-agent Communication Resource Management manners}
\begin{center}
\begin{tabular}{m{0.8cm}|m{4cm}|m{3.5cm}|m{7cm}}
\hline\hline
{\bf Ref.} & {\bf \begin{tabular}{@{}l@{}}
    Train/Execution\\
    Manners
\end{tabular}} & {\bf Optimization Variables}& {\bf RL Approaches and Improvements}\\
\hline
\cite{zhaocontextmab}&{\begin{tabular}{@{}l@{}}
    \multirow{7}{7cm}[-7.2ex]{{\footnotesize $\bullet$} Centralized training\\
    {\footnotesize $\bullet$} Centralized execution}
\end{tabular}} & {\begin{tabular}{@{}l@{}}
    {\footnotesize $\bullet$} Spectrum allocation\\
    {\footnotesize $\bullet$} CPU resources\\
    {\footnotesize $\bullet$} Server connections
\end{tabular}}  & Contextual MAB method has provable regret bound and is time efficient. A two-time-scale execution manner is designed to adapt to the varying environment.\\
\cline{1-1}\cline{3-4}
\cite{moussaidDeepReinforcementLearningbased2018}&&  {\begin{tabular}{@{}l@{}}
    {\footnotesize $\bullet$} D2D pairs connections
\end{tabular}} & Centralized DQN which has a binary-element matrix as the output to determine the connectivity among edge users. The action space is quite large, which does harm learning efficiency. \\
\cline{1-1}\cline{3-4}
\cite{DSAD2DDDQN}&&{\begin{tabular}{@{}l@{}}
    {\footnotesize $\bullet$} D2D pairs connections\\
    {\footnotesize $\bullet$} Spectrum allocation
\end{tabular}}& DDQN has a more accurate estimation of Q-value. The reward function was designed as a nonlinear function of the reward of every single agent, which guaranteed fairness among users.\\
\cline{1-1}\cline{3-4}
\cite{wangmachinelearning}&&  {\begin{tabular}{@{}l@{}}
    {\footnotesize $\bullet$} task offloading\\
    {\footnotesize $\bullet$} transmit power
\end{tabular}} & Multi-stack Q-learning, which has multiple stacks to record the history trajectories to avoid learning the same data samples. This improves the convergence speed.\\
\hline
\cite{liMultiAgentDeepReinforcement2020}&{\begin{tabular}{@{}l@{}}
    \multirow{3}{7cm}[-4.2ex]{{\footnotesize $\bullet$} Centralized training\\
    {\footnotesize $\bullet$} Decentralized execution}
\end{tabular}}&  {\begin{tabular}{@{}l@{}}
    {\footnotesize $\bullet$} Spectrum allocation
\end{tabular}} & A Markov game was modeled for the first time in the multi-agent communication problem. The AC approach runs in CTDE manners, which converge fast in the training stage.\\
\cline{1-1}\cline{3-4}
\cite{vehiclepower}&&  {\begin{tabular}{@{}l@{}}
    {\footnotesize $\bullet$} Power control 
\end{tabular}} & Divide action space into several sub-sets, and a Q-learning-based approach runs individually at each edge server.\\
\cline{1-1}\cline{3-4}
\cite{multiuserDQN}&&  {\begin{tabular}{@{}l@{}}
    {\footnotesize $\bullet$} Spectrum access
\end{tabular}} & LSTM is introduced to aggregate information over time and provide estimated global information of other users.\\

\hline

\cite{yeDeepReinforcementLearning2019}&{\begin{tabular}{@{}l@{}}
    \multirow{4}{7cm}[-3.2ex]{{\footnotesize $\bullet$} Decentralized training\\
    {\footnotesize $\bullet$} Decentralized execution}
\end{tabular}}&  {\begin{tabular}{@{}l@{}}
    {\footnotesize $\bullet$} Power control \\
    {\footnotesize $\bullet$} Spectrum allocation
\end{tabular}} & DQN cannot address partial observable problems. Therefore, the action is taken asynchronously, and the actions of other agents are fed as observation. \\
\cline{1-1}\cline{3-4}

\cite{hassanjointthroughput}&&{\begin{tabular}{@{}l@{}}
    {\footnotesize $\bullet$} Power allocation
\end{tabular}} & The decision is decoupled, and only the clustering of users is determined through RL manners. The clustering decision problem is formulated as a stochastic game, and the selection behavior takes place sequentially.\\

\hline\hline
\end{tabular}
\end{center}\label{table:communication2}
\end{table*}

\subsection{Summary and Lessons Learned}
Within this subsection, we first provide a summary of the insights gained from offloading, caching, and communication tasks. We then explore the distilled information from these topics.
\begin{itemize}
\item\textit{Offloading Scheduling}: 
In this section, some existing RL-based are introduced to solve binary offloading and partial offloading scheduling problems. It aims to minimize task latency and energy consumption. The DQN-based algorithm works well in discrete action space and is suitable for making a binary offloading decision (for example, selecting a server to offload). Policy gradient methods can deal with continuous action space. Thus, they are widely used to optimize partial offloading and resource allocation schemes jointly. Centralized frameworks are easy to implement in simulation. A centralized controller receives global observations in these frameworks and takes global actions. They employed SARL methods (such as DQN, A2C, PPO, etc.) in these frameworks. Decentralized frameworks have better scalability than centralized frameworks. They are convenient for dealing with a large number of user scenarios. They classified decentralized frameworks into fully centralized learning, independent learning, and CTDE frameworks. 

In IQL frameworks, each agent learns a sub-policy model individually and considers other agents as a part of an MEC environment.
However, IQL frameworks are difficult to converge. CTDE frameworks employ advanced MARL algorithms, such as QMIX and MADDPG, etc. These frameworks share a value estimator with all agents and can coordinate these agents to achieve an optimal offloading policy. Some works use meta-RL technologies to enhance the generalization of an approach. Meta-RL technologies focus on diverse MEC environments and improve the adaptability of a traditional RL-based method. Game theory is an essential tool for dealing with a multi-agent system's cooperation and competition. These methods hold a selfish assumption for each agent. They usually aim to achieve a Nash equilibrium to maximize the total utility of a MEC system. RL optimizes the latency in an end-to-end way. RL techniques endow intelligence to edge users that enables them to suit varying task loads and mobility of other edge users.

\item\textit{Content Caching}: In this section, some research discusses different caching schemes with RL-based methods such as cooperative caching, coded caching, game-theory-based caching, and proactive caching. The aforementioned work improves the hitting rate by scheduling the content delivery and placement process. Generally, the edge server or a centralized controller is modeled as an agent. The content delivery phase happens when edge servers try to deliver the requested content of users. In the content placement phase, edge servers decide what to cache to their storage. Cooperative caching is the most commonly considered scheduling scheme. They utilize Q-learning and other simple RL methods to deal with such problems. The scalability of cooperative caching algorithms cannot be guaranteed for more devices. One way is to modify SARL algorithms. Another way is to use distributed RL-based caching. Coding caching is one of the distributed caching approaches. A small-scale coded caching uses Q-learning.

However, a large-scale caching problem utilizes an approximate function of the DRL method. In addition, RL-based cooperative coding caching schemes are proposed in UDN to lessen the caching network pressure. Then, the game-based caching scheme with RL methods balances profits among service providers, mobile network operators, and mobile users in the caching process. Agents could update their value functions to a Nash equilibrium for Stackelberg and Markov games, etc. Finally, based on RL, proactive caching can improve caching efficiency by pre-downloading the content in off-peak times. The RL proactive caching problem includes two RL-based subproblems: content recommendation and delivery problems.

RL-empowered edge servers can actively predict the requirements of edge users, which greatly improves the hitting rate. The advancement of caching schemes guaranteed a high data rate in many intensive applications without exceeding available resources. A great number study of game-theory-based caching phases focuses on the scalability of their RL schemes. Both the distributed manner and state space design can improve the reliability of networks with massive connections.

\item\textit{Communication}: 
This section discusses RL-based wireless communication resource management problems, including spectrum access, spectrum allocation, and power control and allocation. Most spectrum access problems are modeled as discrete optimization problems, using value-based methods to deal with these problems. RL algorithms like DDPG can give continuous output. However, allocating spectrum or transmit power for multiple agents takes a lot of work. A common way to deal with this difficulty is to design a binary channel mode or to schedule the resource for one prime agent and evenly divide the remaining resources. Then other agents will share the same discrete action space, and traditional RL-based methods can be utilized. The RL network with the LSTM technique enhances the communication system performance. In addition, some works study RL-based DSA approaches in decentralized structures. Each agent performs spectrum access individually. Spectrum allocation is one typical dynamic spectrum access, and in some research focuses on D2D communication scenarios in MEC networks, various RL-based methods (such as DQN, DDQN, AC, and game theory) are applied to allocate spectrum. The power is divided into some discrete power values to make the network converge quickly, and then they use a value-based method to find a flexible power allocation solution. Especially in some limited energy applications such as IoV, RL-based power allocation approaches consist of the SARL and the MARL methods designed in a centralized or decentralized manner. 

Furthermore, integrating specific network technologies with RL can enhance the performance of MEC networks. The combination of NFV and RL techniques can improve network flexibility by decoupling network functions from hardware and delivering services through software modules on standardized hardware, thereby catering to diverse user requirements. The integration of SDN and RL enhances the flexibility and convenience of network management and configuration. Slicing combined with RL allows for resource consolidation by partitioning virtual networks.

\item\textit{Why a state space is designed?}: 
Several studies have discussed specific designs of the state space in single-agent scenarios. Typically, the state space design aims to accurately represent the system state at each time slot while minimizing dimensionality \cite{sutton2018reinforcement}. In MEC systems, the commonly used state variables include network information, such as energy levels, channel statuses, throughput, bandwidth, and interference from neighboring users. Additionally, utility functions are sometimes employed as state variables to reduce dimensionality and complexity. In certain MDPs with simple transitions, the mentioned state variables are inherently coupled, reducing the need for numerous variables to characterize the current environmental state. Hence, there is an opportunity to build an MDP based on a utility function with reduced dimensionality.

In multi-agent problems, researchers have also taken into account factors such as the trajectory of the assistant edge server and the mobility of IoT devices. However, most state structures are observed in SARL or centralized MARL approaches, where a concrete or virtual controller can access global information, ensuring Markovian transitions. In distributed scenarios, agents have partial observations and define local state variables, while the states of other agents remain unavailable. The state space of a single agent can be considered a subset of the global state space; however, it is not Markovian. Various approaches can be used to supplement global information for single agents, such as synchronous or asynchronous policies and predefined acting profiles.
\item\textit{Why a NN is chosen?}: 
Deep RL methods leverage the representation capacity of \textit{neural networks} (NNs). However, limited research has explored the potential contribution of modifying NNs to deep RL approaches in single-agent MEC scenarios. This could be attributed to RL's utilization of various techniques to approximate the true value function. In contrast to computer vision or natural language processing, the features embedded in the MEC state space are relatively easy to extract. However, as the complexity of multi-user wireless networks increases, the features or dimensionality of the state space grows rapidly, making NNs increasingly important for improving performance. MEC scenarios commonly employ five NN structures: fully connected NN, CNN, RNN, GNN, and NN with an attention mechanism. Fully connected NNs are extensively employed in standard DRL methods as they map the state space to the value function's domain in an end-to-end manner. CNNs consist of convolution and pooling layers and are applied to sequential data, allowing for dimensionality reduction of input information and model parameter reduction \cite{cnnintro}. 

This characteristic of CNNs reduces computational complexity and accelerates training. However, other modifications, such as GNN, are often considered in conjunction with CNN. GNNs are commonly employed to extract features from data with a graph structure, which is prevalent in complex MEC networks where the state of a structure consisting of BSs and end users can be represented as a graph. RNNs are also employed to process sequential data in MEC systems. LSTM, a specific type of RNN, is frequently used for making predictions based on past observations or incorporating global information from previous moments \cite{lstmreview}. The attention mechanism is consistently employed in the processing of sequential data due to its ability to accelerate training and consistently outperform methods that solely use CNN or LSTM. To reduce computational complexity while preserving the capability to handle sequential data, incorporating an attention mechanism has been explored to enhance the performance of neural networks \cite{vaswani2017attention}. A notable example is the transformer, which can be regarded as a fully connected GNN augmented with an attention mechanism. However, the effectiveness of advanced neural network structures remains to be validated in future research.

\end{itemize}

\section{MEC System with RL for Applications}
\begin{figure*}
    \centering
\includegraphics[width=180mm]{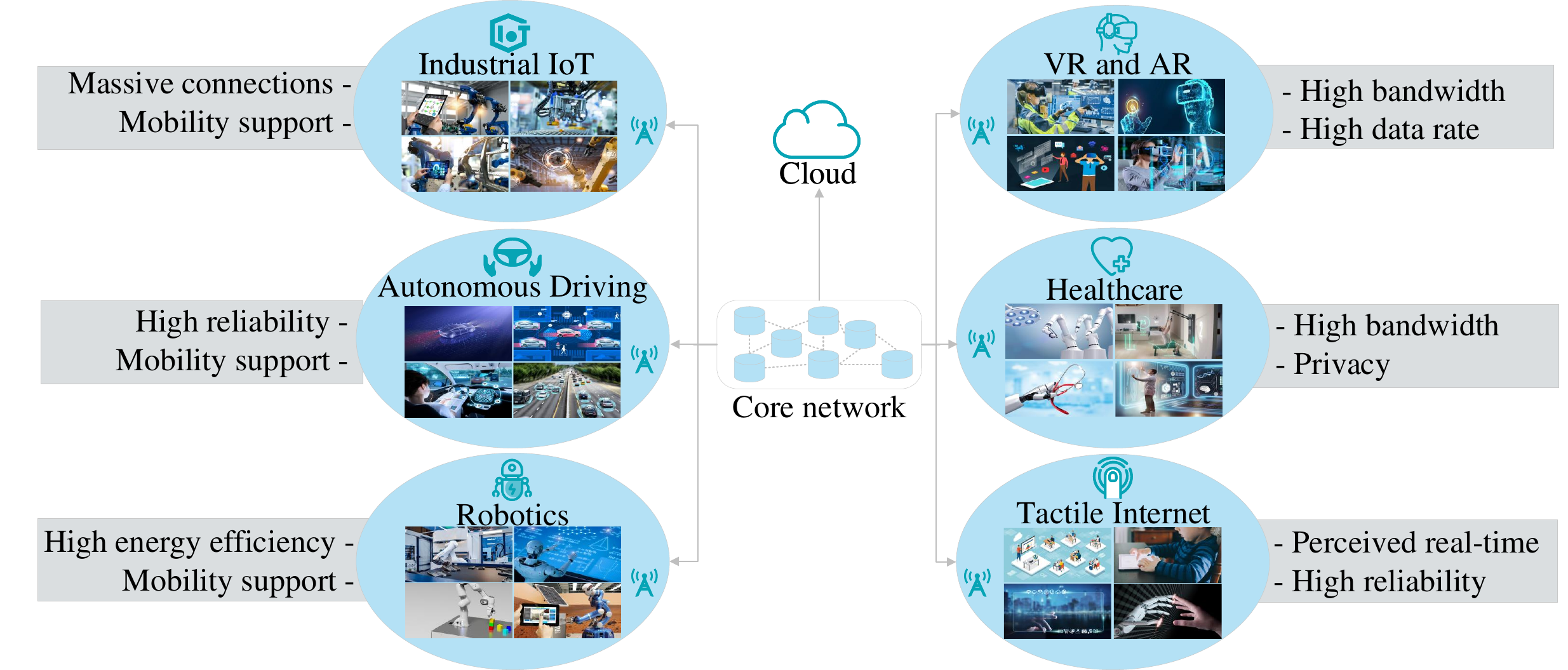}
    \caption{Edge AI Application Scenarios.}
    \label{fig:Application}
\end{figure*}

The MEC network deployed with RL techniques can provide more intelligent services, enhancing intelligent interaction in various applications\cite {zhang2020deep, guan2021customized}, as shown in Fig. \ref{fig:Application}. This section provides ground-breaking applications of the MEC.

\subsection{Industrial Internet of Things}\label{subsec:Industrial Internet of Things}

The industrial revolution has promoted IoT-enabled applications by combining edge AI mechanisms with mobile wireless technologies and maintaining multimedia content transmission \cite{zhang2020deep}. The Industry 4.0-based IoT involves more sensors in the industrial environment, contributing to the efficiency of the traditional industry \cite{yang2018spectrum}. 

However, the growing number of sensors results in a large amount of data, increasing the intensity of the computation tasks. Deep learning technology and edge computing can deal with the intensive data produced by sensors in industrial scenarios. Deep learning methods can help build up the model for the system more efficiently, and edge computing will provide additional computing resources that mitigate the computation burden the industrial IoT devices. Therefore, the integration of AI and MEC technologies facilitates industrial IoT advancement and leads to several challenges for industrial applications: (1) Portable devices' resource-constrained (e.g., battery lifetime and power-hungry) nature; (2) IoT devices or sensors continuously monitor real-time information sent to remote servers or control centers, which causes a large-scale industrial outcome. 

Currently, some research studies wireless resource management with limited resources in MEC networks for real-world industry applications. Most of the research attracted service latency and energy efficiency.
The authors in \cite{dengIntelligentDelayAwarePartial2021} utilized the RL method to optimize offloading scheduling to obtain a minimum delay for the system. The authors considered the limit on computation resources and maximum delay threshold as constraints. They utilized the Q-learning and DDPG algorithms to achieve the minimum delay, where Q-learning reduced the latency by 23\% and DDQG decreased the latency by 40\%.
In \cite{zhouMultiagentRLAided2022}, the authors also focused on the intensive workload of the industrial IoT expert on the industrial end units. However, the user considered two specific communication manners: 5G and WiFi. The security and QoS of 5G outperform WiFi while having a higher cost per bit data delivery and less flexible deployment manners. Thus, coordinating these two communication methods may contribute to the balance of the system's performance and cost. Furthermore, the local task scheduling and power of computing were optimized using a Lyapunov drift penalty method, and the offloading link and transmit power were given by a MARL policy combined with the game theory. MARL schemes reduced the calculation complexity compared with centralized manners.
In \cite{kooOptimalTaskOffloading2021}, the heavy computation workload can be further alleviated by D2D communication. In addition, D2D communication allowed the user to serve the others nearby and greatly optimized the system delay and energy consumption through a Q-learning-based approach.
\textit{Pervasive edge computing} (PEC) supports wireless VR, which allows a portion of tasks to be offloaded to an edge server\cite{ning2020distributed}. PEC contributes to both the reduction of energy consumption and limits on computing resources. With more requirements to guarantee a smooth VR experience, the authors jointly considered the optimization of rendering offloading, computing, and spectrum. The authors utilized a quantum-inspired DQN method to optimize offloading and resource management problems. The utilization of multi-bit quantum action alleviated the effect of the curse of dimensionality and increased the algorithm's scalability.

\subsection{Autonomous Driving}\label{subsec:Autonomous Driving}

Autonomous driving has drawn significant attention in both industrial and academic fields. Autonomous vehicles must frequently interchange information to cooperate, known as \textit{cooperative, connected, and automated mobility} (CCAM) \cite{autonomous5g} to ensure the safety and intelligence of self-driving. The communication that exists in \textit{vehicle-to-everything} (V2X), such as \textit{vehicle to vehicle} (V2V), \textit{vehicle to infrastructure} (V2I), and \textit{vehicle to network} (V2N). These communications are the critical requirements of autonomous driving, which ensure ultra-reliability, high traffic, low latency, and high mobility.

Autonomous vehicles are equipped with many onboard sensors and algorithms to perform data fusion and estimation \cite{autonomousaisurvey} due to the computing capacity being minimal, which makes it non-trivial to meet the computation-intensive and delay-sensitive demands. The authors in \cite{zhaoReinforcementLearningMethod2020} considered using 5G in the communication of vehicles. The 5G communication required high spectrum and energy efficiency, and V2V communication was suitable to meet these requirements. However, mobility of vehicles and a fast-changing communication environment, it was challenging to guarantee the quality of V2V links. The authors then designed a DDQN algorithm to maximize the capacities of V2I links while keeping V2V links within constraints of delay and reliability. Tracking instantaneous channel information used slow fading parameters and statistical information. They regarded each V2V link as an agent. It decided on communication mode and transmission power in a decentralized way. The proposed approach improved the satisfactory probability of V2V links by 8\% compared with centralized approaches, which was very close to the optimal result when the vehicle speed was slow.
In \cite{huEfficientDeepReinforcement2021}, the authors also considered optimizing the V2V communications for vehicle networks. Each V2V agent chose the channel and transmission power in a decentralized way. Different from \cite{zhaoReinforcementLearningMethod2020}, the authors utilized prioritized DDQN to optimize the problem, which guaranteed a more precise estimation of the Q-value. This method outperformed many other baselines and was suitable for real-life scenarios, with a slight drop in the capacity of V2I links and the satisfactory probability of V2V links in the range of 15\%-20\% with high stability.
The authors considered various challenges of utilizing V2X communication in \cite{kimEnvironmentAdaptiveMultipleAccess2021}. Due to the high complexity of V2X communication, standard methods could not deal with the environment's dynamic. Therefore, the authors used RL based method to enrich the knowledge of the environment. First, they reduced the dimension of state variables by filtering the most useful information. Then, they formulated a contextual MAB \cite{lai1985asymptotically} to solve the original problem and utilized TS to learn the parameters. The simulation results outperformed the $\epsilon$-greedy and UCB algorithms. Furthermore, the MAB-based formulation significantly reduced the original problem's computation complexity, and deploying such algorithms required less computational resources, which is friendly to intelligent vehicle systems. 

Some existing work also utilizes RL-based algorithms to schedule the offloading and optimize resource allocation for intelligent vehicles. To address the reliability and latency problem, the authors in \cite{wuResourceAllocationDelaySensitive2021} considered the joint caching and offloading optimization for V2E links. They optimized the energy consumption and delayed tasks using a distributed Q-learning algorithm. The proposed method had an up to 10\% reduction compared to other methods in latency. The authors \cite{ouyang2021task} optimized offloading decisions to maximize the number of computation resources on resource-limited vehicles and minimize energy consumption. The authors formulated a semi-online task distribution model, where each vehicle processed tasks locally or offloaded them to the BS/RSU after collecting energy. They built a semi-online offloading algorithm based on dueling DQN, which fed the continuous state and predicted the behavior of vehicles. The algorithm was operated distributively, and the vehicles used partial information. In the highly dynamic traffic environment, they utilized UAVs to assist in the communication and computation of vehicles. In \cite{maddpgUAVvehicle}, the authors formulated a distributive resource optimization problem while maintaining the QoS. They solved the problem with a MADDPG-based algorithm and achieved better QoS satisfaction ratios than DDPG-based methods and random policies.

Autonomous driving scenarios have the property of high mobility, limited energy, restricted computation capacity, and fading channels, which can benefit from AI-based tools. RL methods are usually utilized to solve problems of resource allocation. Since the safety requirement, the time delay is the most important objective for the above research, while the system consumption is also considered. Other AI-based methods \cite{maCooperativeAutonomousDriving2020} are promising to predict the position of vehicles and to estimate the behavior of other vehicles, which may enhance the performance of the vehicles. 

\subsection{Robotics}\label{subsec:Robotics}

Robotics, especially UAVs, have taken advantage of the beyond 5G communication and MEC techniques. 5G communication has several advantages of ultra-low latency and high bandwidth. However, when meeting some scenarios when the user or equipment units are far away from the core network, the efficiency of the 5G network is greatly damaged. The UAVs are a potential candidate to provide wireless coverage because of their swift mobility and dynamic nature of UAVs. The UAVs can provide network access for long-distance users. The advantages of UAVs include precision trajectory control, lower manufacturing cost, large payload capabilities, efficient energy harvesting, and easier deployment. 

DRL techniques play an important role in handling UAV's intricate problems, e.g., the UAV deployment on the wireless communication platform, UAV trajectories policy, energy consumption, and energy efficiency, etc. Some challenges for the UAV assist MEC networks communication are as follows: (1) UAV communication has several inadequacies and limitations because of channel path loss, antenna designs, frequent handovers, etc.; (2) The traditional research operation only deals with wireless resource optimization given the fixed environment parameters. However, the mobility of the UAV makes the optimal decision time-varying; (3) UAVs may not obtain global information, enhancing the difficulty of wireless resource management. To efficiently address these issues, in particular, the DRL is an efficient tool. 

RL-based UAV control and resource management are widely studied in MEC networks.
In \cite{faraciReinforcementLearningManagement5G2019}, the authors studied an RL-based cooperative job offloading scheme between MEC and UAVs, minimizing the latency and power consumption. Monitored a zone by UAVs that could serve users or communicate with other UAVs. In the model, each UAV could observe the state of the zone and the computer elements queue of one the other. By updating the transition probability matrix, each agent would learn whether to maintain active or turn to another UAV for help. The transition probability and reward function were formulated as matrix equations, reducing the algorithm's complexity and guaranteeing the UAV's real-time performance.
In \cite{wangRLBasedUserAssociation2019}, the authors utilized a Q-learning algorithm minimizing the overall energy consumption considering deploying multiple UAVs in the MEC platform. The UAVs had the benefits of high mobility, low cost, and easy deployment. In contrast, the association between UAVs and edge users led to great challenges to the application of UAVs. The authors formulated a mixed integer non-linear programming problem. Each user could choose whether to offload them to a specific UAV or process these tasks locally, and each UAV could only accept a limited number of tasks. The simulation results showed a better performance than traditional approaches, such as greedy offloading and exhaustive search. The methods above were all centralized methods that a central system that controlled the decision of UAVs and the global objective could be optimized. The proposed manners decreased the energy consumption by up to 80\% compared with traditional methods and was close to the optimal
solution. In real scenarios, the BSs were dominated by SPs, so each BS would focus on maximizing its profits, and the cooperation among BSs would be limited.

Some research mainly focused on distributed control in MEC and UAV communication scenarios. The authors considered UAVs served for multiple service providers in \cite{asheralievaHierarchicalGameTheoreticReinforcement2019}. The authors modeled a hierarchical game into two layers: one layer contained cooperative players, and the other layer had non-cooperative sub-games. They utilized a Q-learning-based method which was modified by the mixed-strategy Nash equilibrium. 
In \cite{asheralievaDistributedDynamicResource2020}, they studied a distributed resource management and pricing strategy considering the IoT system based on Blockchain and UAV-enabled MEC. The BSs were separately controlled, and no information was exchanged among those BSs. The interactions among BSs were represented as a stochastic Stackelberg game, and neither leaders nor followers were able to observe the complete information of others. They then formulated the problem into a POMDP and proved that the best response of followers formed a perfect Bayesian equilibrium. A Bayesian RL was applied to the training of the followers. The Bayesian RL contributed to learning the uncertainty of the unobserved information. The leaders were trained through a polynomial-time DQN framework. They utilized a polynomial-time DQN algorithm and reduced the complexity by approximating the value function in an unsupervised manner.

\subsection{Virtual and Augmented Reality}\label{subsec:Virtual and Augmented Reality}
The successful development of \textit{eXtended Reality} (XR) technology, which includes VR and AR, has led to the revolutionized interactions between users and the virtual worlds. It was anticipated that the requirements for VR devices would be 99 million, and the market would reach 108 billion dollars in 2021 \cite{9411714}. Wireless VR/AR requires high-level energy consumption, heavy computation workloads, high real-time processing demand, and high viewport rendering demand \cite{linResourceManagementPervasiveEdgeComputingAssisted2021}.
XR applications are computing-intensive and require high computing resources to guarantee QoS and QoE. 

Some works use RL algorithms to manage wireless resources improving the QoS or QoE of XR applications in MEC networks. In \cite{9700522}, it proposed a resource management scheme that considered heterogeneous QoS requirements at the server level for SDN-MEC-supported XR applications. They employed an LSTM neural network to build an AC-based policy. The LSTM neural network was good at processing time series data. It can learn the temporal regularity of state space according to the required latency of users, bandwidth, computing, and storage. The simulation showed that the proposed method improved the resource utility by 32.7 compared with SDN CAV\cite{peng2019sdn} and 8.1\% in throughput compared with the greedy scheme. The authors in \cite{9439855} concentrated on congestion control for AR/VR data transmission. They proposed a two-stage RL method to control multipath transmission. First, this method analyzed the \textit{power spectrum density} (PSD) of the AR/VR input stream and used an expectation maximization algorithm to extract the features. Then, the method optimized a Q-table method to obtain a scheme. Furthermore, the authors divided the training process into an ofﬂine and an online self-learning process to improve the training efﬁciency.

Some research focuses on the VR application based on RL methods. In \cite{9411714}, VR devices shifted the \textit{field of view} (FoV) rendering tasks to MEC servers. The authors considered that each user had their FoV preference. They proposed a decoupled learning strategy that decoupled the optimization by separately resolving two subtasks: FoV prediction and rendering MEC association. First, the authors proposed the RNN model, and this model was set as a central controller to predict the requested FoV in the current time slot \cite{cho2014learning}. Then, they proposed four DRL algorithms: centralized DQN, distributed DQN, centralized AC, and distributed AC. Simulation results showed that a centralized DQN had the best QoE and lower latency than other algorithms. They considered a THz MEC system in \cite{9120235} to optimize the QoE in the wireless VR system. They employed A3C to optimize the ofﬂoading decision. The A3C algorithm asynchronously executed multiple policies in parallel, improved learning efficiency, and reduced memory usage. This method achieved higher reward and lower energy consumption of head-mounted displays compared with an AC method and a local rending method. In \cite{9495190}, the authors studied a dynamic caching replacement and ofﬂoading scheme. They converted the optimization problem into a POMDP. They employed an LSTM neural network \cite{hochreiter1997long} to design a DDPG-based algorithm. The authors compared the performances of the proposed method with a random approach, a traditional DDPG approach that removed the LSTM model, and a \textit{normalized advantage functions} (NAF) approach \cite{gu2016continuous}. They studied the performance in different tradeoffs between the transmission delay and the energy consumption. The analysis proved that the LSTM-DDPG-based method had a higher reward and reached a lower delay (or energy consumption) in the latency (or energy consumption) preferred conﬁguration. 

Except for the VR applications, some research is also interested in RL-based resource management in AR applications. For example, in \cite{chen2021energy}, the authors studied an ofﬂoading scheduling and resource allocation approach for AR applications in single-edge and multi-edge node networks. An AR real application has ﬁve subtasks: (1) video source; (2) tracker; (3) mapper; (4) object recognizer; (5) render. They modeled these sub-tasks as a directed acyclic graph and proposed a MADDPG-based scheme to deal with the joint optimization problem for multi-user competition and cooperation. Furthermore, they considered the fairness between users. Specifically, the MADDPG-based algorithm aimed to minimize energy consumption with delay requirements. The energy consumption was reduced by up to 80\% compared with single-agent manners and 88\% compared to the greedy policy. These works show that RL-based methods have significant potential to improve the QoS and QoE of XR Applications. Most works use RNN-class architecture to build policy models, which can learn the temporal regularity of input and fit the optimal policy by combining it with an RL-based algorithm.

A video stream refers to a consistent flow of video data that can be transmitted and processed across a network. MEC offers substantial support for video stream-related applications for the following reasons \cite{jiang2021survey}: (1) MEC possesses robust computing power; (2) By leveraging storage near the client and network edge, MEC mitigates transmission delays; (3) The distributed structure of MEC enables real-time network data acquisition. RL proves effective in addressing resource allocation challenges within video streaming applications. There are resource allocation problems, such as edge caching and computing in the application of video streams, which can be solved by introducing RL technology. In \cite{du2020mec}, the authors proposed a strategy that incorporates RL and employs the A3C algorithm to optimize transmission power control and viewport rendering offloading. This method considers the time-varying nature of the terahertz wireless channel and effectively minimizes long-term energy consumption. In \cite{luo2019adaptive}, the authors introduced a comprehensive method that accounts for both energy consumption and QoE in video streaming processes, including edge caching and video quality adaptation. This method describes the time-varying channel as a \textit{discrete-time Markov chain} (DTMC), converts the problem into an MDP, and subsequently employs the A3C algorithm for problem resolution. In \cite{liu2022delay}, the authors proposed an adaptive bit rate streaming scheme that comprehensively considers caching, computing, and power allocation to optimize both delay and energy consumption by dynamically adjusting bit rates according to user requirements. The proposed manner reduced the weighted sum of delivery delay and energy consumption by 5\%–12\%.

Multimedia encompasses diverse modalities of information representation, encompassing textual, auditory, visual, and other forms. MEC, as a transformative technology, enables the efficient deployment and seamless integration of multimedia applications across network infrastructures. In \cite{dai2021edge}, the authors proposed a framework that divided multimedia files into multiple chunks encoded with varying bit rates to enable adaptive bit rate streaming through caching and transmission processes. Two algorithms based on MAB and DNQ were introduced for handling the individual chunks. While the former algorithm effectively reduces overhead, it exhibits a slower convergence speed. The latter algorithm employs replay memory to enhance convergence speed. In \cite{rio2022deep}, the authors introduced a prediction algorithm based on DRL to adapt to dynamic environments and enhance the QoE in real-time video data transmission. The algorithm employs A2C technology for QoE training, coding quality assessment, and predicting future actions.

\subsection{Healthcare}\label{subsec:Healthcare}
A massive surge for wearables and fitness trackers promotes the development of the healthcare system \cite{ mutlag2019enabling}. Healthcare systems focus on the ubiquitous monitoring of patients detecting disorders, and implementing a suitable treatment plan \cite{qadri2020future}.
It is necessary to provide interoperable platforms and underlying communication technologies for devices, with large-scale sense and analysis data generated every second by massive devices in healthcare systems, burdening beyond the capacity of the network. In addition, MEC is widely adopted to alleviate the latency of services and energy consumption, enabling a faster real-time response.

The critical requirements of healthcare are low latency, low energy consumption, and high reliability and privacy. (1) Low latency: The healthcare system has a time-critical nature. The delay includes end-to-end transmission and processing. An effective way to reduce the delay is by combining RL and MEC technologies. (2) Low energy consumption: Some wearables and implantable devices are deployed in healthcare applications. These wearables can be recharged for several days. However, implants need a long-lasting battery capacity. (3) Security: An important measure is to protect the security of patients in a healthcare system. The misuse of patient data leads to severe medical accidents. However, many attacks compromise the data and privacy of patients. 

For healthcare applications of MEC networks, the current research includes reducing latency, and energy consumption, optimizing network traffic, and enhancing privacy. In \cite{yadav2021smart}, it proposed computation offloading with the RL method minimizing latency and energy consumption and solving energy-hungry and service latency problems. Compared with cloud schemes, they achieved a 45\% decrease in latency and near energy consumption. In \cite{soni2019hmc}, it developed a hybrid RL model compression scheme integrating model-free and model-based RL approaches, deploying networks of medical devices. Privacy protection of users is necessary for healthcare applications. In \cite{min2018learning}, it capitalized on transfer learning methods optimizing the offloading rate to improve each device's computing performance. In \cite{radoglou2021modeling}, it introduced a detection and prevention system to discriminate and mitigate cyberattacks. They reformulated the automated mitigation problem as an MAB problem solved with the RL method. The mitigation accuracy was calculated at 0.923 and the detection accuracy was 0.831. In \cite{mustafa2020rl}, it also presented an RL algorithm for detecting and preventing misdirection attacks for sepsis treatment. In \cite{yu2019deep}, it proposed a deep inverse RL to infer the optimal reward function from some real medical data.

\subsection{Further Tactile Internet Applications}\label{subsec:Further Tactile Internet Applications}
Tactile Internet is a communication infrastructure with low latency, realizing human-machine interaction by combining with AR or VR for sensory controls in various environments \cite{fettweis2014tactile}. Some popular tactile internet application is industrial IoT, autonomous driving, robotics, healthcare, AR, or VR. Additional applications include gaming, education, and manufacturing. The end-to-end delay directly determines the QoE of players in the interaction for games. For real-time applications, tactile users may experience cyber-sickness. For an instant, the user perceives a movement with a slight delay.

Tactile Internet applications require ultra-responsive connectivity, and latency of 1 $ms$ \cite{baker2017internet}. However, the current cellular system and the \textit{wireless local area network} (WLAN) cannot be satisfied. Associates with MEC is an excellent way to reduce latency \cite{simsek20165g}. Privacy is also a challenge, especially in massive connectivity applications in MEC networks.

\section{Future Directions and Challenges}
\begin{figure*}
    \centering
    \includegraphics[width=180mm]{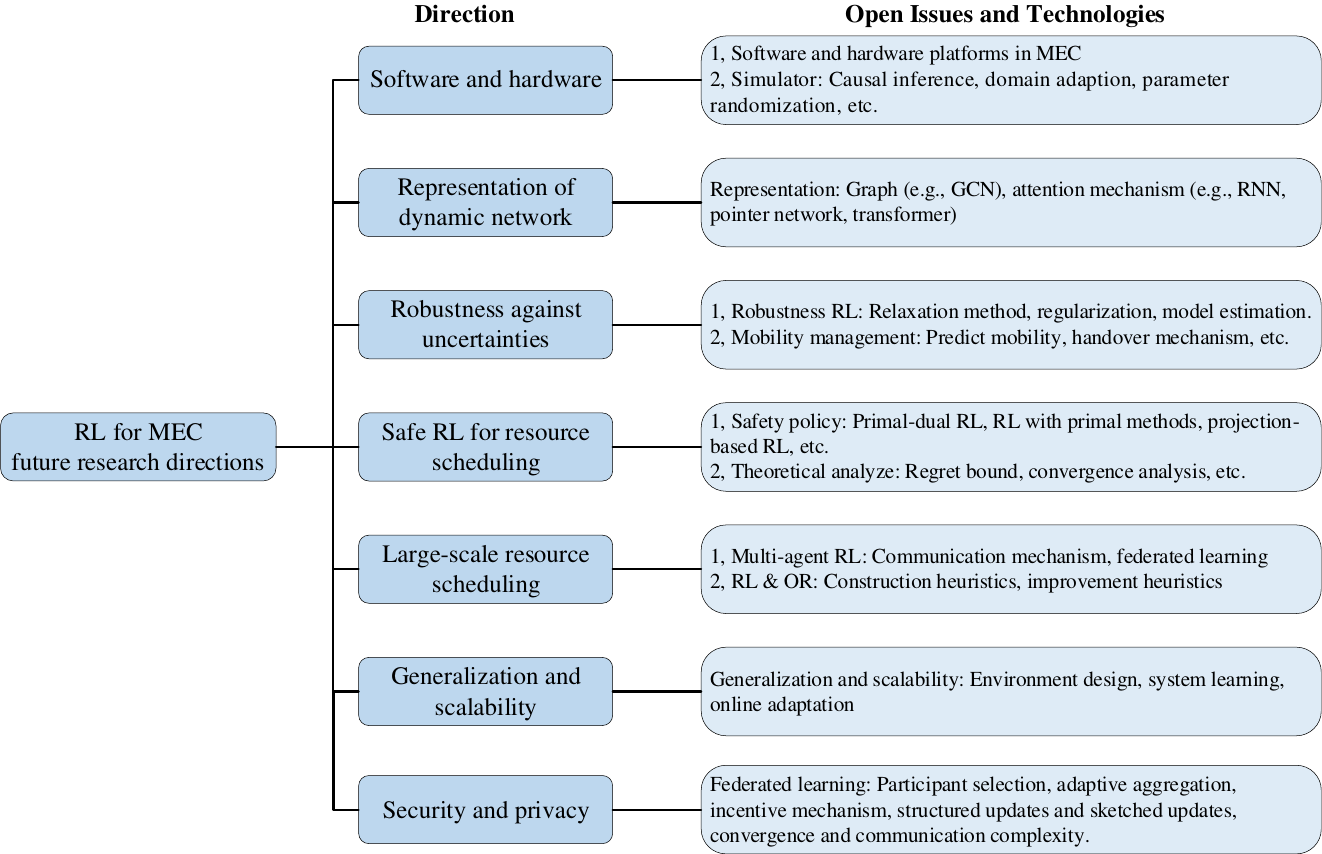}
    \caption{Future research directions for DRL.}
    \label{fig:future work}
\end{figure*}

Apart from the previous work, this section discusses some promising research directions and pending challenges in using and widening the RL in the MEC network \cite{yang2019deep}. A summary of the future research directions, open issues, and technologies is shown in Fig. \ref{fig:future work}.

\subsection{Software and Hardware Platforms}
In recent years, software frameworks and hardware devices have been essential to facilitate the training and implementation of DRL. Therefore, this subsection introduces two parts: (1) The software and hardware platforms in MEC networks; and (2) The software (e.g., simulator) for the DRL model training. These provide an essential platform for connected intelligence in MEC networks. 

Many software can be deployed on edge and provide real-time computing services. For example, Amazon IoT Greengrass, Azure IoT Edge of Microsoft, Cloud IoT Edge of Google, and NVIDIA ``EGX" \cite{zhou2019edge}. In addition, Huawei released an AI platform named HarmonyOS, a next-generation operating system that empowers interconnection and collaboration between smart devices. Computation-intensive applications motivate performing RL-based model training and inference with resource-constrained. Edge AI computing hardware consists of the \textit{graphic processing unit} (GPU)-based hardware and \textit{field programmable gate array} (FPGA)-based hardware, etc. (e.g., NVIDIA ``GPUs," Xilinx ``SDSoC," Google ``TPU") \cite{letaief2021edge}.
Hardware, in the context of RL in MEC, is instrumental in providing the computational power necessary for training sophisticated RL models. However, despite the advancements in hardware capabilities for execution, a significant challenge arises during the training phase due to the sim-to-real gap. The sim-to-real gap refers to the difficulties encountered when transitioning an RL agent from a simulated environment, where it was initially trained, to the actual, physical environment for which it is intended. While hardware enables the execution of learned policies, the discrepancies between the simulated and real-world environments can lead to suboptimal performance or even failure of the RL agent. The necessity of a simulator lies in the further refinement of the learning-from-simulation process to address these challenges and improve the adaptability and robustness of RL agents when deployed in real-world MEC scenarios.

\textcolor{black}{Most recent studies focus on simulation environments, for which interacting with real environments is quite expensive. Besides, the agent receives inaccurate or incomplete data due to missing data, poor connection, and partial observation. Typically, DRL algorithms undergo training within a simulator, a proxy constructed to replicate real-world environmental factors.}

\textcolor{black}{The sim-to-real gap in the context of RL within MEC refers to the disparity between simulated training environments and the complex, dynamic conditions of real-world MEC scenarios. This gap arises due to challenges in accurately modeling the intricacies of MEC networks in simulations, resulting in differences in model fidelity, environmental dynamics, and the representation of sensors and actuators.} Since differences in the distribution of states and unforeseen real-world factors can impact the generalization and performance of RL models in practical MEC applications, it is crucial to bridge the gap to ensure effectiveness and robustness when transferring RL policies from simulation to reality. While certain factors, such as channel fading, interference, multi-user interference, spectrum availability, and allocation, latency, and delay constraints, can be well-accounted for in a standard communication model, other aspects are challenging to model and simulate. RL algorithms must exhibit adaptability and resilience to novel situations not encountered during training. Additionally, they should incorporate mechanisms for fail-safes and error handling to bridge the gap between simulated training and real-world deployment effectively.
The policy, a product of DRL, is developed via data samples extracted from this simulator. Nevertheless, the simulation-to-reality transition is fraught with difficulties, namely (1) discrepancies between simulated and actual environments, leading to ineffective real-world policy applications, and (2) strategies overfitted to the simulator, resulting in poor generalization. Consequently, advanced methodologies such as causal inference \cite{meek2013causal}, domain adaptation \cite{tobin2017domain}, parameter randomization \cite{chebotar2019closing}, knowledge distillation \cite{Gao_2021}, and meta-learning \cite{vilalta2002perspective}, are implemented within DRL to mitigate environmental bias and enhance generalizability.

\begin{itemize}

\item \textit{Causal Inference}: Causal inference is a process of measuring the independent, actual effects of a special phenomenon that is a component of a system. Our objective revolves around constructing a task-offloading DRL model capable of generalizing across an array of MEC environments. We incorporate causal inference to comprehend the invariant prediction of the MEC system model, thereby acquiring a causal representation. This causal feature set is treated as model-independent state abstractions within the DRL model. The intention is to optimize the DRL model's performance across diverse MEC environments sharing a block MDP, given an environment where different states are disjoint yet possess identical latent states (e.g., Task distribution, computing resource utilization, etc.) and a common dynamic (i.e., the same state transition function of different MEC environments) \cite{zhang2020invariant}.

\item \textit{Domain Adaption}: Domain adaptation techniques craft a model by utilizing data from the source domain (e.g., established MEC models) and subsequently enhance the model in a distinct target domain (e.g., novel MEC models) with limited data availability. Specifically, regarding a task offloading issue, it can initially develop a latent state representation network spanning multiple domains (e.g., assorted MEC environments). Building upon the developed state representation network, implementing RL training will obtain an offloading strategy within one source domain \cite{xing2021domain}.

\item \textit{Domain Randomization}: Domain randomization is a potent strategy to enhance simulator fidelity through adaptation. It models the divergence between source and target domains as variability within the source domain itself. During the training phase, domain randomization must present ample variability (e.g., the number of users or servers, server CPU frequency) of MEC network parameters. Following this, we generalize to the real-world MEC network during the testing phase \cite{chebotar2019closing}. 

\item \textit{Knowledge Distillation}: Policy distillation is an extraction process, extracting knowledge to train an efficient, compact network while maintaining comparable expertise levels \cite{Gao_2021}. To alleviate resources-exhausting and time-consuming for intricate MEC tasks, knowledge distillation becomes critical, harnessing historical experiences in this perpetually variable communication system or multi-MEC system. For instance, regarding a task offloading issue, we employ a traditional knowledge distillation algorithm known as the Teacher-Student model. The teacher integrates the output ``offloading decisions" from multiple diverse MEC decision-making models trained separately, and the student receives these ``offloading decisions" represented by a single MEC decision-making model. During the knowledge distillation phase, we train a Net-T (i.e., the network model of the teacher) possessing robust generalizability. We then facilitate Net-S (i.e., the network model of the student) in learning the generalization capabilities of Net-T. Additionally, knowledge reuse is implemented within multi-MEC environments, thereby expediting the training process for new tasks.

\item \textit{Meta-Learning}: Meta-learning, often referred to as a learn-to-learn algorithm, boasts a low sample-efficiency for novel tasks\cite{wang2021meta}. For instance, a meta RL-based resource management methodology can swiftly adapt to fresh MEC tasks with minimal gradient updates and samples. Specifically, we architect multiple MDPs for distinct MEC tasks, each MDP being optimal for a type of MEC task, with learning environments of different MDPs sharing certain similarities. The learning process of the resource management policy is bifurcated: the effective learning of the meta-policy between different MDPs, and the rapid learning of each MDP's specific policy based on a designated MEC task.
\end{itemize}

\textcolor{black}{Besides, performance evaluation of MEC networks plays a pivotal role in optimizing network performance and serving as a decision-making guideline to enhance network reliability.} The performance analysis methods encompass numerical analysis, simulation, testbed experimentation, and real-world network assessments\cite{jiang2021survey}.

\begin{itemize}

\item \textit{Performance Evaluation of the Framework}: The majority of existing approaches rely on simulated data sets for conducting evaluations and simulations. However, in real-world scenarios, obtaining evaluation data poses challenges, and simulations may exhibit deviations and omissions. Therefore, it is imperative to explore collaborative partnerships to acquire more realistic simulation data, thereby augmenting the reference value of simulations and evaluations\cite{luong2019applications}.
     
\item \textit{Balance Between Information Quality and Learning Performance}: Enhancing the quality of information utilized for learning requires incurring additional costs, such as increased delay and reduced learning speed, which may lead to a decline in the overall system performance. Consequently, striking a balance between these factors becomes crucial, aiming to improve the system performance while maximizing the quality of information\cite{luong2019applications}.

\end{itemize}

\subsection{Representation of Dynamic Network}
In an MEC network, data-intensive tasks are pushed toward edge servers in proximity and locally processing these data, reducing traffic process bottlenecks in the cloud and edge networks. In general, such MEC architecture consists of cloud networks and edge networks, and these networks are covered with densely connected devices, and some are covered by multiple networks overlap. Moreover, communication interaction exists among numerous connected devices and multi-edge servers in the uplink and downlink. These characteristics of the networks motivate us to explore the representation-based RL paradigm for MEC networks. The advantages of the RL combined with the representation are as below: (1) representing the relational information between nodes and edge servers with a low-dimensional vector or matrix; (2) improving the performance and efficiency of RL; (3) generalizing to various MEC networks with instances of different sizes.

The graph is a kind of data structure that contains a set of nodes (i.e., edge servers or devices) and their edge servers (i.e., transmission links) \cite{zhou2020graph} capturing the properties of nodes and edge servers. Generally, a wireless resource management problem in MEC networks, such as offloading, caching, and communication, can be formulated as an MDP problem on graphs because of their highly structured nature. In an MEC network, it is necessary to use as few features as possible to represent a complex communication system model. \textcolor{black}{The graph representation learning method represents graph nodes, edge servers, or subgraphs with low-dimensional vectors \cite{Hamilton2020Graph}. Consequently, parameterizing RL networks with graph structures significantly curtails the input information.}

\begin{itemize}

\item \textit{Graph Convolutional Network}: A derivative method of parametric RL networks, employing \textit{graph convolutional network} (GCN) \cite{bresson2017residual}, an amalgamation of CNN and graph embedding\cite{ribeiro2017struc2vec}, facilitates the learning of more intricate networks with extensive applications. Several resource management issues sharing the same structure (i.e., numerous task offloading problems abstracting similar general scheduling problems) are repetitively solved using varying data (i.e., differing task sizes and arrival times). For instance, in a task offloading problem, we model it as a graph where servers and channels are represented as nodes and edges, respectively. Utilizing the graph neural network, we embed the states of servers and channels through convolutional layer processing, obtaining vector representation for each node and edge. Finally, we deploy the greedy algorithm to select the optimal offloading server for each task and then reiterate this step, eventually generating the final offloading solution \cite{dai2018learning}.

\item \textit{Graph Network with Message-Passing}: Furthermore, an efficient message-passing method where vertices communicate with edges for the graph can expedite the RL model learning process in MEC networks \cite{joshi2019efficient}. The procedure resembles the aforementioned steps, with the primary distinction being the representation technique of the GNN. A vertex embedding representation (i.e., encompassing user and server information) can amass more information via an edge-to-vertex adjacency matrix and vertex information transmission function. Similarly, edge embedding representation (i.e., encompassing channel and offloading decision information) employs vertex-to-edge adjacency matrix and edge information transmission function \cite{joshi2019efficient}. This representation method, coupled with message-passing, culminates in significant rewards in solution quality (i.e., it enhances wireless resource utilization and minimizes service delay). 

\end{itemize}

\textcolor{black}{Another parameterized network method, the attention mechanism, is introduced to cope with more complex heterogeneous networks, diverse communication scenarios, and other application requirements.} The attention mechanism requires an encoder mapping the input sequence into a $d$-dimensional space, then using a decoder to map it to the output sequence, which captures the information of interest and suppresses useless information. The RL model with an attention mechanism can remember past actions and know whether actions are good or bad, reducing time and computing costs for MEC networks. RNNs \cite{schuster1997bidirectional} are employed to comprehend sequence decisions with predetermined input and output sizes. However, in the practical implementation of MEC, factors such as terminal mobility, access network uncertainty, and resource dynamics preclude the input or output from maintaining a fixed size. 

\begin{itemize}

\item \textit{Pointer Network}: The RL method integrated with pointer network \cite{Vinyals2015Pointer} offers a straightforward and effective solution to variable length dictionaries, utilizing a softmax probability distribution as a ``pointer". We architect a pointer network model with attention layers, which is composed of an encoder and a decoder. The encoder network applies an attention mechanism to read the MEC environment's input sequences, transforming them into decoder hidden states, which are subsequently passed to the decoder. The decoder network, equipped with a pointing mechanism, generates the probability of each action (i.e., an offloading, caching, or communication decision). This resource management representation model is then trained using REINFORCE or other benchmarks.

\item \textit{Transformer}: A renowned attention mechanism architecture is the transformer \cite{vaswani2017attention}. The Transformer possesses the capability of memorizing previous decisions and leveraging them to make new decisions. For instance, for the task caching issue in the MEC network, we design an augmented representation network based on the Transformer, comprising an encoder and a decoder. The encoder serves to embed the cache information and the decoder to formulate cache decisions. The proposed representation network is employed as the DRL network. 

\end{itemize}

\subsection{Robustness Against Uncertainties in MECs}
Robustness is a vital feature that measures the adaptability of models to uncertain environments. There exist two ways to improve the robustness of the model and algorithm: (1) Robustness RL; and (2) Mobility management.

\textcolor{black}{In MEC networks, deploying DRL in real communication environments is more complicated because of sensor errors, asynchronous states, dynamic environments, etc. Thus, the actual communication system parameters may differ from those in simulations.}
These motivate the development of robust RL. Robust RL methods aim to address the uncertainty in observed states, conducted actions, and sim-to-real gaps.
Several efficacious strategies can fortify robust RL's adaptation to uncertainties, such as relaxation methods \cite{wu2022copa}, regularization \cite{zhang2020robust}, and model estimation \cite{gallego2019reinforcement}. 

\begin{itemize}

\item \textit{Relaxation Method}: Through the relaxation method, the original problem can be translated into a lower bound for the cumulative return. The robustness of the DRL model can be amplified by optimizing this lower bound. For instance, within an offloading scenario, assume that a maximum of $M$ servers are simultaneously chosen to execute tasks. By leveraging Lagrangian relaxation \cite{boyd2004convex}, we mitigate this instantaneous constraint to an average constraint by introducing dual variables and deriving multiple sub-problems. The objective of the sub-problem is transmuted into the reward function of DRL. As the dual variables gravitate towards convergence, the summation of solutions for all sub-problems achieves a lower bound of the solution for the original problem.

\item \textit{Regularization}: A regularizer refines the learned policy, inhibiting drastic alterations attributable to state uncertainty. Various regularization techniques (e.g., entropy regularization, L1 regularization, L2 regularization, and dropout) contribute substantial performance enhancements and are indispensable during neural network training. We incorporate regularization techniques to solve continuous power allocation problems. Through the DRL method, we may elect to regularize different components (i.e., the value network or the policy network) to boost generalization performance across four aspects: sample complexity, weight norm, reward distribution, and noise robustness \cite{liu2021regularization}.

\item \textit{Model Estimation}: Model-free RL algorithms typically necessitate model estimation, the crux of which lies in estimating the reward through the refinement of the corrupted reward. Specifically, a value expansion scheme is utilized to implement MEC resource management control amidst uncertainty (i.e., time-varying MEC environment). We estimate the value over this finite step, thereby enhancing the precision of value estimation. In other words, we use the model to estimate the short-term horizon and deploy Q-learning to estimate the long-term reward \cite{feinberg2018modelbased}.

\end{itemize}

The applications in MEC networks, like wearable devices, are expected to connect to the advanced communication networks directly. Mobility resource management improves robustness and reduces performance degradation due to user mobility. It concerns user mobility and executes mobility-aware and dynamic service requirements. However, the mobility resource still has some exciting work that needs to be further studied. (1) Mobility support enables critical services by designing novel handover mechanisms at high speed. (2) Distributed mobility management achieves seamless handovers and guarantees service continuity. (3) Mobile edge caching techniques provide content caching, predict mobility, and execute caching migration. 
 
\subsection{Safe RL for Resource Scheduling}
As the number of intelligent mobile devices grows, large-scale mobile terminal applications are emerging. However, the available communication resources (e.g., power spectrum) are limited. Besides, mobile devices are also resource-constrained, such as the battery capacity and CPU power. How should we design the RL-based resource management algorithms with resource-constrained for offloading, caching, and communication? Traditional RL methods learn strategies through reward functions and aim to maximize rewards without considering constraints satisfaction. Safe RL, also named constrained RL, explicitly considers the resource constraints and prevents agents from making dangerous actions \cite{xu2022trustworthy}. 

Some advanced safe RL methods are proposed to handle resource management problems with multi-constraints. Notable amongst these include Primal-Dual RL \cite{stooke2020responsive}, RL employing primal methodologies \cite{xu2021crpo}, Projection-based RL \cite{yang2020projection}, and RL incorporating penalty techniques \cite{sootla2022saute}. 

\begin{itemize}

\item \textit{Primal-Dual RL}: Safe RL adopts the direct application of Lagrangian techniques, occasionally resulting in behaviors that infringe constraints during the training evolution in the MEC network. The employment of the Lagrange multiplier update approach, complemented by the derivatives of constraint functions, facilitates the reduction of constraint transgressions. The Lagrangian transformation is executed on the objective function to yield the Lagrangian function. We introduce derivative and proportional control, supporting learning dynamics (i.e., dynamics of the MEC system constraint) through predictive and damping measures \cite{stooke2020responsive}. 

\item \textit{RL with Primal Methods}: In MEC networks, the agent explores the MEC environment maximizing expected total energy or delay while concurrently evading the breach of certain constraints (power, computing resource, and spectrum resource constraints). The objective functions pertaining to resource management dilemmas are typically nonconvex and subjected to multiple nonconvex constraints, making it unattainable to acquire the globally optimal resource management policy. We utilize a primal approach, rectified by constraints, to optimize the policy in lieu of primal-dual structures. This rectification approach alternately updates the resource management policy among objective functions or constraints. Each policy update employs a natural policy gradient or any variant of a policy optimization scheme. It is also imperative to substantiate the theoretical coverage performances of this primal-type framework for resource management issues in the MEC network \cite{xu2021crpo}. 

\item \textit{Projection-based RL}: A projection-based iterative algorithm is devised to confine the resource management policy with feasibility, cost, and fairness considerations. This iterative method, generally, comprises two steps: initially, we affect updates on the objective function, ensuring that the disparity between the preceding and succeeding strategies is less than a specified constraint value, devoid of considering the constraints of the resource management problem. Subsequently, we reconcile the constraint violation of this resource management problem by projecting the infeasible policy into constraint sets \cite{yang2020projection}. 

\item \textit{RL with Penalty Methods}: We introduce safety-augmented (Saute) MDPs to obliterate the constraints of the MEC system model. Saute MDPs incorporate constraints into state spaces, enabling new features and subsequently reshaping the objective function of a resource management problem. We verify that the Saute MDP model for resource management problems adheres to the Bellman equation. This approach, owing to its plug-and-play nature, can ``saute" any RL algorithm. Furthermore, the state augmentation method can conveniently generalize to constraints of various MEC environments \cite{sootla2022saute}. 
\end{itemize}

\textcolor{black}{Despite these advancements, each device may encompass distinct QoS prerequisites and constraints in a convoluted MEC network comprised of heterogeneous multi-connected devices. As a result, we amplify the complexity of safe policy optimization in the MEC network. 
\begin{itemize} 
\item \textit{Question 1: How might we devise a safety RL policy catered to diverse heterogeneous constraints for multi-device systems?}
\end{itemize}}

A viable solution entails designing unique Lagrangian multiplier update operators for disparate heterogeneous constraints. This specifically requires, initially, the analysis of dynamic characteristics of diverse constraints and the design of a particular Lagrangian update operator. Following this, the objective function is transformed into a Lagrangian function, an MDP model for this problem is designed, and then the resource management strategy is updated alongside the Lagrangian multipliers until the model converges. It's noteworthy that the resource management strategy updates at a more accelerated pace compared to the Lagrange multipliers' updates. 

Predominantly, safe RL algorithms focus on primal-dual strategies for single agents. Certain theoretical research establishes the regret boundary, delivers asymptotic convergence analyses, and achieves zero duality gap along with a finite-time convergence rate for general optimization problems. However, such convergence assurances and analyses typically pose significant challenges within MEC networks. 
\begin{itemize} 
\item \textit{Question 2: How might we develop a novel theoretical contribution concerning safe RL in MEC networks?}
\end{itemize}

For the MEC network resource management issue, we derive the theory that procures an optimal policy based on a safe RL strategy. Initially, the characteristics of the resource management problem are analyzed, and a mathematical model is established. Subsequently, constraints are incorporated into the objective function using Lagrange multipliers. The objective function is derived, the derivative of the function is set to zero, and the optimal strategy can be determined. The final step involves substituting the optimal value of the strategy into the objective function to ascertain the calculation formula for the Lagrangian multiplier.

\subsection{Large-scale Resource Scheduling}
An MEC network generally consists of a cloud node and multi-edge nodes, and each edge server provides services for multi-connected devices. The dimension of state and action spaces for resource scheduling problems increases as the number of servers and devices increases. This feature makes it difficult for the RL model to converge and find a flexible solution. 

\textcolor{black}{An efficient approach is to consider the MARL model, in which a multi-agent collaboratively interacts with a shared communication environment to complete resource scheduling by minimizing the delay or energy consumption with limited resources.} The MARL method needs to input an agent's observation instead of the states of all agents, reducing the network size. Given the vastness of state-action spaces, the latency in feedback and rewards, as well as the non-stationary and unobserved environments, effective information exchange amongst multi-agents is paramount to accomplishing consistent performance. Consequently, two prevailing strategies have been devised to navigate these obstacles.

\begin{itemize}

\item \textit{Construction Heuristics}: Construction heuristics engender a comprehensive resource scheduling solution by progressively appending solutions from individual servers or devices at each decision-making junction. Incremental solutions to the issue are obtained using a search algorithm and the values derived from RL. For instance, we utilize construction heuristics to tackle the resource offloading dilemma within MEC networks. Precisely, there are multiple users and servers in the MEC model, and tasks generated by each user can be offloaded to any server for processing. The initial step involves calculating the current moment, with each task that can be offloaded to disparate edge servers corresponding to a Q-value; the subsequent step employs the search algorithm to acquire the matching relationship between the task and the server \cite{xu2018large}.

\item \textit{Improvement Heuristics}: Improvement heuristics augment an initial scheduling solution through iterative searches and enhancements to solution quality. Typically, this model adopts a two-tier architecture. RL makes decisions at a higher level and is accountable for selecting operators, while operators execute specific resource scheduling at a lower level. For instance, we apply improvement heuristics to solve the task offloading issue in MEC networks. The first step involves randomly allocating a server to each different task for processing; the second step is to devise multiple distinctive operators, with each operator being an exchange rule for the matching relationship between tasks and servers; the third step establishes an MDP model, where the state contains the environmental information, and the action is the index of the operator; the fourth step executes the MDP model and enhances the existing solution until the model converges \cite{lu2019learning}.
\end{itemize}

\subsection{Generalization and Scalability of RL in MEC Networks}
For the application of edge RL algorithms, the generalization and scalability of the RL-based method necessary are considered first in MEC networks. Various MEC networks exist in real applications and differ in several ways. For example, the number and location of servers and users, available wireless resources, channel quality, user behavior, etc.
RL model training can cause a lot of time and resource consumption. A big challenge for the well-trained RL model is generalized to various MEC networks that may be an unseen communication environment. 

\textcolor{black}{Enhancing the generalizable and scalable for the deplored RL in edge servers, three perspectives can be considered: (1) Environment design: design different communication environments that guide RL training \cite{tobin2017domain}, (e.g., domain randomization, curriculum learning); (2) System learning: learn the various features of MEC network environments \cite{jouffe1998fuzzy}, (e.g., invariant feature discovery and causal inference); (3) Online adaptation: propose learning algorithms that can adapt fast to diverse network dynamic \cite{hedrick2022reinforcement } (e.g., online identification, meta-learning, and ensemble learning).}
\begin{itemize}
\item \textit{Environment Design}: Environment design encompasses the creation of distinct communication environments to facilitate RL training \cite{tobin2017domain} (e.g., domain randomization, curriculum learning). For instance, within RL methods, domain adaptation techniques are utilized to bolster the generalization of MEC caching models. The fundamental notion is to train RL models with diverse MEC caching system models (i.e., fluctuating numbers of users and cache units) to expose the model to a broad spectrum of MEC environments. Moreover, the principle of curriculum learning can be harnessed to amplify the generalization capability of MEC caching models. The key idea here is to increment the variability of the model progressively, thus enabling the RL model to learn in a phased manner. For instance, instead of augmenting the number of cache units by one at a time, it can be enhanced by two or more, allowing the model to generalize nonetheless. 

\item \textit{System Learning}: System learning focuses on acquiring various features of MEC network environments \cite{jouffe1998fuzzy} (e.g., invariant feature discovery and causal inference). Specific examples have been delineated in Section VI-A. Systematic learning not only diminishes the disparity between simulators and real-world application scenarios but also heightens the generalization and scalability of the model.

\item \textit{Online Adaptation }: Online adaptation recommends learning algorithms that can swiftly adapt to a diverse range of network dynamics \cite{hedrick2022reinforcement } (e.g., online identification, meta-learning, and ensemble learning). A simple methodology involves treating the variations in different models (i.e., fluctuating task sizes, task distributions, channel quality, number of users, server performance, and number of servers) as individual training sub-modules. When the model undergoes alterations, only the changing information needs to be inputted into the sub-module for training, while the hyperparameters of the main module can be reused without necessitating retraining, thereby significantly mitigating training costs.

\end{itemize}

\subsection{Security and Privacy in MEC Networks}
The increasing attention to security and privacy, including data and location privacy, signifies its importance in the current technological landscape. In essence, it entails the identification of potential security threats, the mitigation of these threats, overhead avoidance, and the design of adaptive security mechanisms. MEC transfers some computing tasks to edge servers, which consequently generate privacy-sensitive data. However, the training process of RL methods may necessitate data from other edge servers. Additionally, the transmission of the training data could result in data privacy disclosure. The aspects of collaborative security and distributed privacy protection against widespread attacks warrant further investigation.

\textcolor{black}{Privacy, in its distinct role within the realm of security, emphasizes the protection of users' personal data, which encompasses their location, identity information, and behavioral patterns.} Furthermore, privacy threats may stem from the following areas \cite{ranaweera2021survey}: (1) Authorized and Curious Adversaries: Certain legitimate users may engage in illicit activities, such as data theft, driven by personal interests or other motivations, consequently infringing on other users' privacy; (2) Computational Offloading/Caching: Attackers may monitor the computation offloading/caching process, consequently exposing user location information; (3) Service Migration: Eavesdroppers can easily track users to obtain private data when users access services that involve migration requirements. To tackle the aforementioned issues, we will examine methods to fortify privacy.

Federated learning can keep privacy for devices and only transmit network parameters (e.g., gradient information) instead of the raw training data for aggregation. The federated learning method improves the collaborative training of a DRL model in MEC networks. However, some characteristics of the MEC network, including large-scale edge servers and connected devices, dynamic networks, and time-vary demands, raise challenges of resource allocation and privacy protection in implementing federated learning.

\textcolor{black}{Federated learning is integrated into DRL to enable efficient and secure data transmission.} There are several open issues that warrant further consideration for resource management \cite{konevcny2016federated}:

\begin{itemize}

\item \textit{Participant Selection}: This issue pertains to the selection of devices that participate in training for parameter updates. Aggregating all devices in each round results in extensive training time. Implementing a novel participant selection mechanism can improve training efficiency\cite{wu2020bilateral};

\item \textit{Adaptive Aggregation}: This term refers to the dynamically adjusted aggregation frequency for the parameters of edge servers, balancing the tradeoff between training time and communication overhead. Generally, the majority of existing work employs a fixed interval for the aggregation of global parameters, which can lead to decreased training efficiency\cite{chen2023dap};

\item \textit{Incentive Mechanism}: This involves designing a compensation structure for aggregated edge servers to stimulate participation. In addition, the data of edge servers may be redundant due to information asymmetry among these participants. Thus, incentive mechanisms promote participation and reduce information asymmetry\cite{wang2019profit};

\item \textit{Structured Updates and Sketched Updates}: Structured updates involve directly learning parameter updates from restricted variables (e.g., random mask, low-rank), and sketched updates entail learning all model parameter updates and subsequently compressing them for transmission to the central server (e.g., subsampling, random rotations, and quantization)\cite{deng2020edge};

\item \textit{Convergence and Communication Complexity}: This refers to the measurement of the performance of the proposed federated learning methods through theoretical analysis. It is essential to explore the factors that affect the convergence rate \cite{feriani2021single}.

\end{itemize}

Intrusion detection serves as a pivotal approach, playing a crucial role in strengthening the resilience and integrity of network security. This process involves identifying attackers, unauthorized users, and security policy violations by analyzing log files, system and network data, and user behavior.

Intrusion detection methods are commonly categorized as either anomaly detection or signature detection based on their detection mechanism\cite{otoum2019empowering}. Anomaly detection compares data or behavior against a normal behavioral model to detect possible intrusions. In contrast, signature detection identifies known attack patterns and searches for similar ones. However, anomaly detection might yield false positives by misidentifying normal behavior as an intrusion, while signature detection cannot identify unfamiliar intrusion patterns.

\textcolor{black}{To overcome the limitations of anomaly detection, we can integrate RL with traditional methods capable of handling large-scale anomaly detection.} Hierarchical RL breaks down the anomaly detection process into multiple layers, each comprising distinct decision tasks, thus improving the detection of new anomalies\cite{pateria2021hierarchical}. Through interaction with the environment, each layer's decision-making strategy can be continuously updated and optimized, thereby enhancing the detection of new anomalies. The central concept of this method is to establish a hierarchical RL model. In this model, the RL agent chooses the operator of the anomaly detection scheme and uses traditional heuristics or rules for specific anomaly detection optimization.

\textcolor{black}{The integration of transfer learning and RL can enhance signature detection's ability to handle the increase in intrusion types.} The specific program steps include: Step 1 - Identifying the source and target domains, selecting the relevant or similar domain to facilitate the learning of similar features; Step 2 - Analysing the differences between domains, reflected in every element of the MDP; Step 3 - Evaluating transferable knowledge, such as expert trajectory, strategy, value function, and environmental model law, which assist in determining the similarity between source and target tasks; Step 4 - Selecting methods and transferring knowledge to apply the expertise gained from the original task to the target task, this includes reward design, example imitation, knowledge distillation, and representation transfer; Step 5 - Retraining the migrated model in the target domain; Step 6 - Testing and verifying the model.

\textcolor{black}{Contrary to traditional defense methods that react to attacks by implementing countermeasures post-detection, active defense proactively resists and may even counterattack the intruder.} Active defense can be enacted through three strategies: Firstly, by detecting and blocking malicious information to prevent network attacks, though this process can result in network performance degradation due to misjudgment. Secondly, by leveraging past attacks to anticipate potential attack patterns and adjust network defense strategies, although this might limit the defense system's scalability and its sensitivity to specific attacks. Lastly, by executing a shuffle operation to augment the system's unpredictability, eliminating malicious spies sent by attackers, and hence averting certain attacks entirely\cite{zhou2020endogenous}.

\textcolor{black}{To mitigate the issue of misjudgment, reward shaping can be employed to set more accurate rewards.} By meticulously designing and adjusting reward-shaping techniques, we can reduce the potential for misjudgment, leading to enhanced performance and decision accuracy across various domains and applications of RL. For instance, the reward obtained by the agent is bifurcated into intermediate and final rewards. The former refers to a continual guidance reward system, while the latter dictates the desired end state for AI.

\textcolor{black}{Additionally, transfer learning can be utilized to bolster the system's resilience against various types of attacks.} Robust RL can enhance the system's unpredictability by introducing uncertainty. This can be achieved through methods such as introducing noise and considering uncertainties in the reward function\cite{moos2022robust}. The specific steps include: Step 1 - Clearly defining and analyzing the agent's operating environment, identifying uncertain factors in the environment and their impact on the agent; Step 2 - Selecting the algorithm based on the scene and disturbance factors, which might involve information extraction, the addition of regular items, disturbance confrontation, and more; Step 3 - Training the RL agent according to the overall structure and methods employed; Step 4 - Analyzing algorithm robustness, for instance, handling noise and environmental changes; Step 5 - Model optimization and deployment.

\section{Conclusion}
Driven by the unprecedented surge in computational demands, the transition from cloud computing to MEC is facilitated by leveraging nearby computing and caching resources. As future communication technologies anticipate the implementation of edge AI, it presents significant opportunities for introducing RL into MEC to cater to the demanding requirements of ultra-low latency and energy-limited applications. However, the deployment of RL introduces challenges in network design, algorithm structure, and optimization attributable to the inherently dynamic and resource-constrained nature of the networks.

This paper presents a thorough survey of RL-based scheduling schemes for wireless communication, caching, and computing resources within MEC networks. Initially, it explores the motivations and challenges associated with MEC and why RL can emerge as a promising technology in the forthcoming era of mobile edge networks. Subsequently, it reviews and analyses RL-based approaches to offloading scheduling, content caching, and communication. It further elaborates on the advanced applications of RL in MEC networks, including industrial automation, autonomous driving, robotics, VR, AR, and others. Lastly, it discusses the opportunities and potential future directions of RL applications in MEC networks.

This comprehensive overview of RL-based MEC in this survey aims to serve as a valuable reference and guideline for further research and application in edge AI.

\newpage

 





\bibliographystyle{IEEEtran}

\end{document}